\documentclass[journal=jpcbfk,manuscript=article]{achemso}
\setkeys{acs}{articletitle = true}
\usepackage{amsmath, amsthm, amssymb, amsfonts}
\usepackage[sort&compress]{natbib}
\usepackage{natmove}
\usepackage{graphicx}
\usepackage{gensymb}
\usepackage{xcolor}
\usepackage{changes}

\author{D. Thirumalai}
\email{dave.thirumalai@gmail.com}
\affiliation{Department of Chemistry, The University of Texas, Austin, TX 78712, United States of America}
\author{Changbong Hyeon}
\email{hyeoncb@kias.re.kr}
\affiliation{Korea Institute for Advanced Study, Seoul 02455, Republic of Korea}
\author{Pavel I. Zhuravlev}
\affiliation{Biophysics Program, Institute for Physical Science and Technology and Department of Chemistry and Biochemistry, University of Maryland, College Park, MD 20742, United States of America}
\author{George H. Lorimer}
\affiliation{Biophysics Program, Institute for Physical Science and Technology and Department of Chemistry and Biochemistry, University of Maryland, College Park, MD 20742, United States of America}

\title{Symmetry, Rigidity, and Allosteric Signaling: From Monomeric Proteins to Molecular Machines}

\begin{document}

\begin{abstract}
Allosteric signaling in biological molecules, which may be viewed as specific action at a distance due to localized perturbation upon binding of ligands or changes in environmental cues, is pervasive in biology. Insightful phenomenological MWC and KNF models galvanized research in describing allosteric transitions for over five decades, and these models continue to be the basis for describing the mechanisms of allostery in a bewildering array of systems. However, understanding allosteric signaling and the associated dynamics between distinct allosteric states at the molecular level is challenging, and requires novel experiments complemented by computational studies. In this review, we first describe symmetry and rigidity as essential requirements for allosteric proteins or multisubunit structures. The general features, with MWC and KNF as two extreme scenarios, emerge when allosteric signaling is viewed from an energy landscape perspective. To go beyond the general theories, we describe computational tools that are either based solely on multiple sequences or their structures to predict the allostery wiring diagram. These methods could be used to predict the network of residues that carry allosteric signals.  Methods to obtain molecular insights into the dynamics of allosteric transitions are briefly mentioned. The utility of the methods is illustrated by applications to systems ranging from monomeric proteins in which there is little conformational change in the transition between two allosteric states to membrane bound G-protein coupled receptors, and multisubunit proteins. Finally, the role allostery plays in the functions of ATP-consuming molecular machines, bacterial chaperonin GroEL and molecular motors, is described. Although universal molecular principles governing allosteric signaling do not exist, we can draw the following general conclusions from a  survey of different systems. (1) Multiple pathways connecting allosteric states are highly heterogeneous. (2) Allosteric signaling is exquisitely sensitive to the specific architecture of the system, which implies that the capacity for allostery is encoded in the structure itself. (3) The mechanical modes that connect distinct allosteric states are robust to sequence variations.  (4) Extensive investigations of allostery in Hemoglobin and more recently GroEL, show that to a large extent a network of salt-bridge rearrangements serves as allosteric switches. In both these examples the dynamical changes in the allosteric switches are related to function.
\end{abstract}
	\maketitle

\section{Introduction}
 
Response to a local perturbation, such as binding of a ligand to an enzyme, which is sometimes amplified at larger length scales for functional purposes, is common in biology. 
Similarly, perturbations that trigger responses involving multiple enzymes through signaling networks enable cells to respond to environmental stresses. These are examples of allosteric signaling that occur on diverse length and time scales. A well known and well studied example of control at the molecular level in the structural transitions, triggered by oxygen binding to Hemoglobin, is the sequential model, first considered by Pauling \cite{Pauling35PNAS}. Much of the investigations on allosteric  (a term first introduced in 1961~\cite{Monod61ColdSpringHarb}) signaling has been on molecular systems,  which were galvanized principally with the publication of the seminal paper by Monod, Wyman, and Changeux (MWC) \cite{Monod65JMB} over fifty years ago. 
Subsequently,  the induced fit mechanism, proposed by Koshland, Nemethy, and Filmer \cite{Koshland:1966p3058}, has also provided fundamental insights into the kinetics of allosteric transitions in a number of systems. The quaternary structural changes accompanying oxygen binding to hemoglobin was elegantly described using the MWC model. Refinement and reexamination of  the accuracy of the MWC model, inspired by increasingly precise measurements on carbon monoxide binding to hemoglobin, continues to be a topic of abiding interest\cite{Viappiani14PNAS,Eaton1999NSB}. Although the basic ideas of allostery in the MWC are no doubt important, experiments and increasingly computer simulations reveal nuanced views of allosteric signaling including the applicability of the so-called induced-fit mechanism \cite{Koshland:1966p3058}.  

Although it is often discussed in terms of propagation of disturbance, leading to signaling at the molecular scale, allosteric transitions also occur on the cellular length scale, which was already described in another study launching the field of system biology \cite{Monod63JMB}. On the $\sim \mu m$ length scale, cells process information about their environment and transmit the signals by amplification by stimulating receptors on the cell surface through elaborate pathways that are frequently accompanied by post-translational modification. 
An example is phosphorylation of protein kinases whose activation results in the modification of downstream targets. 
In this instance, signaling occurs through a cascade of activation and deactivation reactions  in noisy environments with fidelity \cite{Thattai02BJ,Heinrich02MC}.
The molecular level response to ligand binding resulting in control at the cellular level are also examples of allostery, which broadly refers to a global response of  proteins or complexes to external perturbations.

The most common example of allostery involves conformational change of protein or multi-protein conformations upon binding of a ligand.
Another is allosteric regulation, when binding of a ligand influences the catalytic activity of an active site, located far from the binding site. These phenomena imply signal propagation across the protein molecule, which we refer to as allosteric signaling. 
A more subtle example of allosteric signaling is a change in catalytic activity upon mutation of a spatially distant residue, even though the wild type protein does not manifest allosteric properties  \cite{Datta:2008p4543,Hilser:1998p4276,Benkovic:2003p2946,Gianni:2006p4098,Zhuravleva:2007p2904,Lenaerts:2008p3539}. 

From a certain perspective, molecular allosteric activation has similarities to the propagation of perturbations across an infinite system that is well-known in crystals. For example, the elementary collective excitations (phonons), which  propagate throughout  the sample of a crystal, can be computed as vibrational modes of a solid. Although in these situations the theory is well understood, there are difficulties in applying them in a straightforward manner to biological systems.  Proteins (and their complexes) are  finite and heterogeneous objects, which makes it difficult to generalize the quasi-particle (phonons) picture easily.  Just like in crystals, allosteric propagation in biological molecules can also be thought of as excitations, which spread across the entire complex in an anisotropic manner in order  to  execute specific functions. Because of the link between structural responses and function, the problem of allostery continues to attract a great deal of attention \cite{Monod65JMB,Nussinov16ChemRev,Changeux13NRMCB,Cui08ProtSci,Tsai14PlosCompBiol,Horovitz16ChemRev}.

The objectives of this review  are: (1) We first describe the essential requirements for allosteric signaling, which are independent of  the precise allosteric mechanism operative in specific systems. We suggest that allosteric systems, whether they are monomers,  multi-domain proteins, or genuine energy consuming molecular machines, must have approximate symmetry (aperiodic arrangement of atoms in the language of Schr{\"o}dinger \cite{schrodinger1943life}) in the functional state. The presence of symmetry or approximate symmetry implies that at least a portion of the complex must be  rigid \cite{Anderson:1997} but not overly so, which we define as the ability of these elements to resist mechanical (at least modest) force. Absent this requirement signal propagation cannot occur. In the next section, we further elaborate on this idea by adopting concepts in the study of stress bearing capacity in crystals. (2) We then apply this notion to answer a few questions: Can we decipher the allosteric wiring diagram (AWD), defined as a network of residues that are responsible for signaling, and if so, can they be determined from either the sequences or structures for systems of interest? We review some of the computational techniques used to determine the molecular details of AWD in general allosteric system. (3) We then describe methods used to simulate the dynamics of  transitions between two allosteric states, which are nominally the {\it apo} and the ligand-bound states. (4) The utility of these concepts and methods are illustrated using a few examples, covering monomeric and multisubunit proteins, as well as ATP-consuming machines. The perspective is concluded with an outlook on some of the outstanding issues in this field.   

Two comments about the terminology used here are worth making. (1) Historically, the term allosteric signaling (or regulation) was used to describe the consequences of binding an effector molecule to an enzyme that is distant from the active site. However, over the years (including in the MWC model) the term allostery has been used to characterize the response in a protein in regions far from the site to which a ligand binds.  In this expanded view, conformational changes that occur in a monomeric protein driven by ligand binding would also be an example of allosteric signaling. In this review we adopt the generalized definition of allostery. (2)  It is customary to use  the term allosteric states, $T$ (taut or tense) and $R$ (relaxed), nominally for quaternary structures in allosteric systems. For example, the differences in oxygenated ($R$) and deoxygenated ($T$) quaternary structures in Hemoglobin are due to oxygen binding to the heme group. Of course, the $R$ and $T$ states might be populated even in the absence of oxygen in Hemoglobin, which is written as $T_0 \rightleftharpoons R_0$ in the MWC model (the subscript $0$ refers to absence of oxygen). The extent to which the $R_0$ state is populated depends on the system and the external conditions. Here, we refer to $T$ and $R$ as allosteric states, regardless of whether we are describing tertiary or quaternary structural changes. With this terminology, we describe ligand-induced conformational changes monomeric proteins as allosteric transitions, even when the structural differences between the two end states are negligible.

\section*{Symmetry and Rigidity requirements}
Allosteric signaling could be pictured as arising from propagation of local strain generated by ligand binding \cite{Zheng05Structure,Dutta18PNAS,Mitchell16PNAS}. The resulting strain might be thought of as inducing excitations around one of the allosteric states. In analogy with problems dealing with propagation of excitations in ordered solids, we postulate that allosteric proteins must satisfy two requirements. (1) Certain regions of allosteric proteins must be stiff, and hence resist mechanical force. More precisely, the network of residues that transmits allosteric signals must be capable of bearing ligand-induced  strain over almost the whole complex. The need for this requirement can be explained using an analogy to the transmission of local disturbance in regular solids with long-ranged translational order. In solids vibrations of atoms in lattice sites are propagated throughout the sample by phonons, which are the elementary excitations. The propagation of excitation is possible  because of the stiffness or rigidity of the solid with long-range translational order.  Stress  propagation cannot occur in dense liquids, which have only have short range order. (2) The requirement that there be rigid regions in a biological complex implies that the allosteric states must have lower symmetry than the disordered regions, permitting them  to transmit signals across the complex.  In solids translational symmetry is broken, thus lowering the symmetry with respect to liquids. As a consequence,  the ordered state is described by elastic constants. In the same vein, in finite-sized biological complexes the structural parts  must accommodate excitations (aptly termed ``allosterons" by McLeish \cite{Mcleish2018PTRB}) across the length scale of the structure. Therefore, at least a portion of the allosteric protein must be structured.  Unlike in periodic solids in which phonons propagate isotropically, this is not so in allosteric proteins. Here, allosteric signaling occurs, in a directed manner, reflecting the architecture of the protein or the complex, which already suggests that the allosteric mechanism at the molecular level is system dependent. The arrangements of strands and helices connected by loops are suggestive that the ordered states are like "nematic droplets" in which strain propagation is necessarily anisotropic. Because the architecture of the biological molecules that are capable of resisting stress vary greatly, it is logical that the molecular mechanisms of allosteric signaling also vary substantially. The few examples provided here attest to the diversity of molecular mechanisms in allosteric signaling.   (3) There is a major difference between ordered solids and allosteric proteins. In the former, the relative positions of atoms and the unit cells are preserved. However, in proteins the relative positions of the strain-resisting structural elements change in response to binding of ligands. Nature has evolved structures in which movements of ordered regions could be accommodated through the motions of hinges and flexible loops \cite{Papaleo16ChemRev} so that the distinct allosteric states could be accessed one from the other.

\begin{figure}[h!]
\includegraphics[width=0.6\textwidth]{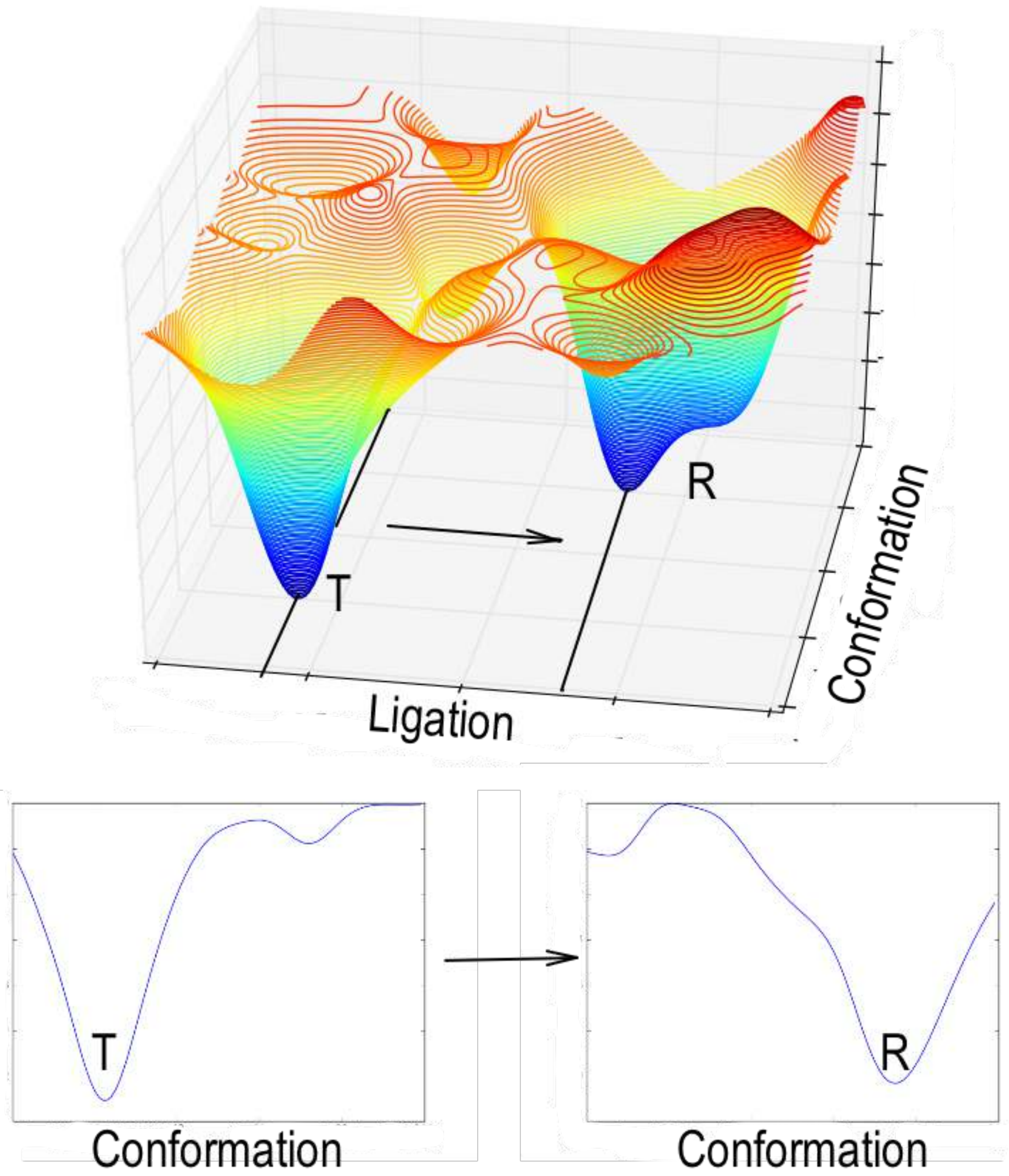}
\caption{The joint energy landscape for ligand binding and conformational transition between the allosteric states $T$ and $R$  is shown schematically. The free energy gap between the $T$ and $R$ states determines the mechanism of allosteric signaling.  If ligand binding is rapid compared to conformational transitions, which is often the case, then it is possible to use the two disconnected basins of attraction corresponding to the $T$ and $R$ states to predict the molecular basis of allostery. The lower figure is a one dimensional projection. 
\label{fig:landscape}}
\end{figure}

\section*{Scenarios for Allostery}
Since the beginning, two extreme scenarios for allosteric transitions have been envisioned. 
(1) Preexisting equilibrium between the allosteric states, or population shift or conformational selection (CS) mechanism assumes that the protein switches between the two allosteric states ($T$ and $R$)  on biological (or laboratory) timescale even in the absence of the ligand.  One of the states has a lower free energy (Fig.~\ref{fig:landscape}). In this scenario, the ligand binds preferentially to state $R$, which is populated even in the absence of the allosteric effector. 
Recall that in the MWC model (for a simple derivation using the grand canonical partition function see Hess and Szabo \cite{Hess79JChemEd}) the $T$ and $R$ states are in equilibrium even the absence of oxygen, which is the very tenet of the CS mechanism.
Upon binding of the ligand, the $R$ state becomes lower in free energy, and the observed conformational change is  manifestation of the population shift between the two states\cite{Monod65JMB}.  
This picture is natural, since proteins are not static, but exist in an ensemble of interconverting conformations ~\cite{McCammon1977b,fraunfelder,Henzler-Wildman2007a,femimore,Honeycutt1992}.  
(2) The induced fit \cite{Koshland:1966p3058}, or reaction front \cite{Zhuravlev:2010p11378} mechanism is precisely the opposite: the conformation changes as the perturbation (consequence of ligand binding to state $T$) spreads from the initial point (ligand binding site) through the rest of the protein \cite{Koshland:1966p3058}, but the $R$ state is typically not visited in the absence of the ligand. Pauling \cite{Pauling36PNAS} alluded to the induced fit  picture in his attempt to understand cooperative binding of oxygen to hemoglobin.

The two extreme scenarios can be visualized using an energy landscape picture (Fig.~\ref{fig:landscape}). If the free energy gap $\Delta G_{TR}(=G_R-G_T)$ between the two allosteric states $T$ and $R$ is large, then it is unlikely that the $R$ state (ligand bound) can be sampled within biological timescales in the absence of a ligand.
In the opposite limit, the $R$ state could be readily accessed even in the absence of ligand binding. Several NMR experiments show that typically only a small percent ($ \lesssim 5\%$) of the ligand-bound $R$ state (see for example \cite{boehr2009NCB}) is populated in the absence of a ligand.
The CS mechanism assumes that the protein switches between the two allosteric states on a biological (or laboratory) timescale even in the absence of the ligand, but one (the {\it apo}) of the states ($R$) has a lower free energy (Fig.~\ref{fig:landscape}) at zero ligand concentration. 
Upon binding of the ligand, the $R$ state becomes lower in free energy, and the observed conformational change is  a manifestation of the population shift between the two states~\cite{Monod65JMB,boehr2009NCB,Hammes09PNAS}. 
The population shift hypothesis implies that the structural elements that bear allosteric signals are encoded in the {\it apo} state, and could in principle be determined from the structures in the absence of the ligand. However, understanding the dynamical changes in the conformation of the {\it apo} state during the allosteric transition requires knowledge of the barrier (not unrelated  to the free energy gap between the two states), and the on and off rates of ligand binding to the $T$ and $R$ states \cite{Hammes09PNAS}.  

\begin{figure}[t]
\includegraphics[width=0.5\textwidth]{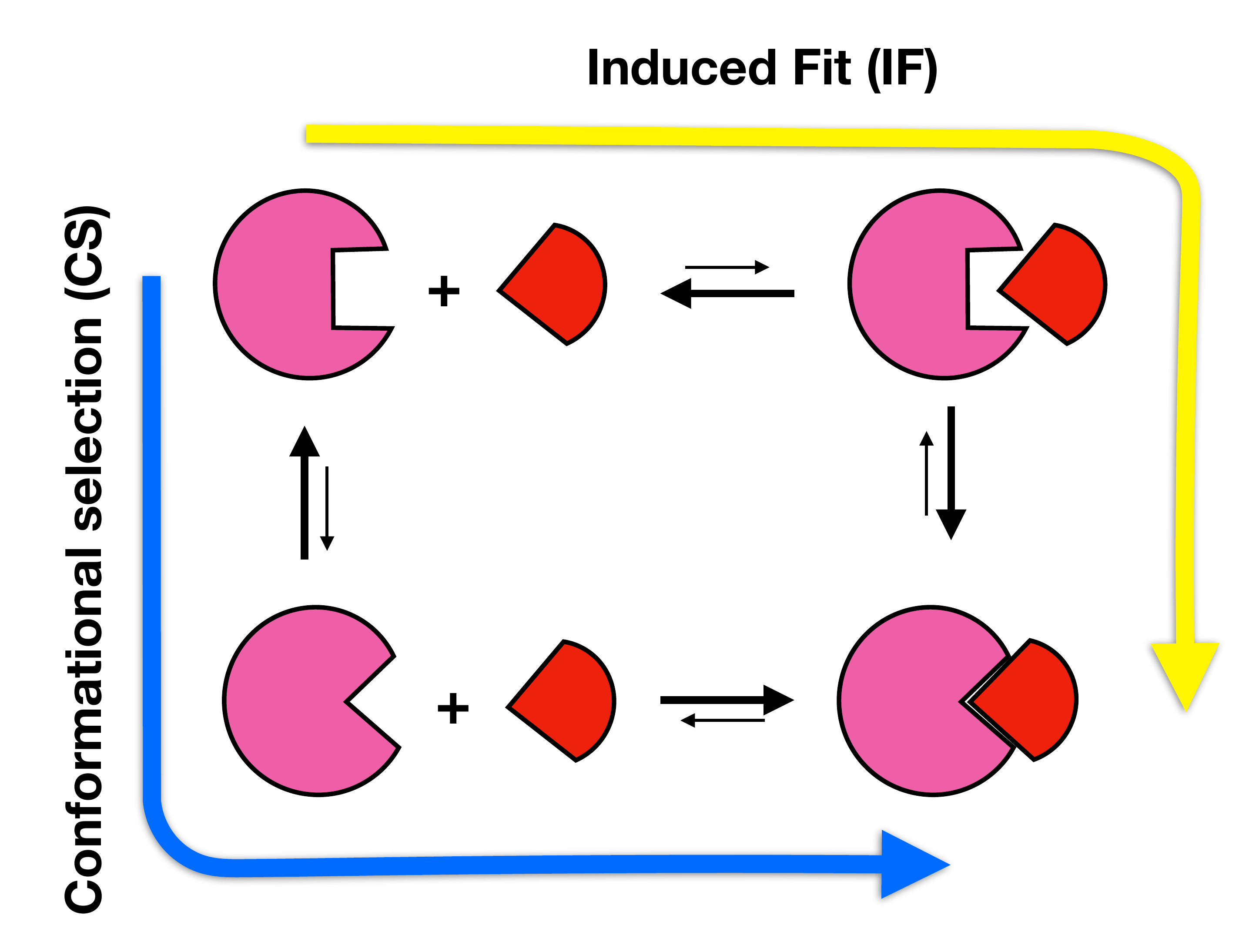}
\caption{Schematic of the two extreme scenarios for allosteric transitions. The conformational selection (CS) picture is indicated in the lower part and the induced fit (IF) is displayed on the top. The allosteric protein is in pink and the ligand is in red. The figure, drawn based on the theory in \cite{Hammes09PNAS}, is meant to show that both IF and CS mechanisms might be present in the same enzyme depending on the ligand concentration and free energy difference (Fig. \ref{fig:landscape}) between the two allosteric states (see Eq. \ref{CSIF})}
\label{HammesFig}
\end{figure}

The induced fit likely holds if $\Delta G_{TR}/k_BT$ ($k_B$ is the Boltzmann constant) is large enough that the population of the ligand-bound $R$ state is negligible. 
The conformation of the protein is forced to change as the perturbation spreads from the ligand bound region through the rest of the protein\cite{Koshland66Biochem}.  The discussion in terms of the $\Delta G_{TR}/k_BT$  makes clear that it is the flux between the two major pathways  that distinguishes between CS and IF mechanisms, as was described in a most insightful article by Hammes, Chang, and Oas\cite{Hammes09PNAS}. Their arguments clearly show  that both the CS and IF scenarios could  simultaneously exist in a single enzyme depending on the magnitude of $\Delta G_{TR}/k_BT$ (Fig. \ref{HammesFig}), and the second order $k_{on}^T$ (rate of ligand binding to the $T$ state), and $k_{on}^R$, the corresponding on rate for the $R$ state. In the limit of small ligand concentration, the ratio of the fluxes $J_{CS}/J_{IF}$) through the CS and IF pathways is, 
\begin{equation}
\frac{J_{CS}}{J_{IF}} \approx \frac{k_{on}^R}{k_{on}^T} \exp{(-\Delta G_{TR}/k_BT)}.
\label{CSIF}
\end{equation}
The above equation assumes that the reverse ligand unbinding reaction rates from both the $T$ and $R$ states are small.  

The CS and IF scenarios described above could be illustrated using allosteric inhibition in the gene regulatory protein, catabolite activator protein (CAP) consisting of a  cyclic guanosine monophosphate (cGMP) domain that is coupled to a DNA binding domain. Using NMR studies it was established \cite{tzeng2013NCB} that in the wild type (WT) CAP there is a dynamic equilibrium between the inactive and the active DNA binding states. The latter has $\approx$ 7\% population implying that a small free energy difference ($\Delta G_{TR}/k_BT = 2.58$) separates the two states.  However, binding of cGMP to a  double mutant (CAP$^*$) suppressed the DNA binding activity although it had no effect on the WT CAP.  Such a dramatic suppression of allosteric activity was interpreted in terms of loss of stability of the active state in CAP$^*$ that exists only at 7\% level in the WT CAP. From the perspective given above, we can surmise that the difference in the free energy, $\Delta G_{TR}/k_BT$  between the active  and inactive states is far greater than unity in the case of CAP$^*$. The inaccessibility of the active state on the time scales of milliseconds renders the CAP$^*$ protein inactive when bound to cGMP, which in turn suppresses allosteric activity. Comparison of the NMR structures of CAP and CAP$^*$ structures shows that a length of a helix \cite{tzeng2013NCB}, likely involved in signaling, increases by a half a turn, thus making it more rigid, which perhaps leads to an increase in $\Delta G_{TR}/k_BT$. This example suggests that for efficient allosteric signaling by the CS mechanism there may be an optimal value of $\Delta G_{TR}/k_BT$ -- an idea that warrants further investigation. From a structural perspective, it also follows that although rigidity is a requirement for allosteric proteins, the various load bearing structural elements that resist force cannot be too rigid. 
 \\
 
{\bf The Shifting Ensemble View:} The two extreme scenarios and the analysis of allosteric suppression in CAP$^*$ is simplistic because both the states $T$ and $R$ are ensembles of conformations, and the $T \rightarrow R$ transition likely occurs through a number of pathways (Fig.~\ref{fig:landscape}). 
Hence, the allosteric mechanism could be modulated by altering the free energies of the ensembles, as articulated recently by Hilser and coworkers \cite{Motlagh2014Nature} who view  the thermodynamics of allostery as arising from shifting nature of ensembles in response to effector binding.  
Such a picture naturally emphasizes the entropic contributions to allosteric signaling, as was first realized by Cooper and Dryden\cite{Cooper84EBJ}. The thermodynamic ensemble  view of allostery, which is an alternative to the  CS and the IF mechanisms, arises if the scenarios are described in terms of free energy changes due effector binding. 
In the ensemble view, entropic contributions to the thermodynamics of allostery is emphasized \cite{Motlagh2014Nature}.  In practice, NMR experiments could be used to assess the importance of  conformational entropy \cite{whitley2009CPPS,wang2005JACS,Law2009JACS,tzeng2013NCB}. Contribution to conformational entropy arises due to several types  of motion, including rigid body movements in a multi-domain protein, side-chain dynamics  in PDZ domain allostery \cite{fuentes2004JMB,Law2009JACS}, backbone dynamics like in CAP \cite{tzeng2012nature,tzeng2013NCB}, or even folding/unfolding of a part or the whole protein upon the ligand binding \cite{wu2008BJ,roca2008pnas,ferreon2013Nature,garcia2010Cell,sevcsik2011JACS,Hyeon06PNAS}. In some of these cases, allostery may be purely dynamical, that is, the response of the molecule to ligand binding, or mutation, would manifest itself not as a change in the structures between two allosteric states, but in the change in the \textit{dynamics} of certain parts of the molecules (see the example of ligand-induced transition in the enzyme, Dihydrofolate Reductase discussed below). 

Using the thermodynamic treatment of ensemble of the relevant states of the system (protein(s) + ligand(s)), it is possible to anticipate a variety of possibilities. In this approach, the states of every object (protein, protein domain or ligand molecule) in the system has a free energy, and if binding/dimerization takes place, there is also a free energy of the resulting interaction added to the mix, for each pair of interacting partners.  The presence of multiple free energy scales elicits distinct behavior.  	

\begin{figure}[t]
\centering
 \includegraphics[width=0.9\textwidth]{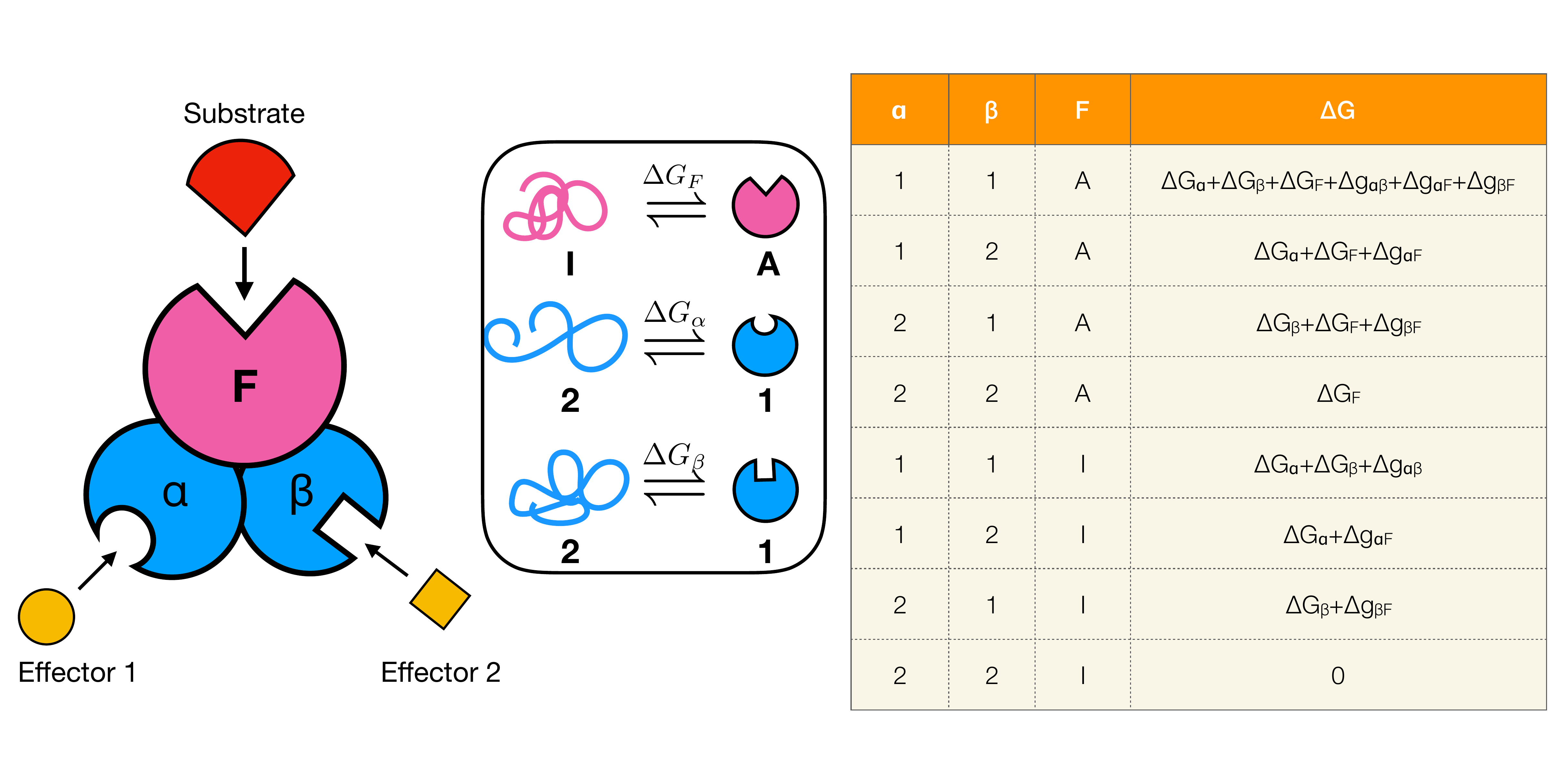}
   \caption{{\bf Allostery of trimeric enzyme.} 
   The allostery in a trimeric enzyme \cite{Motlagh2014Nature}, consisting of the functional domain ($F$) with two other domains ($\alpha$ and $\beta$) to which the Effectors 1 and 2 can bind. Each domain may transition from disordered to ordered state (see the sketch in the box in the middle).  The $F$ domain switches between the $I$ (inactive) and $A$ (active) with the free energy being $\Delta G_F$. The $\alpha$ and $\beta$ domains exist in state 2 (inactive) and state 1 (active) with free energy changes $\Delta G_{\alpha}$, and $\Delta G_{\beta}$, respectively. 
  Effector 1 and Effector 2 bind specifically to the ordered form of $\alpha$ and $\beta$ domains. 
   The Table on the right gives the values of the total free energy changes for all the eight states for the enzyme. 
   The inter-domain interaction free energy is given by $\Delta g_{XY}$, where $XY$ could be either $\alpha\beta$, $\alpha F$, or  $\beta F$. The scenarios for allostery emerging for the trimeric enzyme are discussed in the text. 
\label{trimer}}
\end{figure}

For instance, if $R$ is the active (referred to as $A$ in Fig. \ref{trimer}) state, $T$ is the inactive ($I$ in Fig. \ref{trimer}) state, then ligand binding could shift the equilibrium  $A \rightleftharpoons I$. If $I$ dominates the ensemble before ligand binding, and the equilibrium shifts towards $A$ upon binding, there would be a measurable change in both the activity and the conformation.  
On the other hand, if the $A$ state is easily accessible in the absence of the ligand (Fig.~\ref{fig:landscape}), then the conformational change may not be detectable, but there would still be a change in activity.  The largest change in activity, with almost no change in conformational entropy, would be in the case when $I$ is just slightly more populated before the binding event. In this situation,  $\Delta G_{TR}/k_BT$ is expected to be small, as in the CAP example. Upon ligand binding, $A$ becomes more populated. This is the CS limit. The two behaviors are obviously distinct.  However, the mechanism is the same. The only variable that changes  is the free energy difference $\Delta G_{IA}^\text{apo}$ between the active and the inactive states in the absence of ligands.
	
Consider the case of  two, or three interacting domains and subunits, each with distinct binding sites (Fig. \ref{trimer}).  Statistical mechanical analyses of this commonly occurring situation reveals fascinating behavior, such as a switch from positive to negative cooperativity due to effector binding \cite{Motlagh2014Nature}. Let us imagine a multisubunit protein with three binding pockets (Fig. \ref{trimer}). We assume that one (domain $F$) binds a substrate in a functional state, and the other two (domains $\alpha$ and $\beta$) bind ligands (effectors) without chemically altering them. Each domain could have two conformations, with free energies, $\Delta G_{F}$, $\Delta G_{\alpha}$, and $\Delta G_{\beta}$ associated with the conformational transition. The  activity may be measured when domain $F$ reaches the functional state.  The domains $\alpha$ and $\beta$ are assumed to be in conformation ``1'' (active) or ``2'' (inactive), and $F$ could be in $A$ (active) or $I$ (inactive) state. The eight possible states for the $\alpha\beta F$ trimer are: 11A, 12A, 21A, 11I, 12I, 21I, 22A, 22I.
If $\Delta g_{\alpha\beta}$,$\Delta g_{\alpha F}$ and $\Delta g_{\beta F}$ are the interaction free energies between the domains when they are in conformations 1 or A, then it is straightforward to calculate the Boltzmann statistical weights for each state listed above\cite{Motlagh2014Nature,Viappiani2014PNAS}. For example, if the free energy of 22I is 0 then the free energy of the state 12A would be $\Delta G_{\alpha}+\Delta G_{F}+\Delta g_{\alpha F}$, and for 11I it is $\Delta G_{\alpha}+\Delta g_{\alpha\beta}$. The observed activity will be proportional to the sum statistical weights of the conformations with domain $F$ in state A: 11A, 12A, 21A, 22A (see Fig.~\ref{trimer}). 
	
This simple picture  leads to a few possibilities, all dictated by the relative statistical weights of the various states. One can imagine a situation, where without any ligands, the activity is small because the states 11A,12A,21A and 22A have a low statistical weights. However, binding of the effector 1 (to $\alpha$ domain) could shift the statistical weights (by changing $\Delta G_{\alpha}$), such that the enzyme becomes active. In this case effector 1 is an activator.
Effector 2 binding to $\beta$ domain (in the absence of effector 1) may also increase the weight of states with the functional domain in the active form. However,  if effector 1 binds after that, it may decrease them again, playing in this case the role of repressor. Thus, effector 1 switches from being an activator to being a repressor due to binding of the effector 2. In addition, changing temperature could also influence the free energy of domain conformational transitions and the interaction free energies, thus triggering a switch in the binding behavior (e.g., from positive to negative cooperativity). Thus, the shifting ensemble view of allostery allows for a  description of complicated dynamical behavior in a unified framework. However, for practical applications, one needs to know all the free energy parameters in the model.

Although a number of global observations on a variety of unrelated systems could be rationalized using the picture of shifting ensembles in response to different environmental cues, such as ligand binding, or post-translational modifications, a molecular description requires a combination of experiments and computations that probe the dynamics of transition between the multiple states that affect allosteric signaling and function.

\section{Methods for computing the Allostery Wiring Diagram (AWD)}

In the last couple of decades, several computational methods have been introduced to capture the network of residues, here referred to as AWD, that is most dominant in facilitating allosteric signaling. Among these include the purely sequence-based Statistical Coupling Analysis (SCA) method \cite{Lockless99Science} and the related Direct Coupling Analysis.\cite{Weigt09PNAS} Most of the other methods rely on the structures of the relevant systems as a starting point to predict the AWD \cite{Zheng05Structure}. We should point out that the precise connections between the predictions using theory and computations  and the experimental outcomes are often difficult to make because isolating allosteric activities from the usual thermal fluctuations, and perhaps other factors,  requires precise experiments.  

\subsection{Sequence based Methods} 
There has been considerable interest in deciphering co-evolution of residues in protein families.\cite{Neher94PNAS,Fodor04JBC} This has led to the development of many methods based on statistical analysis of sequences. In the context of allosteric signaling these methods could be useful only if it is established that the co-evolving residues also drive allosteric transitions. However, to our knowledge this has not been explicitly shown in computations or experiments. Nevertheless, sequence analysis in combination with structural information is likely to be increasingly important in predicting the AWD. 

In a series of interesting papers, Ranganathan and coworkers \cite{Lockless99Science, Hatley03PNAS, Suel2002NSMB} introduced the statistical coupling analysis (SCA) method to identify the relevant energetically coupled coevolving residues. The first step in this method is to create a family of related sequences  using multiple sequence alignment (MSA) algorithm. The basic premise of the SCA method is that the coevolution of positions, either for structural or functional reasons, could  be captured by comparing the statistical properties of amino acids in the full MSA and its statistically significant sub-alignment created from the MSA.  This is typically expressed in terms of a SCA matrix, whose elements are computed using the probabilities of observing a particular amino acid in the sub-alignment and in the full MSA. The sub-alignment is created by constraining  the probability, $p_{j,S}^x$ ($S$ stands for sub-alignment) of observing one type of amino acid ($x$) at position $j$ to be unity. This probability is compared with the probability of observing amino acid $x$ at position $i$ ($i \ne j$) in the MSA. By analyzing a pseudo free energy matrix in terms of $p_i^x$ and $p_{j,S}^x$, the energetic coupling between sites $i$ and $j$ could be calculated \cite{Lockless99Science}. The energetically coupled residues are deemed to be coevolving and are associated with protein "sectors", which were postulated to be functionally significant \cite{halabi2009cell}. 
Although this method is doubtless interesting, its usefulness has been questioned in several studies \cite{Fodor04JBC,Liu09Proteins,Chi08PNAS} for the following reasons. 
(1) From a technical view point, any meaningful information about such  functionally relevant sectors, perhaps even in allosteric enzymes, requires a large number of sequences in the MSA as well as in the sub-alignments \cite{Dima06ProtSci}. In addition, SCA cannot be applied in instances in which a particular residue is fully conserved in the fMSA, which is the case in the G-protein coupled receptors discussed below. 
(2) It appears that much of the information from the SCA could be extracted using standard sequence conservation, perhaps allowing for the chemical identity of amino acid residues in the calculation of sequence entropy \cite{Dima06ProtSci,Tecsileanu15PLoSComp}. 
(3) It is likely that coevolution might involve variations between three or more residues, which are simply difficult to capture by sequence gazing alone because the number of sequences in the MSA are not usually very large. Despite these limitations, SCA might be a useful starting point to infer some aspects of relevant allosteric residues \cite{McLaighlin12Nature}. 
(4) There is no theoretical or physical basis for SCA, which makes it difficult to assess its strengths and weaknesses independent of experiments. 

\subsection{Structural Perturbation Method (SPM)}
As explained above, the physical basis for determining the AWD using the SPM is inspired by concepts in solid state physics. In a crystal lattice, the propagation of excitation is only possible because of rigidity in the crystalline phase, described using appropriate elastic constants. There are two crucial differences between crystals and biological molecules with well-defined structures. 
(1) In contrast to crystals, biological molecules, including large oligomeric structures, are finite. As a result  the analogue of phonons and dispersion relations are hard to define in the strictest sense (However, the fractal nature of the low frequency density of states in proteins \cite{Elber86PRL}, and the Lindemann criterion to describe solid and fluid-like behavior in different parts of the structure of proteins \cite{Zhou99JMB} were studied using solid state physics concepts).  
Nevertheless, the insights obtained using the low lying normal modes into the dynamics of proteins obtained using Elastic Network Models (ENMs) suggest that such concepts might be adopted in the study of allostery. 
(2) More importantly, although folded proteins can be thought of as  "nematic droplets"  the absence of quantifiable symmetry breaking makes it difficult to express their pliant nature quantitatively.  The persistence length and the resistance to mechanical forces needed to disrupt the allosteric structures are the only measures of rigidity in proteins. 

Despite these caveats, it is possible to computationally obtain the AWD using the analogy to phonons in crystals \cite{Zheng05Structure}.
Imagine a perturbation at a particular residue, which can be realized by mutations of residues or binding of ligands or mechanical force.
The consequences of such a perturbation do not (typically) propagate uniformly but do so in an anisotropic manner \cite{BaharBJ01}. In other words, there are certain residues or structural elements that are affected to a greater extent than others.
The network of residues across the enzyme, which carries the excitations due to local perturbations, is the AWD. It should be emphasized that changes in the conformations during the allosteric transitions are likely to be heterogeneous  even if they occur through a reasonably well-defined AWD. 

The analogy to phonon propagation in solids was used to propose the SPM by Zheng and coworkers \cite{Zheng05Structure,Zheng06PNAS} in order to determine  the AWD. 
Related ideas in biophysics and other areas may be found elsewhere \cite{Kumar15BJ,McLeish15BJ,Townsend15JBC,Flechsig17BJ}. The same or variants of the SPM have been used as a computational tool for the design of materials with specific responses \cite{Yan17PNAS,Rocks17PNAS}. The first step in the SPM  is the representation of the structure of a given allosteric state (for example the $T$ state) as an elastic network of connected springs. 
In the ENM\cite{TirionPRL96,Bahar05COSB,Bahar2010ChemRev,Tama06ARBB}, the structure is represented as a contact map, which is usually computed from the coordinates  of the C$_{\alpha}$ atoms of each residue in a given structure.  A contact between two residues implies that the distance between their  C$_\alpha$ carbon atoms, whose sequence separation exceeds a minimum number,  is less than a cutoff distance, $R_c$. In certain applications, described below, generalization of the standard ENM \cite{Bahar2007COSB,Bahar10ARB} is used by representing each residue by two beads \cite{Tehver08JMB}, one representing the C$_{\alpha}$ carbon and the other the center of mass  of the side chain (SC)  with Glycine being an exception. The center of mass is determined using the side-chain heavy atoms. For Gly, only the C$_{\alpha}$ carbon atom is used. It is worth pointing out that ENM has been proven to be useful in  structure refinement \cite{Tama04JSB,Schroeder14Acta,Schroeder07Structure}.  

Following the insightful studies by Bahar and coworkers, who pioneered the applications of ENM and its variants to a variety of systems \cite{Bahar05COSB,BaharPRL97,BaharBJ01,Bahar2010ChemRev}, a harmonic potential is imposed between all the interaction sites (C$_{\alpha}$ carbon atoms and/or the SCs) that are within $R_c$ in the given allosteric structure. The potential energy in the ENM is,
\begin{equation}
E_S = \frac{1}{2} \sum_{\substack{i,j \\ d_{ij}^0 < R_c}} \kappa_{ij} (d_{ij} - d_{ij}^0)^2,
\label{ENM}
\end{equation}
where $d_{ij}$  is the distance between the interaction centers $i$ and $j$, $d_{ij}^0$ is the corresponding distance in the starting allosteric structure, and $\kappa_{ij}$ is the spring constant. The sum is over all the pairs of sites that are in contact ($d_{ij} < R_c$). The value of $R_c$ is chosen  to ensure  that the B-factors calculated using Eq. \ref{ENM} and the experimentally measured values are as close as possible \cite{Zheng07BJ,Tehver08JMB}.  The residue-dependent spring constants, $\kappa_{ij}$, are chosen  to reflect the physical properties of the protein under consideration. The simple function $E_S$ potential given in Eq.\ref{ENM}, reflecting the contact map of the system, can be readily used to calculate the spectrum of normal modes numerically.
The associated eigenvalues and eigenvectors are used to calculate the overlap reflecting the conformational changes between two states with known structures \cite{zheng03PNAS}.

The AWD for a given structure is obtained using the response of the normal modes to local perturbations.
Consider the effect of a point mutation of a residue $k$ at location $n$ on a functionally relevant mode $M$, identified as having a large overlap \cite{Zheng05Structure,zheng03PNAS} 
between two allosteric states. For such a perturbation the response of the residues in contact with $k$ is calculated using,
\begin{equation}
	\delta\omega(M,n) = \nu_M^T\cdot \delta H \cdot \nu_M,
	\label{SPM}
\end{equation}
where $\nu_M$ is the eigenvector of the mode $M$. The Hessian matrix $\delta H$, arising from perturbation of $E_S$ (Eq. \ref{ENM}),	is
\begin{equation}
	\delta H = \frac{1}{2}\sum\delta \kappa(d_{nj}-d^0_{nj})^2,
\end{equation}
where $\delta \kappa$ is a small perturbation associated with residue $k$. 
The magnitude of the response, $\delta\omega(M,n)$, is proportional to the contribution to the elastic energy of mode $M$ arising from springs that are connected to $k$. 
The high $\delta\omega(M,n)$ values correspond to residues that play functionally important roles in the allosteric motions.
The SPM , which may be implemented using any energy function besides Eq.\ref{ENM}, has been successfully used to identify the AWDs in a number systems such as DNA polymerase\cite{Zheng05Structure}, helicases \cite{Zheng12BJ}, molecular motors \cite{Tehver10Structure,Zheng12JCP}, the bacterial chaperonin \cite{Zheng06PNAS}. 

There are several major advantages in using ENM to investigate allosteric transitions. 
(1) The simplicity of the ENM makes it easy to obtain the dynamical modes easily.
(2) It is found that for many large multisubunit complexes the observed conformational changes are dominated by only a few lowest ENM modes.  This is likely to be a consequence of the finding that domain movements in a large complex are robust. 
(3) Excursions along the vectors, which were used to probe the plausible existence of low lying excitations in soft glassy systems \cite{Rosenberg89JPCM} as well as proteins \cite{Miyashita03PNAS}, have given insights into the extent of conformational changes that occur during allosteric transitions.
(4) The SPM in conjunction with ENM  is a powerful method in the determination of the AWD, as we show below.

\subsection{Evolutionary interpretation and robustness}
In order to assess if the AWD predicted by the SPM, which reflects signals in the  dominant modes, is robust to sequence variations one has to incorporate evolutionary information into the local force constants $\delta C_{ij}$s.
We devised a way to include sequence information by computing $\delta C_{ij}$ using probabilistic arguments \cite{Zheng06PNAS}. Using the MSA it is possible to construct the statistical interaction for contact ($i,j$) due to interchange of a pair of residues in two homologous sequences in the MSA, provided the contact is conserved. A similar assumption is made in the Direct Coupling Analysis used to predict protein structures \cite{Weigt09PNAS,Schug09PNAS,morcos2014PNAS}. For such an exchange of the residues, the variation in the context-dependent force constant $\delta C_{ij}$ is computed using $\delta C_{ij} = \langle \delta E(\lambda_1,\lambda_2)\rangle$ where the average is over all pairs of sequences in the MSA and $\lambda_1$ and $\lambda_2$ are the residue pairs that are swapped between a sequence pair in the MSA. Mode robustness due to sequence variation is expressed using,
\begin{equation}
	f_{\delta E} = \delta E/E,
\end{equation}
where $\delta E = \sum\limits_{i,j}(\delta C_{ij}/C)E^M_{ij}$ and $E=\sum\limits_{i,j}E^M_{ij}$ is the total energy due to variation in $C$, where $C_{ij}$, the force constant for contact $(i,j)$, is taken to be a constant $C$ for all contacts ($C_{ij}=C$). 
Modes with the smallest values of $f_{\delta E}$ are most robust to sequence variations. Thus, SPM in conjunction with evolutionary information could be used to asses if AWD is encoded in the architecture of enzyme families.

\begin{figure}[h!]
\centering
 \includegraphics[width=0.4\textwidth]{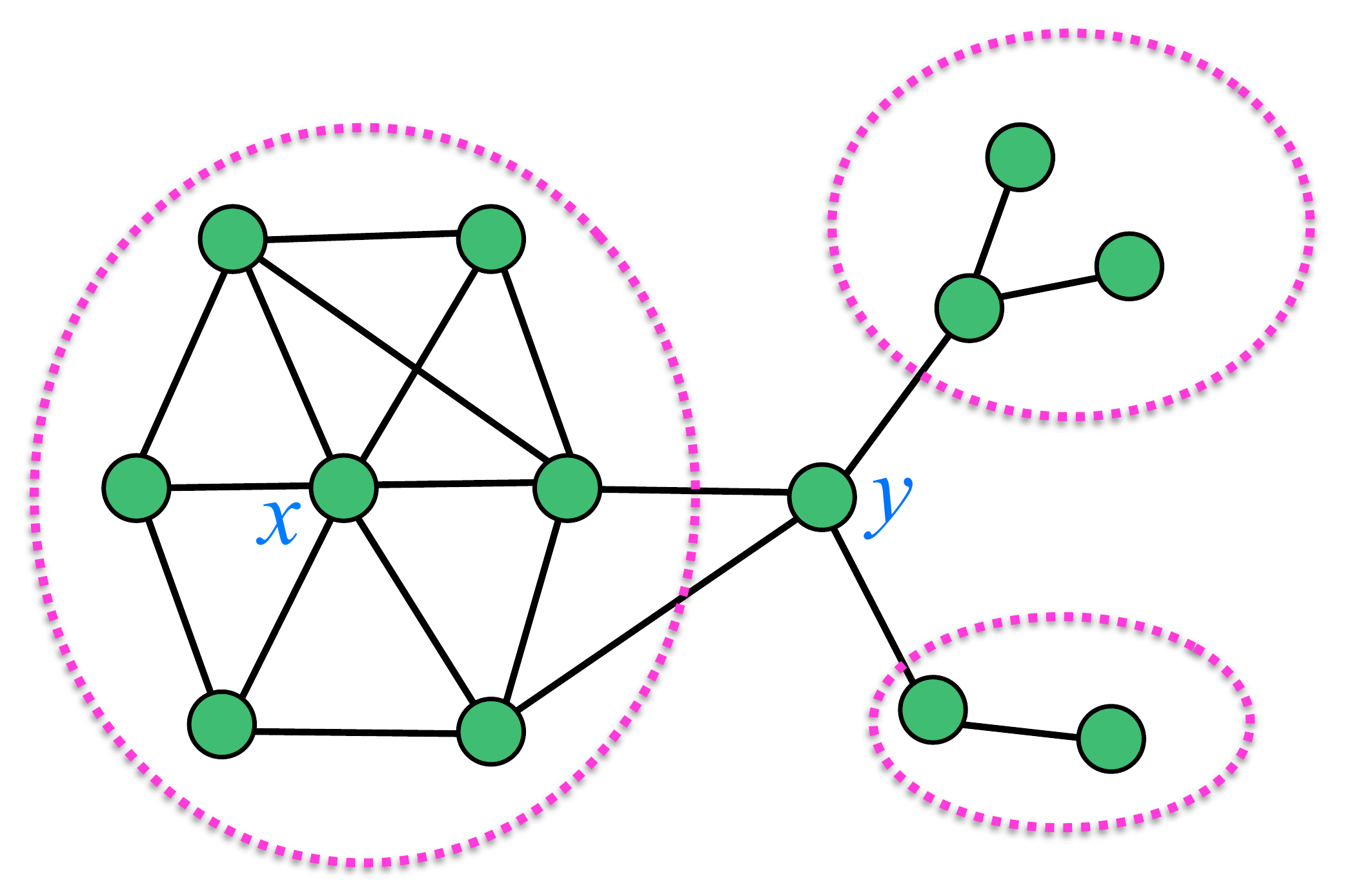}
   \caption{ Betweenness centrality to identify allosteric hotspots.
   A graph explaining the concept of betweenness centrality. From the perspective of signal transmission over the entire network, the node $y$ is more important than $x$ although the node $x$ is highly connected to other nodes. 
Removal of $y$ makes the original graph disjoint into three disconnected graphs, and hence its importance. 
\label{graph}}
\end{figure}

\subsection{Graph theoretical methods}
Network theory (or graph theoretical method) has recently been used, especially in the physics community, to address a series of problems \cite{strogatz2001Nature,Jeong01Nature,jeong2000Nature,albert2000Nature}. Recently, such methods have found applications in allosteric signaling with the view towards controlling allostery at the molecular level \cite{Dokholyan16ChemRev,Proctor15ChemSci}. 
In network theory, a system of interest is simplified into a set of nodes (vertices) and links (edges). For a given network, the importance of a node is assessed using a number of measures, referred to as centralities. For example, the degree centrality measures the number of edges linking a node of interest, and the closeness centrality measures the inverse of the average minimum distances to a node of interest from all other nodes.  Depending on the context, one can quantify the importance of a node by using different centrality measures.  
To identify an important node for signal transmission, it is convenient to use a measure referred to as  betweenness centrality, which is defined using the average ratio of the number of minimal paths between the nodes $s$ and $t$ passing through the node $v$ ($\sigma_{st}(v)$) in comparison to the number of all the minimal paths between the nodes $s$ and $t$ ($\sigma_{st}$): 
\begin{equation}
C_B(v)=\frac{2}{(N-1)(N-2)}\sum_{s=1}^{N-1}{\sum_{t=s+1}^N{\frac{\sigma_{st}(v)}{\sigma_{st}}}}. 
\label{betweenness}
\end{equation}
Fig.~\ref{graph} illustrates the concept of betweenness centrality. The $C_B(v)$ measure shows that node $y$ is more important than the node $x$ in terms of the information flow or allosteric signaling. 
If the node $y$ is removed then the entire network would become disjoint, splitting  into three distinct clusters with each cluster unable to  communicate with the other. However, removal of the node $x$, even if it has the highest degree centrality, does not impair the information flow across the entire network. 

In practice, a network is constructed by choosing a physically reasonable $R_c$ value for establishing a link between two residues if either backbone-backbone, or sidechain-backbone or sidechain-sidechain distance is within $R_c$.    
It is also possible to consider extending the definition of a link by including water mediated contacts \cite{Lee2016BiophysJ} or assigning a differential weight to the links based on the nature of residue contact pairs. 
Approaches similar to the one described here using graph theoretical method have been proposed to identify the allosteric hotspot or signaling pathways.  For example, 
the community detection approach was proposed for intermolecular allosteric communication network of tRNA-protein complexes and the algorithm to compute community network has been incorporated into an analysis module in NAMD \cite{Sethi09PNAS}. 
Calculating the energy flux across the intra-molecular residue network was proposed to relate allosteric signaling with vibrational energy transfer \cite{ishikura2006CPL}. It should be noted that just like SPM, the graph theoretical approach is based on the structures of the allosteric proteins.

\subsection{Atomically detailed Molecular Dynamics (MD) simulations}
Multiple global/local variables that can faithfully represent the conformational dynamics of a biomolecule 
can also be used to construct the AWD of the allosteric protein of interest.  
Specifically, provided that each of the $N$ variables displays a state-dependent  two-state-like transition, 
it is straightforward to define the conformational state of the molecule in terms of $N$ binary switches, $s_1$, $s_2$, $\cdots$, and $s_N$. 
In this representation there are total $2^N$ possible microstates. 
The $\alpha$-th macrostate $\Psi_{\alpha}$, where $\alpha$ refers to functional macrostate of the molecule, say, active or inactive state of a receptor or an enzyme, can be delineated using a linear combination of each microstate with an appropriate weighting factor, such that $\Psi_{\alpha}(t)=\sum_{i=1}^{2^N}c_i^{\alpha}(t)\phi_i$ with $\sum_{i=1}^{2^N}c_i^{\alpha}(t)=1$. The time-dependence of the weighting factor is made explicit 
because of the presence of time-evolving conformational dynamics even in the absence of apparent conformational transition from one functional macrostate to another \cite{Lee2015PLoSComp}. 

In principle, as long as the time trajectories generated from MD simulations are long enough to ensure adequate sampling of the conformational ensembles of all the available functional states and the force fields are reasonably accurate, 
the collective dynamics of binary switches in systems like GPCR could be used to describe the allosteric signaling associated with the transition from one functional state to another. 
However, physical time scales covered by MD simulations ($\sim \mu$sec) are often too short to explore the conformational transition between two functional states ($\gtrsim 1$ ms). Nevertheless, in the GPCR case there are constant fluctuations in the binary switches within a single functional basin of attraction, and their fluctuations are not independent of one another in the presence of spatiotemporal correlation. 
By quantifying the cross-correlations between the fluctuations of all the switches, 
\begin{align}
C_{ij}=\frac{\langle\delta s_i\delta s_j\rangle}{\sqrt{\langle(\delta s_i)^2\rangle}\sqrt{\langle(\delta s_j)^2\rangle}}
\end{align}
where $\delta s_i\equiv s_i-\langle s_i\rangle$, one can capture the signature of allosteric communication between the switches and visualize their  role  using the AWD for each functional state \cite{Lee2015PLoSComp}. 

\section{Dynamics of Allosteric Transitions} 
Although the AWD could be determined using equilibrium methods, described in the previous sections, distinguishing between the various allosteric mechanisms requires a dynamic description of the transition between the allosteric states. Because of the inherent difficulties in using straightforward atomically detailed MD simulations, several alternate ways have been proposed to calculate the dynamical prowesses that occur during the transition from two distinct allosteric states, say $T$ and $R$. Typically, the two states differ structurally, which could be reflected in the absence of certain contacts in the $R$ state that are present in the $T$ state and/or the changes in the positions of loops connecting the secondary structural elements that contain the  AWD residues. The task is to generate an ensemble of trajectories that connect the $T$ and $R$ states from which the dynamics could be quantified. The Targeted Molecular Dynamics \cite{Schlitter94JMG,KarplusJMB00}, the Hamiltonian Switch Method (HSM) \cite{Hyeon06PNAS}, milestoning \cite{Elber17QRB,Kirmizialtin15JPCB,Elber10PNAS},  and the Two Basin Model \cite{Zuckerman04JPCB,Zheng07Proteins,Tehver10Structure,Noid13JCP} are the common computational techniques used to investigate the dynamics of the $T \rightarrow R$ transition. Here,  we describe the HSM, which has been successfully used in conjunction with coarse-grained models of allosteric proteins \cite{Mugnai17PNAS,Tehver10Structure}. Although not the focus of this review we wish to point out that highly innovative methods for simulating allosteric transitions using atomically detailed MD simulations have been introduced by Elber \cite{Elber11COSB,Elber17QRB,ElberCOSB05}.  
\\

{\bf Hamiltonian Switch Method (HSM):} The basic assumption of the HSM\cite{Hyeon06PNAS,Chen07JMB}  is that the local strain accumulated during ligand or cofactor binding propagates faster than the rates of the global conformational transitions between the  $T$ and $R$ states. This assumption has been justified {\it a posteriori} by successful applications of the HSM to a variety of molecular systems.  To illustrate the general idea of the HSM, we assume that the system evolves as described by the Brownian dynamics. Consider a constant field $f$ (due to local strains caused by ligand binding) applied to an allosteric system. If the field is weak, the changes in Hamiltonian, describing the allosteric states, is a linear function of $f$, and is given by $\langle H(x) \rangle_f- \langle H(x) \rangle_0=fx$,
where $\langle H(x)\rangle_f$ is the Hamiltonian when $f \ne 0$, and $\langle H(x)\rangle_0$ is the equilibrium value in the absence of the field (\cite{Edwardsbook,GardinerBook}). The effect of the field on the allosteric protein is expressed by a potential such as  $H_{ext}(x)=fx$, where the variable $x$ is conjugate to the field $f$. The relaxation of the protein conformation $x$ in response to $f$ is related to the response function $\mu(t)$, 
\begin{equation}
\label{eq:response}
\langle x(t)\rangle_f-\langle x\rangle_0=\int^{t}_{-\infty}{dt^{\prime}\mu(t-t^{\prime})f}
\end{equation}
In the context of allosteric proteins, the Hamiltonian for a closed (ligand unbound) state is $H^c$ at $t=(-\infty, 0)$. At time $t$, an external force, $-\frac{\partial H^o}{\partial \vec{r}}$, is added. Here, $H^o$ is the Hamiltonian for the open (ligand-bound) state. According to the fluctuation dissipation theorem \cite{GardinerBook}, protein conformation will evolve from that of the $T$ state to the $R$ state such that at $t \rightarrow \infty$ the swarm of trajectories would reach the $R$ state, thus providing the entire dynamics of the $T \rightarrow R$ transition \cite{Hyeon06PNAS,Chen10PNAS,Xiong15Proteins}. It is important to note that the switch in the Hamiltonian could be made rapidly, which would elicit non-equilibrium response, or adiabatically to capture allosteric signaling under near equilibrium conditions. 
\\

\section{Applications to single domain and multisubunit proteins and ATP-consuming molecular machines}

As discussed in the previous sections, there may only be a few general principles that govern allosteric transitions in monomeric or multisubunit proteins.  Results from recent experiments show that the venerable MWC and KNF models have to be modified to explain new data even in well-studied systems.  For example,  in a recent study of carbon monoxide (the differences in the kinetics of binding of various ligands is discussed by Szabo \cite{Szabo78PNAS}) binding to hemoglobin \cite{Viappiani14PNAS}, it was found that in addition to the MWC quaternary states the tertiary structures in the $T$ and $R$ states were needed to explain the experiments.   Similarly, using native mass spectrometry it has become possible to measure individual ligand binding constants to distinct subunits. Analyses of these beautiful experiments \cite{Gruber18PhilTranRoySocB,Gruber16ChemRev,Dyachenko13PNAS} further extend the theoretical underpinnings of allosteric signaling, and offer a platform for understanding distinct mechanisms using models that go beyond both MWC and KNF pictures. Of course, if the goal is to control allosteric signaling by design then it becomes necessary to obtain the molecular basis using experiments as well as computations. The applications to specific systems, described here and elsewhere, using the methods described above, illustrate the bewildering range of molecular mechanisms used by nature to effect signaling.     The major conclusions that emerge from applications to diverse systems are that the molecular mechanisms exquisitely depend on the architecture of the allosteric systems as well as external cues, which could greatly alter the mechanisms. In this section we describe the nature of allosteric signaling using a variety of increasingly complex systems as case studies.

\begin{figure}[t]
\includegraphics[width=0.6\textwidth]{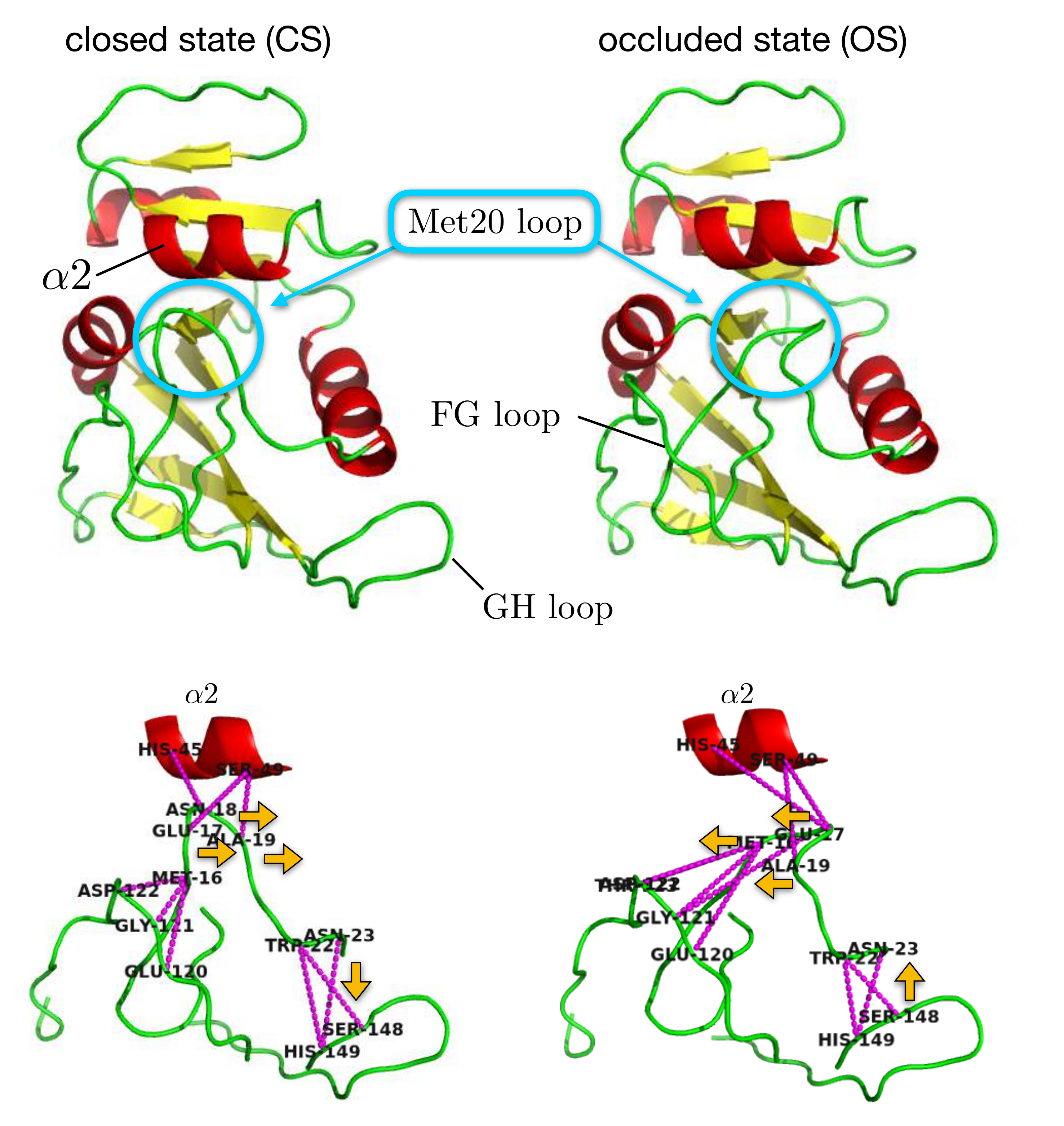}
\caption{Structure of DHFR in the closed (CS on the left) and the occluded states (OS on the right). 
The region around the Met20 loop that displays the most significant conformational change is highlighted. In the bottom  the coordinated changes in the distances of residues that accompany the sliding motion in the CS $\rightarrow$ OS and OS$\rightarrow$CS transitions are shown. The arrows indicate the direction in which the changes occur.  In the forward transition the Met20 loop is pulled by the $\beta$G-$\beta$H loop, which results in it being pushed away from the $\beta$F-$\beta$G loop. The push-pull results in the sliding of the Met20 loop without significant conformational changes in the overall structure of the enzyme. 
\label{DHFR}}
\end{figure}

\subsection{Dihydrofolate Reductase (DHFR)}
The intensely investigated single domain enzyme DHFR (Fig. \ref{DHFR}) catalyzes the reduction of 7,8 dihydrofolate (DHF) to 5,6,7,8 tetrahydrofolate (THF)  \cite{schnell2004ARBBS,boehr2006science}.  By binding the cofactor, nicotinamide adenine dinucleotide phosphate (NADPH), hydride transfer from NADPH to protonated DHF leads to the production of NADP$^+$ and THF$\cdot$DHFR. This is required for normal folate metabolism in prokaryotes and eukaryotes and plays an important role in cell growth and proliferation.   As a result of the clinical importance of this process, allosteric transition in DHFR has been studied extensively using a wide range of experimental and theoretical methods.

High-resolution crystal structures show that the 186-residue {\it E. coli} DHFR enzyme has eight $\beta$-strands and four $\alpha$-helices interspersed with flexible loops that connect the secondary structural elements (Fig. \ref{DHFR}). The structure of DHFR can be partitioned into adenosine-binding and loop subdomains. In the catalytic cycle, the Met20 loop changes conformation between the closed (CS) and occluded (OS) states (Fig. \ref{DHFR}). Interactions through a hydrogen bond network with the $\beta$FÐ$\beta$G loop stabilize the CS. The crystal structures of {\it E. coli} DHFR complexes in the catalytic cycle have given a detailed map of the structural changes that occur in the enzyme. In addition, the conformational changes in {\it E. coli} DHFR in response to ligand binding have been inferred using a variety of experimental techniques, including X-ray crystallography, fluorescence, and NMR. In discussing the ligand-induced conformational changes in DHFR, we view the CS and OS as the two distinct allosteric states, which differ mainly in the position of the Met20 loop. The conformational change of  CS$\rightarrow OS$ closes the active Met20 loop during the catalytic cycle.

Much of the discussion of NADH binding to DHFR has been in terms of the CS \cite{boehr2006science} mechanism although it has been argued  that there is flux through the IF pathway as well  with the CS mechanism dominating at low ligand concentration.\cite{Hammes09PNAS}. The concentration of the ligand determines the ratio of flux between these two pathways. Using Eq. \ref{CSIF} and the values of $k_{on}^{CS}$ and $k_{on}^{OS}$ at low ligand concentration 
(estimated  to be $k_{on}^{CS} \approx 10^7 \text{ Ms}^{-1}$ and $k_{on}^{OS}\approx 10^9\text{ Ms}^{-1}$ in Reference \cite{Hammes09PNAS}), we find that roughly two-thirds of the DHFR molecules follow the CS pathway and the remaining molecules reach the OS state through the IF route. The ratio of $\approx$ 0.7 estimated here is close to $\approx$ 0.8 obtained in \cite{Hammes09PNAS} using a more elaborate analysis of the parallel allosteric pathways (Fig. \ref{HammesFig}). At high ligand concentration, the IF pathway dominates. Such a conclusion could be visualized using the illustration in Fig.~\ref{fig:landscape} based on the observation that a small fraction of DHFR is in the OS state even without the ligand.\cite{boehr2006science}. However, it does not provide a molecular picture accompanying the CS$\rightarrow$OS transition. 
\\

{\bf Inferring the AWD from sequences:} 
The reformulated statistical coupling analysis (SCA) method \cite{Dima06ProtSci} was used to predict the AWD in DHFR. Based on 462 sequences generated using the MSA the network of key residues that  drive the CS$\rightarrow$OS transition \cite{Dima06ProtSci} was determined. As pointed out earlier, the SCA might merely reflect, to a great extent, sequence conservation. One way to exclude such (conserved) residues from the AWD is to estimate the degree of conservation using sequence entropy, which is calculated as $S_i=-\sum\limits_{x=1}^{20} p_i^x \log p_i^x$. It has been shown \cite{Chen07JMB} that the  chemical sequence entropy or $S_{CSE}$, which only discriminates between 4 classes of residues (hydrophobic, polar, positively charged and negatively charged), 
\begin{equation}
S_{CSE}(i)=-\sum\limits_{x=\{H,P,+,-\}}p_i^x \log p_i^x
\label{CSE}
\end{equation}
 is more informative. The residues that remain in the AWD,  after excluding the highly conserved residues (e.g., $S_{CSE}<0.1$), may be connected to allosteric motions. For example, they are in the hinge regions, close to the active-site cleft or form new contacts in OS state\cite{Chen07JMB}. The residue dependent $S_{CSE}(i)$ (see Fig. 2 in \cite{Chen07JMB}) shows that, besides the high degree of conservation in the chemical identity of residues near Met20, there is a  network around M42, that is spatially separated from the Met20 loop,  which could dynamically drive the CS$\rightarrow$OS transition. This prediction using $S_{CSE}$ combined with SCA has been validated using NMR studies \cite{Mauldin10Biochem}.
\\

{\bf Dynamics of the CS$\rightarrow$OS transition:} Understanding  the transition dynamics between the two allosteric states is necessary to uncover the molecular mechanism driving the conformational change in response to ligand binding. To this end, HSM was used to probe the dynamics associated with the major residues that drive the  CS$\rightarrow$OS transition using Brownian dynamics simulations\cite{Chen07JMB}.  Static analyses determined the contacts that are  broken in the CS and new ones that  form in the CS$\rightarrow$OS transition\cite{radkiewicz2000JACS}.  By monitoring  the local movements of the Met20 loop in the coarse-grained Brownian dynamics simulations, it was discovered \cite{Chen07JMB} that the Met20 loop slides along helix $\alpha$2 in the adenosine-binding domain (see Fig.~\ref{DHFR}). By dissecting the events at the residue level, it was found that the rupture of the contacts in the CS (Asn18-His45, Asn18-Ser49, and Ala19-Ser49) and formation of Glu17-Ser49 during the CS$\rightarrow$OS transition facilitates the sliding of Met20 along $\alpha$2. In the loop subdomain, the flexible Met20 loop interacts simultaneously with both the $\beta$FÐ$\beta$G and $\beta$GÐ$\beta$H loops (Fig. \ref{DHFR}). In order to dissect the order of events that occur in the CS$\rightarrow$OS transition, Chen and coworkers computed the kinetics of breakage and formation of a number of contacts involving the two loops (see the results in Figs. 6aÐc in the study be Chen\cite{Chen07JMB}). By monitoring the time-dependent changes in the formation and rupture of various contacts, it was established that  the rupture of contacts between the Met20 loop and $\beta$FÐ$\beta$G loop in the CS and formation of contacts between residues in the Met20 loop and $\beta$GÐ$\beta$H loop occur nearly simultaneously. Only subsequently the interaction between Glu17 (in the Met20 loop) and Ser49 (in $\alpha$2) that exists only in the CS, takes place. Thus, the sliding of the Met20 loop on $\alpha$2 requires coordinated motion of a number of residues in the loop domain as illustrated in Fig. \ref{DHFR}.  It is gratifying to note that many of the residues identified in the dynamics of the sliding of the Met20 loop are also predicted based on the chemical sequence entropy, $S_{CSE}(i)$ given in Eq. \ref{CSE} and the SCA. Thus, there indeed is a connection between the AWD and allosteric dynamics, which is recurring theme in the examples described in this review.

It is surprising that there is an elaborate residue network in the CS$\rightarrow$OS transition given that the root-mean-square deviation (RMSD) between the closed and occluded crystal structures is only 1.18 \AA. Despite the high structural similarity between the two states,  substantial conformational rearrangements are needed to facilitate the transition between the CS and OS states. Clearly, this is an ideal  example of allostery without significant conformational change \cite{Cooper84EBJ,Nussinov15COSB,McLeish13PhysBiol}. Because the two states are structurally similar, it follows that the ligand-induced transition is likely driven by entropy changes\cite{Kornev18BiochemSocTrans}, which is a necessary condition for observing the Cooper-Dryden scenario \cite{Cooper84EBJ}, and is an over riding principle in much of the ensemble view of allostery \cite{Motlagh2014Nature}.

\subsection{PDZ Allostery} 
\begin{figure}[h!]
\includegraphics[width=1.0\textwidth]{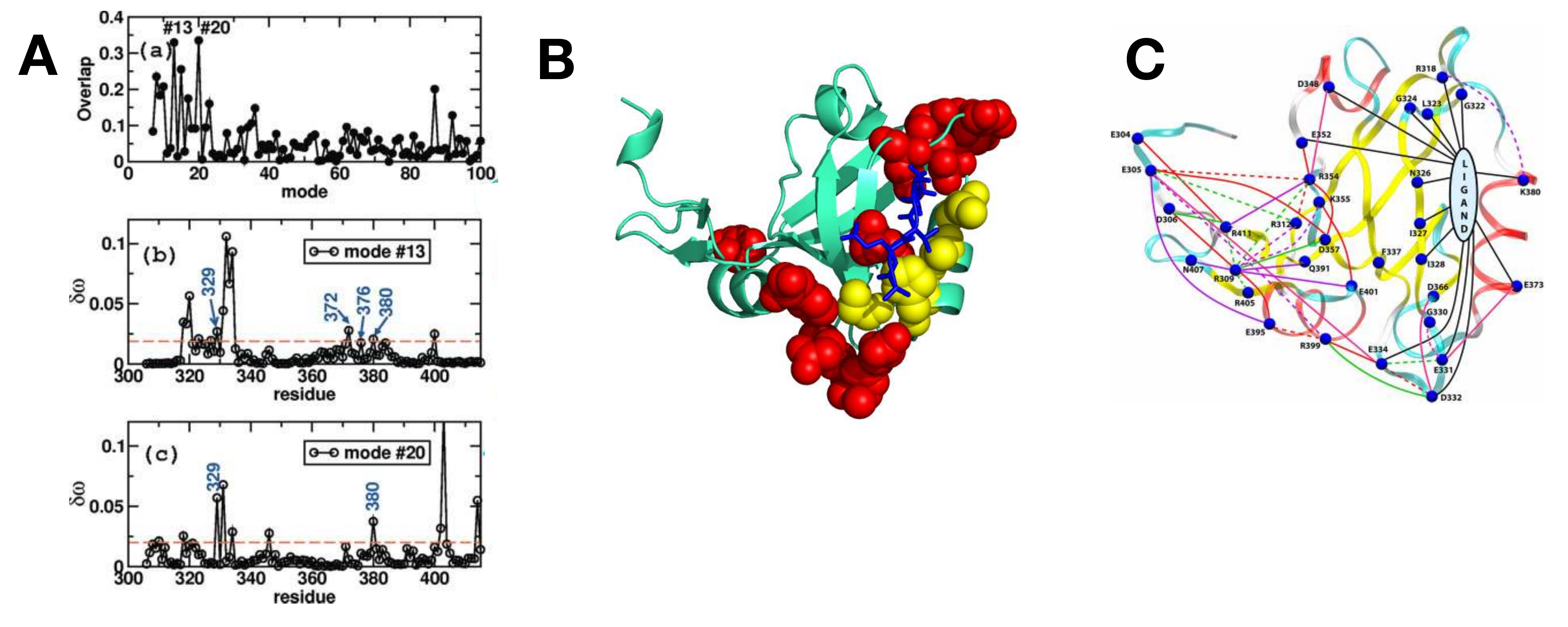}
\caption{AWD for the PDZ  domain.
(A) Application of the SPM to the PDZ domain. (a) The overlap of ENM modes are calculated up to 100 modes using the eigenvectors  defining the transition  between the two allosteric states (with and without bound ligand). 
Although multiple modes participate in the transition, modes \#13 and \#20 display the largest overlap. 
(b), (c) Responses to the perturbation assessed in terms of  $\delta\omega$ values (Eq.\ref{SPM}) 
for the mode \#13 and and \#20.   
(B) Hotspot residues predicted using the SPM are highlighted in the structure with red spheres. 
The hotspot residues identified in experiments are shown in yellow spheres. 
(C) A network view of the perturbation in pairwise electrostatic interaction energies, $\Delta E_{ij}=\langle E_{ij}\rangle_{\text{bound}}-\langle E_{ij}\rangle_{\text{unbound}}$. 
(i) The blue spheres are residues with $|\Delta E_i^{\text{total}}|(=|\sum_{j}\Delta E_{ij}|)$
 or $|\Delta E^{\text{ligand}}_i|>6$ kcal/mol. 
 A few residues with large $|\Delta E_{ij}|$, but $|\Delta E_i^{\text{total}}|<6$ kcal/mol (R312, R354, K355, R399, E401, and R411) 
 are also shown as spheres. 
 (ii) Connections with negative and positive $\Delta E_{ij}$ values are indicated with solid and dashed lines, respectively, i.e., a solid (or dashed) line indicates a contact more (or less) favorable in bound (or unbound) state. 
 (iii) The connections are colored on the basis of magnitude of $|\Delta E_{ij}|>10$ kcal/mol (red), $> 6$ kcal/mol (green), $> 4$ kcal/mol (purple), and $> 3$ kcal/mol (pink). 
 The black lines represent interaction between peptide ligand and residues with 
 $|\Delta E_{ij}|>6$ kcal/mol. The interaction with $|\Delta E_{ij}|>3$
 kcal/mol were considered only for residues that are directly perturbed on ligand binding. 
 The panels (A) and (C) were taken from References \cite{Liu09Proteins,Kumawat17PNAS}.
 \label{PDZ1}}
\end{figure}

PDZ domains are also examples in which allosteric signaling is transmitted without obvious detectable conformational changes \cite{Petit09PNAS,Buchli13PNAS,Buchenberg17PNAS,Kumawat17PNAS,Liu17PNAS,Stock18PhilSocRoySocB}. PDZ domains are a large family of globular proteins that mediate protein-protein interactions and play an important role in molecular recognition \cite{Harris01JCellSci,Kim04NatRevNeurosci,Jemth07Biochem}. 
PDZ domains  generally have between 80--100 amino acids with structures consisting of a mixture of six $\beta$ strands and a couple of $\alpha$-helices (Fig. \ref{PDZ1}).
The folding of PDZ domains have been investigated using both experiments \cite{Gianni05PEDS,Gianni07PNAS} and simulations \cite{Liu16JPCB} but the allosteric transitions in PDZ in which there is little detectable conformational change have elicited different interpretations. In particular, there is debate on whether allostery in PDZ is dominated by entropy or by enthalpy \cite{Kumawat17PNAS,Liu17PNAS}.  
\\

{\bf AWD using SPM and Molecular Dynamics Simulations:} The normal modes of the PDZ domain using the PDB structure (code 1BE9 corresponding to the third PDZ domain) were determined by representing the protein by $C_{\alpha}$ side chain ENM with a cutoff distance for the contact being $R_c=8$ \AA. The overlap of the 100 lowest frequency modes for the bound to unbound transition in the PDZ domain shows that this transition can be accurately described using the 13$^{th}$ and 20$^{th}$ modes (Fig. \ref{PDZ1}A-a). 
The SPM analyses for the two modes identify a set of residues that have the largest response to local perturbations (Fig. \ref{PDZ1}A-b and \ref{PDZ1}A-c). 
Among the key residues, coupling involving Gly329, His372, and Ala376 have been deemed to be important in experimental studies \cite{Chi08PNAS}.  In addition, there are charged residues as well as pH sensitive His residues in the AWD.  The complete list of residues predicted by the SPM are mapped onto the structure of PDZ domain (Fig. \ref{PDZ1}B). 

In a recent, atomically detailed MD simulations with and without bound ligand \cite{Kumawat17PNAS}, a vast network of residues dispersed throughout the structure is identified as hot spots.  It has been suggested that there are rearrangements, without global conformational change in the protein, in the electrostatic AWD (Fig.\ref{PDZ1}C).  However, in this case the transition is proposed to occur by differential changes in the enthalpy rather than entropy, as is likely the case for DHFR. The shift in the electrostatic AWD involves hydrogen bond network rearrangement, which has been used to propose that allosteric signaling proceeds by a population shift \cite{Kumawat17PNAS,Liu17PNAS} or the CS mechanism. Although the CS is the postulated mechanism \cite{Kumawat17PNAS}, it is difficult to rule out the IF mechanism unless the entire kinetic rates for ligand binding and unbinding as a function of ligand concentration to the two states are measured.\cite{Hammes09PNAS}

NMR experiments were used to assess the importance of side chain motion in allosteric signaling of the PDZ3 domain \cite{Petit09PNAS}, which contains an extra $\alpha 3$ helix that apparently has no influence on the function because it is spatially far from the ligand binding site. Deletion of $\alpha 3$ decreases the binding affinity by about twenty fold, which in free energy terms is not that significant. However, the absence of $\alpha 3$ enhances side chain motions throughout on a pico to nano second time scale. These results were used to suggest that allosteric signaling occurs by an entropic mechanism, which is different in interpretation from the recent MD simulations \cite{Kumawat17PNAS}. 
\\

\begin{figure}[t]
\includegraphics[width=1.0\textwidth]{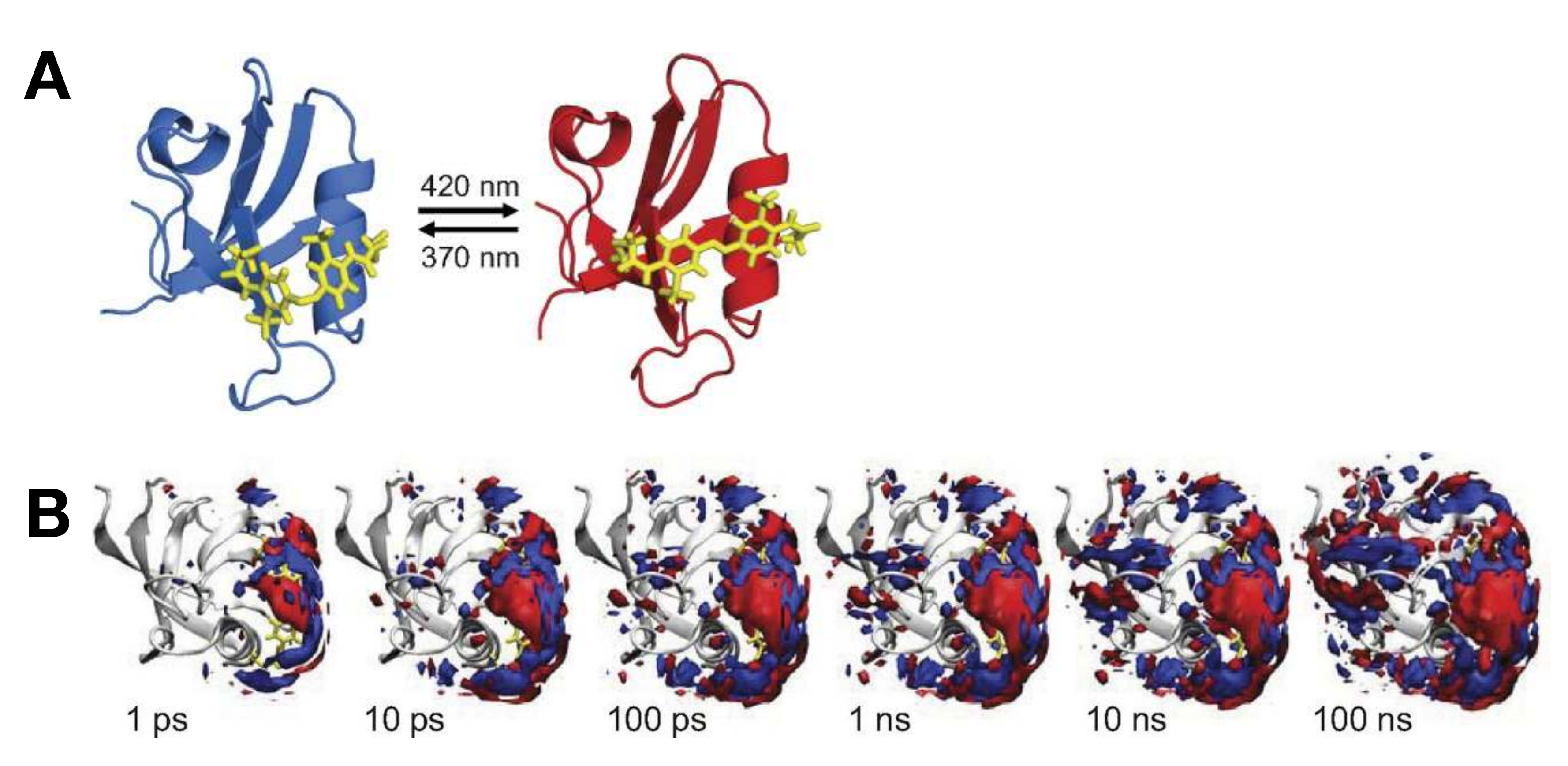}
\caption{Dynamics of allosteric transitions in PDZ2.
(A) Structures of the photoswitchable PDZ2 domain between the {\it cis} and {\it trans} conformations of the azobenzene ligand.
(B) Change of water density with simulation time upon photoswitching. 
Increased and decreased densities are depicted in red and blue, respectively. 
The contour surfaces correspond to changes of $\pm 0.01$ water$/\text{\AA}^3$ 
The protein (gray ribbon) is shown with the photoswitch (yellow, visible only in part). 
Figure was taken from Ref.\cite{Buchli13PNAS}. 
\label{PDZ2}}
\end{figure}

{\bf Non-equilibrium response in PDZ domain to perturbation:} In an interesting experiment, Buchli \cite{Buchli13PNAS} covalently attached an azobenzene derivative, which  photo switches between {\it cis} and {\it trans} conformations by exciting with different wavelengths. The non-equilibrium switch from the {\it cis} and {\it trans} states complemented by molecular dynamics simulations produced a picture of the propagation of the disturbance created by the excitation. The major finding in this study is that the binding groove to which the  azobenzene derivative is attached opens on a time 100 ns, which involves a rearrangement of network water molecules, as revealed by MD simulations \cite{Buchli13PNAS}. The long times needed for cooperative rearrangement of water molecules is not only surprising but also implies that the solvent might play a crucial role in allosteric signaling.  

Recently, extensive atomic detailed non-equilibrium MD simulations were performed \cite{Buchenberg17PNAS} for the photoswitchable PDZ in order to provide a molecular movie of the conformational changes as a complement to the experiments \cite{Buchli13PNAS} (see Fig.~\ref{PDZ2}). By non-equilibrium it is meant that the isomerization dynamics in the azobenzene derivative occurs on a time scale that is much faster than the time for the conformational changes in the PDZ domain.  There is a hierarchy of time scales ranging from ps to $\mu s$ in which the protein responds to the photo switch. Of particular note is that there is a great variation in the time dependent changes in the distances between various residues (see Fig.3 in  \cite{Buchenberg17PNAS}). That there is a great deal of heterogeneity (no two trajectories are alike) in the allosteric spreading dynamics, which was also previously shown for GroEL \cite{Hyeon06PNAS} using non-equilibrium coarse-grained simulations, might be a very general feature of the dynamics of allosteric transitions. The diversity of pathways identified in the simulations does not imply that there are no hot spot residues. Rather, the AWD determined using the SPM based on ENM provides the most likely load bearing regions in the protein. These methods based on equilibrium dynamics, principally hinging on the properties of low frequency modes, already indicate that signal propagation occurs in an anisotropic manner.  It would be most interesting to classify the multiple pathways observed in  \cite{Buchenberg17PNAS} by suitable unsupervised clustering methods to assess if there are certain dominant paths that carry greater weight than others. In so doing, which requires an analyses of the existing trajectories, one could learn if in a probabilistic sense certain residues are more important than others in transmitting allosteric signals.

\subsection{Allosteric transitions in a multisubunit enzymes}  
{\bf S-adenosyl homocystein hydrolase:}
S-adenosyl homocysteine hydrolase (SAHH) is a key enzyme for S-adenosyl-L-methionine (SAM)-dependent methylation pathway. 
The SAM-dependent methylation pathway produces a by-product, S-adenosyl-L-homocysteine (SAH), which itself acts as  
a strong inhibitor of the pathway.  In eukaryotes, SAHH is the only enzyme that hydrolyzes SAH into adenosine and homocysteine, relieving the accumulation of SAH.  
As the virus requires methylation for replication, 
accumulation of SAH in the absence of SAHH function leads to shutting down the SAM-dependent methylation pathway and consequently inhibiting virus replication. 
Therefore, SAHH - a key host enzyme for virus - is an important antiviral drug target \cite{Clercq02NRDD}.

It may appear that targeting the active site using a drug, where specific enzyme - substrate contacts are formed, is a straightforward strategy for enzyme regulation (or inhibition). On the other hand, allosteric modulation from the distal residues dispersed over the whole enzyme architecture, is also essential because it is known that influence from a remote site can impair the precise positioning of catalytic elements in the active site \cite{knowles1991Nature}.
Here, we show that the  multifaceted computational approaches are particularly useful in establishing the link between the conformational dynamics, allosteric signaling in mutlisubunit enzymes, and enzyme function \cite{Lee11JACS}. 
\\

\begin{figure}[ht]
\includegraphics[width=1.0\textwidth]{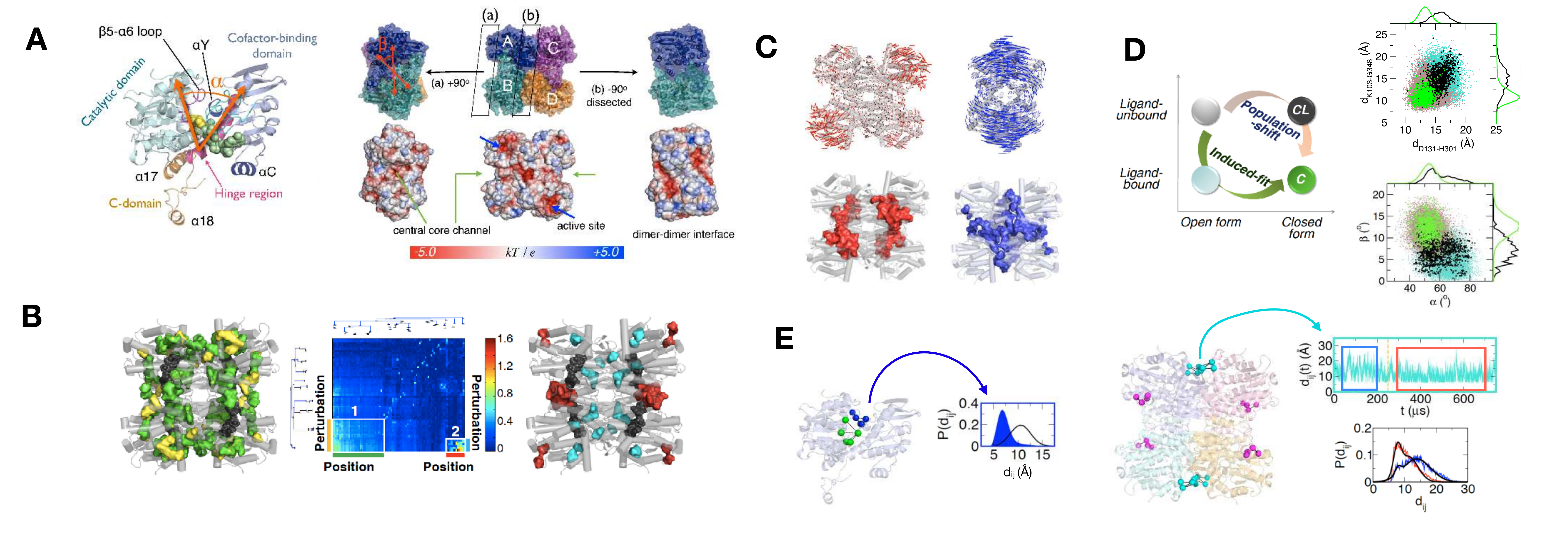}
\caption{\footnotesize
{\bf Allosteric dynamics in SAHH. }
(A) Structure of SAHH. A single subunit structure is shown on the left, and homotetrameric structure is on the right. 
The angles $\alpha$ and $\beta$ are used to probe the substrate binding-induced closure of SAHH.  
(B) Results for the AWD from the SCA. 
For cluster1 (left), perturbation and position clusters are colored in yellow and green, respectively. 
For cluster2 (right), perturbation and position clusters are colored in cyan and red, respectively.
(C) Depicted are the high $\delta\omega$-residues extracted from the mode 7 (the ENM mode that exhibits maximal overlap with the displacement vector from O to C state) of open (left) and closed (right) structures. 
(D) Simulation results projected on to ($d_{K103-G346}$, $d_{D131-H301}$) (left panel)
and ($\alpha$, $\beta$) (right panel) before (cyan) and after (brown)  ligand binding. Ensembles of CL and C states are in black and green, respectively, with their histograms shown on the right. 
The partial overlap between the histogram (green and black) indicates that allosteric transition occurs via a hybrid mechanism involving both the CS and IF. 
(E) (Left) The histograms before (black line) and after (blue) the ligand binding calculated for the distance of an intra-subunit residue pair (blue) in the ligand binding cleft. 
(Right) A time trace of the distance between an inter-residue pair (cyan) at the AC subunit interface displays a bimodal hopping transition, whose population change is shown in the bottom panel. The histograms before and after the ligand binding (blue and red lines, respectively) fit to a double Gaussian function (black lines) indicate a population shift upon substrate binding. The figure was adapted from Reference \cite{Lee11JACS}. 
\footnotesize
\label{SAHH}}
\end{figure}

{\bf Determination of the co-evolving residue network of SAHH using SCA:} 
Since SAHH functions as a tetramer consisting of identical subunits (Fig.~\ref{SAHH}A), it can be argued that the co-evolving clusters of residues revealed from the monomer sequence are also involved in allosteric communication, which would be dispersed throughout  the tetrameric architecture. 
Most of the SCA-identified co-evolving residues are not well-conserved in the SAHH family except for those in the C-terminal domain and 
Cys79 that is in direct contact with the substrate (SAH). SCA identifies two co-evolving clusters of residues. 

(i) Cluster 1: The residues in the green clusters surrounding the active site are influenced by the perturbation of residues in the yellow clusters (Fig.~\ref{SAHH}B, left).
It is remarkable that the residues coevolving with those in the remote site are distributed  around the binding cleft.  Moreover, the central core channel (see Fig.~\ref{SAHH}A, right) is also detected as a part of the main AWD. 

(ii) Cluster 2: Two spatially disconnected sets of residues are identified in the C-terminal domain (F425 -- Y432) and the residues (G122, P123, and D125) in the catalytic domain distal from the active site. 
According to mutation studies, 
K426 and Y430 in the C-terminal region are the critical residues involving the cofactor affinity and/or the assembly of the tetrameric structure \cite{li2007BC,ault1994JBC}. 
More recently, another mutagenesis study has highlighted the important role of H-bond forming D239 with N27 in the C-terminus, validating the prediction from SCA on SAHH \cite{wang2014JBC}. 
Furthermore, although it is not detected in the SCA, it has been reported that 
a mutation of R49 in direct contact with D125 dramatically reduces the enzymatic activity of SAHH \cite{vugrek2009HM}. 
\\

{\bf Identifying the AWD of SAHH using the SPM:}
SPM was also used to determine the ``hot spot" residues controlling the of open-to-closed conformational transition. For the closure dynamics dictated by the lowest mode (mode 7) of the open form of SAHH, residues with high $\delta\omega$ (see Eq.(3)) are mainly in the hinge region (Fig.~\ref{SAHH}C, left). 
This finding is in agreement with the time-resolved fluorescence anisotropy measurements that monitored the domain motions of the mutant SAHH, where importance of residues at the hinge region was highlighted \cite{wang2006Biochemistry}. 
On the other hand, for the dimer-dimer rotation dynamics described by mode 7 of the \emph{closed} form, high $\delta\omega$ residues are found at the interfaces between subunits (Fig.~\ref{SAHH}C, right). 
Key residues controlling the next higher frequency mode (mode 9) are distributed mainly at the interfaces between the subunits, contributing to the concerted inter subunit dynamics.
\\

{\bf Substrate binding mechanism of SAHH:} 
The two limiting mechanisms, IF and CS, were examined for the closure dynamics of SAHH upon substrate binding. 
For SAHH, the relative importance of IF vs CS can be addressed computationally
by projecting the population of accessible conformational states onto a low dimensional energy landscape, and comparing the two populations obtained from the holo and ligand-bound conditions. 
If the overlap between the closed state populations under the conditions of holo and ligand-bound states is substantial, then the CS mechanism is more likely, as discussed before (Eq. \ref{CSIF}). In contrast, if there is little overlap between the statistical ensembles, then IF mechanism is favored.
SAHH structure in the open form  transiently visits the ``closed-like (CL)" structural ensemble even in the holo form (i.e., in the absence of ligand). 
For the CS mechanism to be operative, it is required that the enzyme recognize the ligand solely by selecting a pre-existing conformational ensemble.  A distribution of scattered plot using inter-residue distance pair ($d_{K103-G346}$, $d_{D131-H301}$) 
or angle pair ($\alpha$, $\beta$) as surrogate reaction coordinates should show  overlap of the CL ensemble with C the ensemble. 
In contrast, under the IF mechanism, we expect negligible overlap between the distributions of CL and C. 
The C and CL structural ensembles probed in terms of the above mentioned order parameter pairs ($d_{K103-G346}$ and $d_{D131-H301}$) and ($\alpha$,$\beta$) show clear distinction but with a certain degree of overlap, giving evidence that for this enzyme both IF and CS mechanisms are likely to be operative (Fig.~\ref{SAHH}D). 

We note that one of the two mechanisms could become more dominant, which not only depends on the concentration of ligand \cite{Hammes09PNAS} but also on the position of the attached probe \cite{Hammes09PNAS}.
For SAHH, the CS mechanism becomes apparent for the dynamics probed for the residue pairs at the inter-subunit interface. This is revealed by the dynamics of $d_{AC(BD):W17-I321}$ where one of the bimodal basins corresponding to the closed structure at $d_{AC(BD):W17-I321}\approx 8$ \AA\ becomes more populated upon the ligand binding (Fig.~\ref{SAHH}E, right). In contrast, the modulation of the conformational landscape, which alters the positions of the energy minima, is more dominant in the dynamics of intra subunit residue pairs (Fig.~\ref{SAHH}E, left).

\begin{figure}[h!]
\centering
 \includegraphics[width=0.9\textwidth]{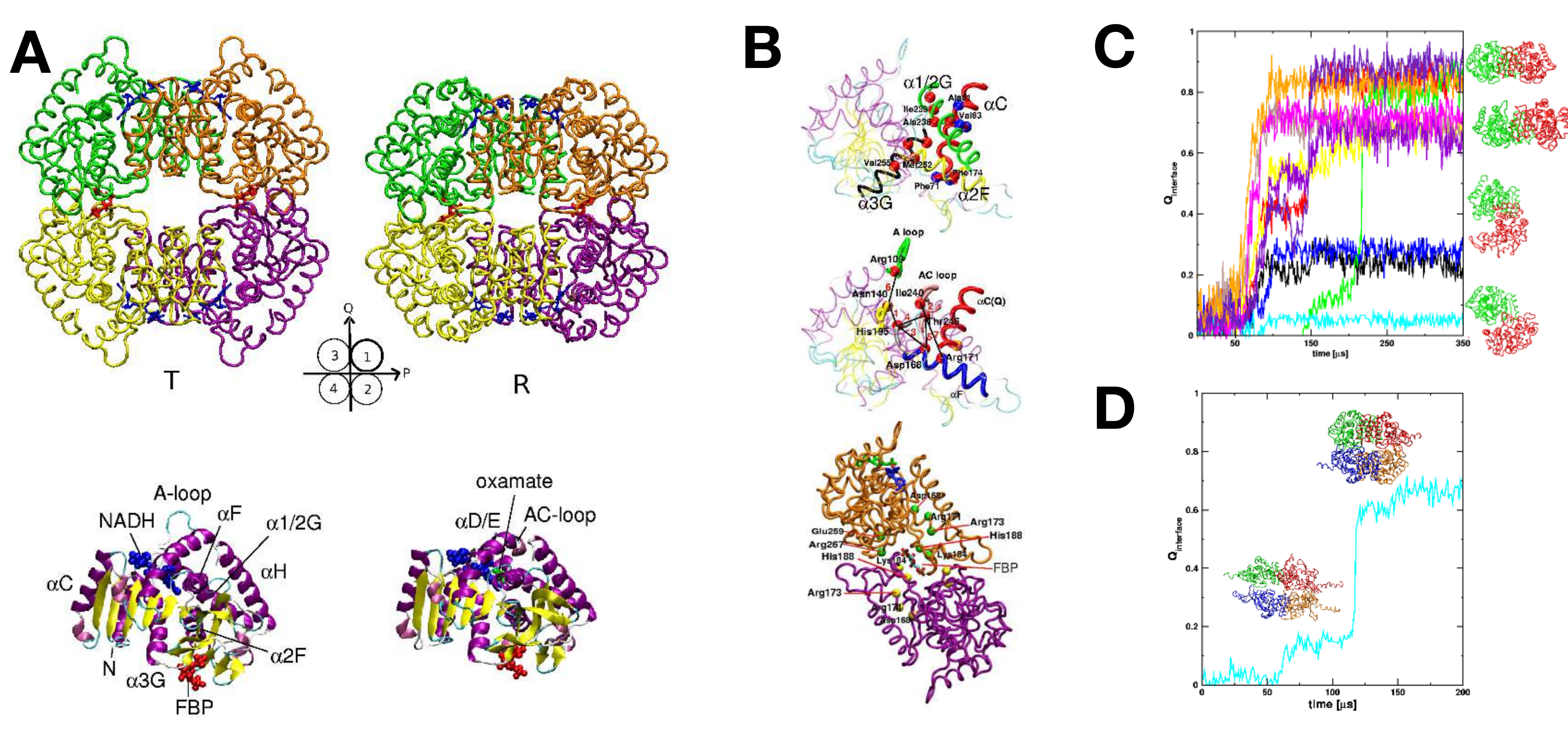}
   \caption{
   \small
  {\bf Allosteric dynamics of LDH.} (A) Structural representation of the T (left) and R (right) states of the  tetrameric L-Lacatate Dehydrogenase LDH. The structures are colored for different chains (the organization of the four subunits is shown schematically in the middle in circles) with bound ligands in licorice (red for FBP and blue for NADH). Cartoon representations of the subunits of the T and R states are given in the lower part.  Ligands NADH and FBP are displayed using blue and red spheres.
   (B) Changes in the interface hydrophobic interactions at the Q-interface in the $T\rightarrow R$ transition. (Top) A snapshot from the Brownian dynamics simulations shows that helix $\alpha$C belonging to subunit 3 is in red with hot spot hydrophobic residues displayed in blue spheres. Helices $\alpha$2F, $\alpha$1/2G and $\alpha$3G of subunit 1 are shown in yellow, green and black, respectively. The AWD residues  are shown in blue spheres.  (Middle) Formation and rupture of contacts at the active site inside subunit 1 during the $T\rightarrow R$ transition. A few key residues are labeled and the dynamical sequence of changes of interactions are indicated using numbers: 1 indicates the first event and 7 is the last event. The numbering of the residues follows the conventional LDH numbering system. (Bottom) Changes in the electrostatic interactions on the P-interface in the $T\rightarrow R$ transition. Subunits 1 and 2 are shown in orange and purple with key charged residues shown as green and yellow spheres. All the residues are labeled and FBP is explicitly shown.
   (C) The time-dependent changes in the fraction of interface contacts, $\langle  Q(t)_{interface} \rangle$, the same 11 trajectories during the allosteric transition. The heterogeneity in the trajectories is evident. 
   Structures of the trajectories at the end of the assembly process are shown from top to bottom. Subunits 1 and 3 are colored in red and green, respectively.
   (D) Time-dependent changes in $Q(t)_{interface}$ of the tetramer during the assembly process is plotted for the cyan trajectory. A snap shot of the metastable intermediate with $Q(t)_{interface}\approx  0.2$ shows that the individual subunits have 
   nearly adopted native-like structures rapidly but the overall assembly is far from complete. Assembly process is highly cooperative when the correct orientation registry is achieved, which in this trajectory occurs at $t\approx  125$ $\mu$s, which about three times longer than needed for folding of individual subunits. The subunits 1--4 are colored by red, orange, green and blue, respectively. 
   The figures were based on results reported in Reference \cite{Chen18JPCB}. 
   \normalsize
   \label{LDH}}
\end{figure}

\subsection{AWD in L-Lactate Dehydrogenase also drives Tetramer Assembly}

The important enzyme L-lactate dehydrogenase (LDH) \cite{Sada15Science,Changeux05Science}, which has received surprisingly little attention from the biophysical community, has been used as a case study to illustrate the role AWD plays in the assembly of this tetrameric multisubunit protein. LDH catalyzes the interconversion of pyruvate and lactate \cite{Rossmann77PNAS}. Some bacterial LDH are allosteric enzymes that are activated by fructose 1,6-bisphosphate (FBP) \cite{Iwata1994NSMB,Fushinobu96JBC,Uchikoba02Protein}. In contrast to non-allosteric vertebrate LDHs, allosteric LDHs assemble as  tetramers \cite{Rossmann77PNAS,Fushinobu96JBC,Uchikoba02Protein,Arai2002ProtEng} 
composed of identical subunits related through three twofold axes labeled P and Q (Fig. \ref{LDH}A) and R (not to be confused with the notation used for the allosteric state $R$), 
following the Rossmann convention \cite{Rossmann77PNAS}. 

The crystal structures of LDH from {\it Bifidobacterium longum} (BLLDH)  in the low ($T$)  and high ($R$) affinity states  have been determined \cite{Iwata1994NSMB} (Fig. \ref{LDH}A). 
The FBP binding sites consist of four positively charged residues, R173 and H188 from the two 
P-axis-related subunits. Upon FBP binding, the charge repulsion between R173 and H188 from the two P-axis-related subunits is diminished.  As a result, the open conformation of LDH in the $T$ state changes to the closed conformation in the $R$ state. Comparison of the crystal structures of the two allosteric  states  shows that the T$\rightarrow$R transition 
involves two successive rotation of the subunits: first rotation is $\approx 3.8^o$ about an axis close to H188, caused by binding of FBP on the P-axis-related subunit interface. The second is a $\approx 5.8^o$ rotation about the P-axis, which creates  changes in the hydrophobic interactions  on the Q-axis related subunit interface to which
 the substrate would bind. Changes in the tertiary structure and cooperative interactions are also amplified to buffer the quaternary structural changes. By comparing the known crystal structures of LDH, the conformational changes of the structural units are found in the sliding of  helix $\alpha$C on the Q-axis-related subunits interface,  causing helix $\alpha$1/2G  to kink leading to the shift in the carboxy terminal out of the active site (see Fig. \ref{LDH}A). Therefore, $\alpha$C sliding is thought to  control the affinity of the substrate \cite{Iwata1994NSMB}.These quaternary structural rearrangements reflect the allosteric control
of LDH, triggered by switching between the  two distinct states. The  mechanism underlying this transition, which is assumed to preserve the symmetry of the allosteric states, is in accord with the MWC model \cite{Monod65JMB,Changeux05Science} in two important ways. First,
the LDH regulatory proteins are oligomers formed by identical subunits with underlying symmetry. Second,  
the interconversion of oligomers between the discrete conformational states is independent of the presence of ligands, which  is the basis of conformational selection.  Of course, in the presence of ligands the transition occurs more readily leading to enhanced stability of the $R$ state.
\\

{\bf Sequence of Molecular Events in the Allosteric Transitions in LDH:} How allosteric signal transmission spreads from the effector site to the active site was probed by focusing on  the dynamics of the charge interactions on the P-interface around the FBP binding sites, the dynamics of 
the hydrophobic interactions on the Q-interface around the active sites, and the interactions inside a single subunit. The hierarchy of time scales  associated with a cascade of important events were monitored by measuring the time-dependent changes between the residues that occur during the $T \rightarrow R$ transition. The Brownian dynamics simulations \cite{Chen18JPCB}  show that at the Q-interface the distances exhibit a fast and slow phases. In the fast phase, the 
hydrophobic interactions are modified before the interactions change at the active site. In contrast, the slow phase lasts 
until the charge repulsion is neutralized by FBP binding to the P-interface. 
In detail, the hydrophobic interactions on the Q-interface between $\alpha$C(Q) and the active control loop, 
$\alpha$C(Q) and $\alpha$2F change in the fast phase, which result in contact formation between 
Arg171-Tyr190 and Asn140-His195 in the active site. Subsequently, the active control loop changes conformation resulting in a contact between 
Asp168-His195 forms, and then Ile240 moves out of the substrate binding pocket. These events result in the  closing of the active loop.  

In the slow phase, interactions on the Q-interface are altered.  Electrostatic  repulsion between residues 
Arg173 and His188 is reduced by binding of the ligand FBP, which leads to proximity of the two P-interface related subunits (see middle panel in Fig. \ref{LDH}B). At long times,  
the eventual decrease in the distances between  charged residues Asp168-Lys184, Arg170-Lys184 and Lys184-Lys184 decrease, thus stabilizing the P-interface by FBP do not increase monotonically.  The results established that $\alpha$C(Q) slides along the Q-interface and changes the active area conformation that is stabilized by binding of FBP to the  P-interface. Once again, in the tetrameric LDH we find that  the salt bridges acts as switches, driving the allosteric transitions.

In order to assess the relevance of the residues involved in the dynamics of allosteric signaling, the kinetics of assembly of LDH was investigated. Formation of a dimer, simulated using just two subunits, showed that in the predominant assembly pathway, prefolding of the subunits is required prior to dimer formation. However, the fully folded subunits were arranged in an incorrect orientation, which results in pausing of the assembly in metastable  intermediate. In this state the fraction of the interface contacts, $Q(t)_{interface}$, is in discordance with the value it adopts in the fully assembled tetramer.   The incorrectly oriented structures  show that one monomer of the dimer is in the position where a P-interface related subunit in the tetramer LDH should be, which suggests that the P-interface interactions contribute to the correct orientation of the Q-interface related subunits 1 and 3. An implication is that the symmetric tetramer structure should be more stable than the dimer. 
To establish this point, additional simulations were performed for the incorrectly oriented structures (cyan trajectory in Fig. \ref{LDH}C) by including the P-interface related subunits (2 and 4). In Fig.~\ref{LDH}D, $Q(t)_{interface}$ in the cyan trajectory  increases dramatically   from 0.05 to 0.7 (Fig. \ref{LDH}C). The structures of the intermediate state of the assembly shows that the incorrectly oriented structures anneal  in the tetramer assembly in contrast to the dimer formation in which the assembly is kinetically trapped due to energetic frustration. It is once again   the electrostatic interactions associated with the P-interface that stabilize the interaction on the Q-interface.   Interestingly, the dynamics of allosteric transitions also involve the interplay between charged and hydrophobic in traction across the P and Q interfaces respectively, leading to the surprising conclusion that the network of residues controlling allosteric transitions and the assembly of LDH are nearly the same. 
\\

{\bf Link between assembly, allosteric signaling, and AWD:}
The links between the slow phase in the allosteric transition, salt bridge formation and allosteric signaling were investigated by using the SPM to predict the AWD for LDH. 
 As in other systems a single mode was  found  to have the largest overlap, implying that it contributes most to the conformational changes in the
T$\rightarrow$R transition. Focusing on mode 7, site mutation effects for each residue of subunit 1 using the SPM were performed \cite{Chen18JPCB} in order to determine the AWD connecting the $T$ and $R$ states using the SPM equation (Eq. \ref{SPM}). The effect of mutation is 
evaluated by calculating the residue-dependent elastic energy change associated with mode 7 under perturbation. Seven  peaks  were associated with the most important residues in the AWD (see Fig. 6 in \cite{Chen18JPCB}).  Interestingly, all these residues are around helix $\alpha$D/E, $\alpha$F, and $\alpha$H, which are the structural elements  that undergo slow dynamical transitions between the $T$ and $R$ states. Most interestingly,  the allosteric transitions are governed by governed by salt bridge switching. 
It is surmised that these residues drive global motions in the allosteric transition.  Taken together, the residues that are involved in allosteric signaling also drive the dynamics of allosteric activation as well as tetramer assembly in LDH. 

It is interesting that the prominent feature in the allosteric transitions in the tetramers (SAHH and LDH) involve rotation of the subunits. In LDH this has implications for the assembly as well (Fig.~\ref{LDH}D). From the principle of parsimony it is tempting to speculate that in symmetric multisubunit proteins assembly is preceded by the formation of the individual subunits, with relative rotation being the rate limiting step. If this were to hold then it may well be the case that the interface residues play the crucial role in driving the allosteric transitions as well. Whether this conclusion holds for other multisubunit allosteric enzymes is an open question. The importance of interface residues in allosteric signaling in SAHH lends further credence to our findings in the LDH assembly.

\begin{figure}[h!]
\centering
 \includegraphics[width=0.7\textwidth]{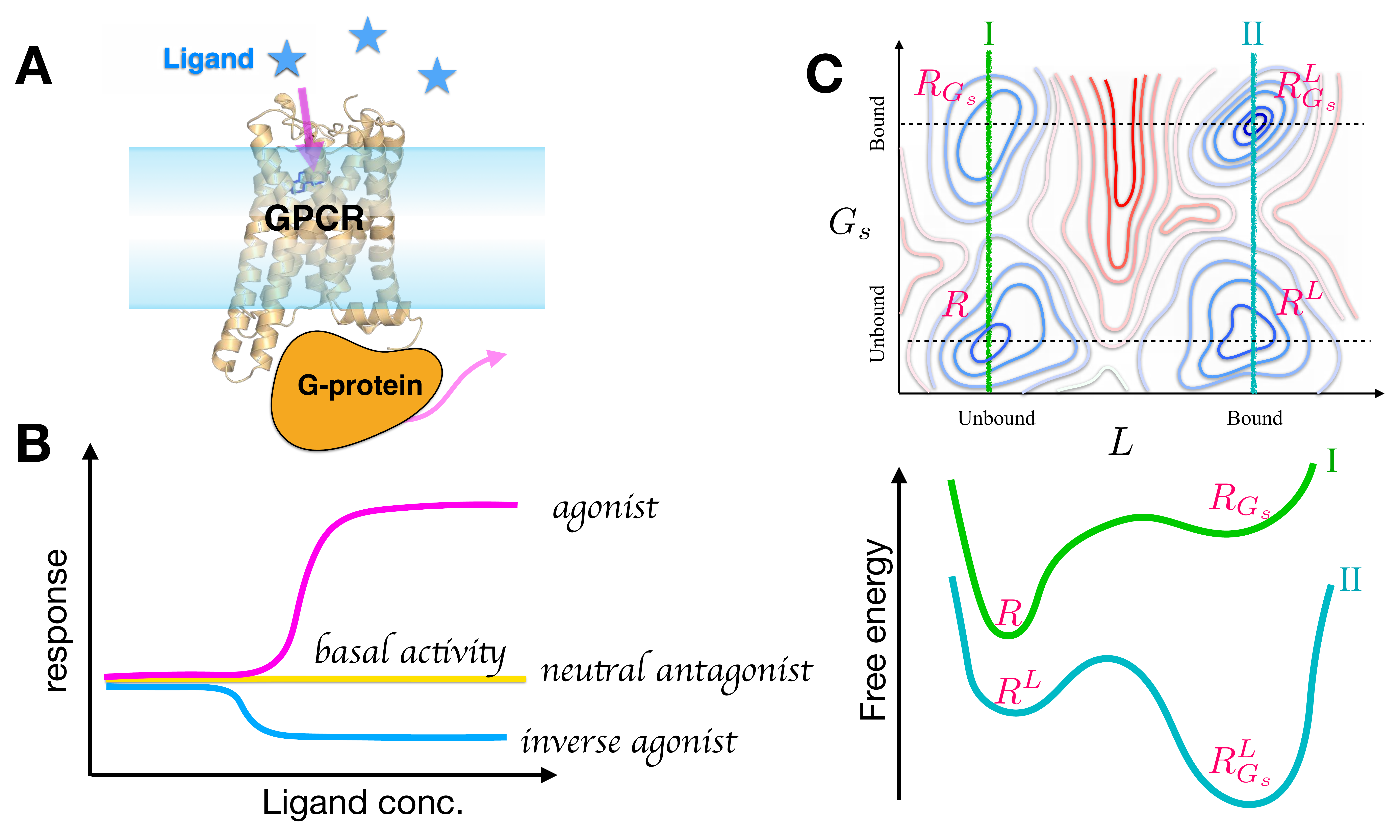}
   \caption{{\bf Signaling in GPCR.} 
(A) A cartoon of GPCR signaling. Ligand (agonist) binding to GPCR results in the accommodation of G-protein, which elicits further downstream G-protein response.  
(B) Typical biological response with increasing concentration of the cognate ligand. 
Agonist binding induces full biological responses above the basal line, whereas binding of inverse agonist or neutral antagonist suppresses signaling. 
(C) Hypothetical free energy surface describing GPCR activity. $R$ is a free GPCR with neither the ligand nor the G-protein. 
$R^L$ represents a ligand bound form of GPCR. 
$R_{G_s}$ is G-protein bound GPCR with no ligand. 
$R_{G_s}^L$ is G-protein and ligand bound GPCR. 
If $L$ is an agonist, the maximal signal from GPCR activity is expected to relay from the form of $R_{G_s}^L$. 
The relative stability of the receptor configuration without (I) and with ligand $L$ (II) is displayed at the bottom. 
Figure adapted from Reference \cite{jang2017JPCB}. 
\label{GPCR_cartoon}}
\end{figure}

\subsection{Allosteric activation in G-protein Coupled Receptors (GPCR)}
G-protein coupled receptors (GPCRs), are one of the key membrane protein families responsible for a plethora of sensory and physiological processes, such as vision, olfaction, cardiovascular functions, allergic/immune responses, and so forth. \cite{Rosenbaum2009Nature,schertler1993Nature,palczewski2000Science,rasmussen2011Nature}.  
The class A GPCR families share a common structural architecture consisting of seven transmembrane helices connected by three intracellular and three extracellular loops.   
Residing in plasma membranes, GPCRs serve as gatekeepers of extracellular signals, transmitting them into the intracellular domain with high precision and regulating the ensuing G-protein signaling pathways (Fig.~\ref{GPCR_cartoon}A).

High concentration of agonist boosts the activity of GPCR, facilitating the accommodation of G-proteins. In contrast, binding of inverse agonist (or neutral antagonist) suppresses the downstream signal even below  the basal level
 \cite{farrens1996Science,Rosenbaum2009Nature,rasmussen2011Nature,lebon2011Nature} (Fig.~\ref{GPCR_cartoon}B). 
 Without ligand binding or with neutral antagonist in the orthosteric binding pocket, GPCRs relay a basal level downstream signal involving the G-protein signaling pathway.  
 The existence of the basal signal implicates an occasional conformational fluctuation of the receptor in its active state to which G-protein can bind. A hypothetical free energy surface of GPCR, for a given ligand and G-protein concentration, is useful in visualizing this possibility (Fig.~\ref{GPCR_cartoon}C) \cite{jang2017JPCB}.  
Even in the absence of the ligand, GPCR could bind to the G-protein and form a holoenzyme complex ($R_{G_s}$). However, such a complex is thermodynamically unstable compared to the {\it apo} form of the receptor ($R$) (see the free energy profile along the line labeled with $I$). 
On the other hand, ligand (agonist)-bound form of the receptor ($R^L$) is likely to gain further stability by binding G-protein to form a holoenzyme complex ($R^L_{G_s}$). 
  
Given that almost 40 \% of the currently available drugs are targeted to GPCRs, a detailed knowledge of the physical underpinnings of GPCR function, more specifically the nature of the AWD and allosteric hotspots that regulate the function of GPCRs would be of critical importance for efficacious  rational drug design.    
A signaling molecule sensed at the orthosteric or allosteric ligand binding pocket elicits a long-range signal transmission across the trans membrane (TM) helices, thus giving rise to the conformational adaptation of the TM helices, which is required to accommodate the G-protein.  
The regulation of GPCR function by cognate ligands could be considered as a good example of allostery.   

However, compared with the allosteric transitions in prototypical systems that display large conformational changes, the structural alteration of GPCRs upon ligand binding is only minor. For example,  the root mean square distance (RMSD) between active and inactive conformations is as small as $\lesssim 3$ \AA.  When the active and inactive forms of GPCR are overlapped, the most conspicuous change is  the 10$^o$ tilt angle between the intracellular side of TM5, TM6 helices. 
Thus, similar to PDZ domain and DHFR \cite{Cooper84EBJ,Buchli13PNAS,Chen07JMB} the nature of allosteric signaling  in GPCRs is also  more dynamic than a simple change in conformations \cite{Lee14Proteins,Lee2015PLoSComp}.

Although the presence of signaling ``pathways" is questioned from the ensemble perspective of allostery \cite{pan2000PNAS,Motlagh2014Nature}, the foci of protein engineers often lie in identifying the key residues involved in long-range allosteric regulation and the pathways associated with them.
Concerning the allosteric hotspot of GPCR regulations, biochemical or site-directed mutagenesis studies have suggested 
fingerprint residues, called microswitches for rhodopsin-like class A GPCR family. 
It is well known that CWxP, DRY, ionic-lock, and NPxxY motifs, play key roles in orthosteric (allosteric) regulation \cite{nygaard2009TPS, katritch2013ARPT,Lee14Proteins,Lee2015PLoSComp}, where orthosteric regulation is a terminology used in GPCR research for the conventional regulation due to binding of a ligand to the  binding site in the vestibule formed by the seven TM helices in the EC domain.  
When the active and inactive structures of GPCRs are overlaid, the microswitches show changes in their side chain orientations although their backbone positions remain almost identical. 
A relay of signal propagated through the change in the rotameric states of the microswitches is deemed responsible for the aforementioned 10$^o$ outward tilt of TM5 and TM6 helices leading to a conformational transition from an inactive to active state, which enables the accommodation of the G-protein \cite{farrens1996Science,Rosenbaum2009Nature,rasmussen2011Nature}. 
Since the microswitches are critical in the allosteric dynamics of GPCRs they serve as benchmark residues for predicting the allosteric hotspots and the AWD of GPCRs.  

However, the microswitches of GPCRs as allosteric hotspot residues is not easy to identify  using frequently used computational methods due to the following reasons \cite{Lee14Proteins}.  
The structural fine-tuning around them from the inactive to active state is localized. As a result the stiffness of GPCRs is large, as revealed by localized structural transitions. Thus,  it is difficult to detect the principle microswitches using low frequency modes from normal mode analysis and the response to perturbations of these modes \cite{Zheng05Structure}. In this instance, it may be necessary to take into account coupling between low and high frequency modes, as was done in the context of myosin II\cite{Zheng09BJ}. 
In order to detect the overlap of the normal mode with the change in rotamer angle of the microswitch, one has to look for high frequency modes \cite{Lee14Proteins}.  
Second, the extent of sequence conservation, quantified using the MSA pf GPCR related sequences, shows that the sequences of microswtich residues in GPCRs are highly conserved. 
As a result, the SCA method that uses sequence covariation as a basis of capturing the evolutionary signature of allosteric communication \cite{Lockless99Science,Suel2002NSMB} cannot be utilized in identifying the microswtiches, which are the allosteric hotspots in GPCR.  

To cope with this unusual situation encountered in deciphering the GPCR allostery, 
one can represent the protein structure in terms of a connected network, and apply  graph theoretical methods to study the allostery.  
A series of centrality measures  could be used to analyze a protein structure represented by a graph, which is inferred from the structure. 
Among them, the betweenness centrality, $C_B$ (Eq.\ref{betweenness}), which measures at each node the amount of information to be transmitted for a given network structure (see Methods) \cite{Lee14Proteins}, is contextually the most suited to detect the allosteric signaling in GPCR. 

For the class A GPCRs, 18 microswitches (N24, D52, D101, R102, Y103, W129, P189, Y197, E228, C245, W246, P248, N280, S281, N284, P285, Y288, F295) \cite{Nygaard2013Cell,katritch2013ARPT} have been suggested to be responsible for allosteric regulation.   
They are part of three well known motifs: 
DRY (D101$^{3.49}$, R102$^{3.50}$, and Y103$^{3.51}$) in TM3, 
CWxP (C245$^{6.47}$, W246$^{6.48}$, and P248$^{6.50}$) in TM6, 
and NPxxY (N284$^{7.49}$, P285$^{7.50}$, and Y288$^{7.53}$) in TM7 \cite{suwa2011Pharmaceuticals,hofmann2009Hofmann} 
(where ÒxÓ stands for any amino acid residue and the numbers in the superscript are based on the Ballesteros Weinstein numbering system \cite{ballesteros1995MethodsNeurosci}). 
Historically, the importance of these residues was first recognized either due to the extent of sequence conservation among the class A GPCR family or by comparison between two GPCR subtype structures in different states. 
Their functional importance was subsequently confirmed by mutagenesis studies \cite{nygaard2009TPS,Rosenbaum2009Nature}. 
Thus, a receptor belonging to the class A GPCRs is expected to utilize many of these 18 microswitches for allosteric signaling. Although one should bear in mind  that the functional role of these microswitches have not been verified for all the GPCR subtypes, the 18 microswitches can be used as benchmark residues to assess the performance of a prediction tool in identifying the allosteric hotspots in GPCRs. 
\\

\begin{figure}[h!]
\centering
 \includegraphics[width=0.8\textwidth]{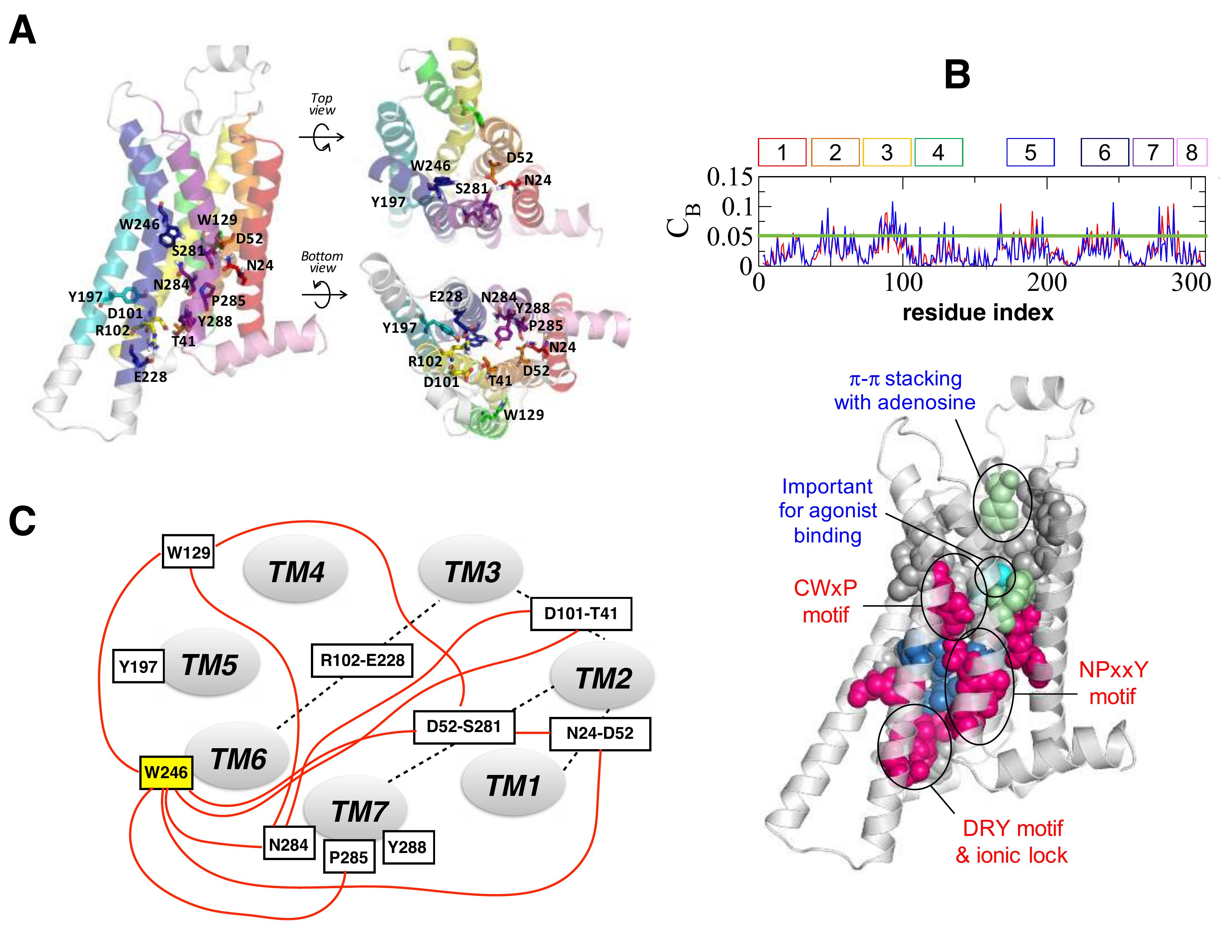}
   \caption{{\bf Intramolecular signal transmission of GPCRs.} 
(A) The structure of A$_{\text{2A}}$AR with annotated microswitch residues.  
(B) Betweenness centrality value ($C_B$) for the residues constituting the structure of A$_{\text{2A}}$AR. Highlighted are the residues with top 10 \% $C_B$ value, among which 11 microswitches are identified. 
(C) Cross-correlation among the local dynamics of residues around the 10 microswitches. 
W246 senses the presence of the agonist ligand in the orthosteric ligand binding site and actuates its signal to the rest of the receptor structure; and cross-correlation analysis of dynamic trajectories identifies W246 as the key hub residue of allosteric signaling pathways. The figure adapted from Reference \cite{Lee2015PLoSComp}. 
\label{GPCR}}
\end{figure}

{\bf Glimpses into GPCR allostery using the correlated fluctuation among microswitches:} 
The graph theoretical analyses of the crystal structures of A$_{\text{2A}}$ adenosine receptor suggests that 11 out of 18 microswitches, which have the top 10 \% of the $C_B$ values for all the residues, are the central hubs in the intramolecular signaling network \cite{Lee14Proteins} (Fig.~\ref{GPCR}C). 
A modification of these hub residues by  mutations is likely to impair the allosteric signaling of A$_{\text{2A}}$AR.

When the state of the molecular interaction is coarse-grained as a binary representation, $s_i =0$ (OFF) or $1$ (ON), in reference to those found in the agonist-bound active state,  
the cross-correlation $C_{ij}=\langle\delta s_i\delta s_j\rangle/\sqrt{\langle(\delta s_i)^2\rangle}\sqrt{\langle(\delta s_j)^2\rangle}$ where $\delta s_i\equiv s_i-\langle s_i\rangle$, calculated based on atomically detailed MD simulation trajectories, reveals that
the local dynamics of these microswitches depend on the type of ligand bound to the orthosteric site. 
In particular, they are highly correlated in the agonist-bound active state. 
Residue W246 located at the deep bottom of the orthosteric ligand binding cleft serves as an actuator of the ensuing intra-molecular signaling as well as a sensor of agonist ligand (Fig.~\ref{GPCR}C). 
\\

{\bf Water molecules in allosteric signaling of GPCRs: } 
Water molecules trapped inside biomolecules and identified in their crystal structures, often dubbed as crystal or biological waters, have been spotlighted in a series of studies. We have already highlighted their plausible importance in the allostery of the PDZ domain. 
Crystal waters have also been identified in GPCR structures, 
and their functional role, if any, in the allosteric signaling of GPCRs 
is an interesting topic to address. 
Although the TMs of GPCRs are generally hydrophobic, assembled inside bilayer lipids, the narrow channel formed by TMs is comprised of an array of polar residues, which allow water molecules to stay inside and to flow through, otherwise  impermeable, bilayer membranes.
Water molecules at the location where crystal waters are identified in the GPCR structure have low mobility, 
but there are also many water molecules with high mobility occupying the TM channel of the GPCRs.

Possible functional roles of internal water molecules inside the TM domain in GPCR activation have been discussed in a number of MD simulation studies  
for various GPCR subtypes: 
A$_{\text{2A}}$AR \cite{Yuan2014NatCommun,Yuan2015Angewandte,Lee2012BJ}, 
$\beta$2-adrenergic receptor \cite{bai2014PCCP}, 
rhodopsin \cite{Sun2014JPCB,Leioatts2014Biochemistry,grossfield2008JMB,Jardon2010BJ}, 
dopaminergic receptor \cite{Selent2010PLoSCompBiol}, 
and $\mu$-opioid receptor \cite{Yuan2013Angewandte}.  
These studies have shown that the TM pore of GPCRs is hydrated, and the volume of water occupying the pore changes with the receptor state. 
Among others,  
Leioatts \emph{et al.} \cite{Leioatts2014Biochemistry} have counted the number of water molecules inside TM domain and demonstrated that upon an elongation of the retinal, which corresponds to the retinal configuration seen in the active-like crystal structures, the influx of water increases inside the hydrophobic core of the protein TM domain and in turn induces a concerted transition in the highly conserved Trp265$^{6.48}$ residue.  

Dynamical properties of the internal water molecules over the entire architecture of GPCR have not fully been discussed.
While x-ray crystal structures provide a glimpse of ordered water molecules interacting with the interior of GPCRs \cite{Yuan2013Angewandte,burg2015Science,liu2012Science,Venkatakrishnan2013Nature},  these waters are  static, but have finite lifetimes. 
Furthermore, the roles of mobile water molecules with relatively fast relaxation kinetics is unknown. 
According to Lee \emph{et al.} \cite{Lee2016BiophysJ}, who have studied water dynamics in A$_{\text{2A}}$AR using $\mu$sec MD simulations, 
the average dwell time of water on water-GPCR interface, which include the interior of pore, varies over three orders of magnitude timespan from $\sim \mathcal{O}
(10^2)$ ps to $\sim \mathcal{O}(10^2)$ ns,  
depending on the location. Interestingly,  the water molecules of the TM channel in the agonist-bound active state flow three times more slowly than those in the antagonist-bound inactive state. 
In particular, they found that water molecules exhibit unusually slow relaxation ($\sim \mathcal{O}(10^2)$ ns) around the microswitch residues in the active state.

\begin{figure}[t]
\centering
 \includegraphics[width=0.7\textwidth]{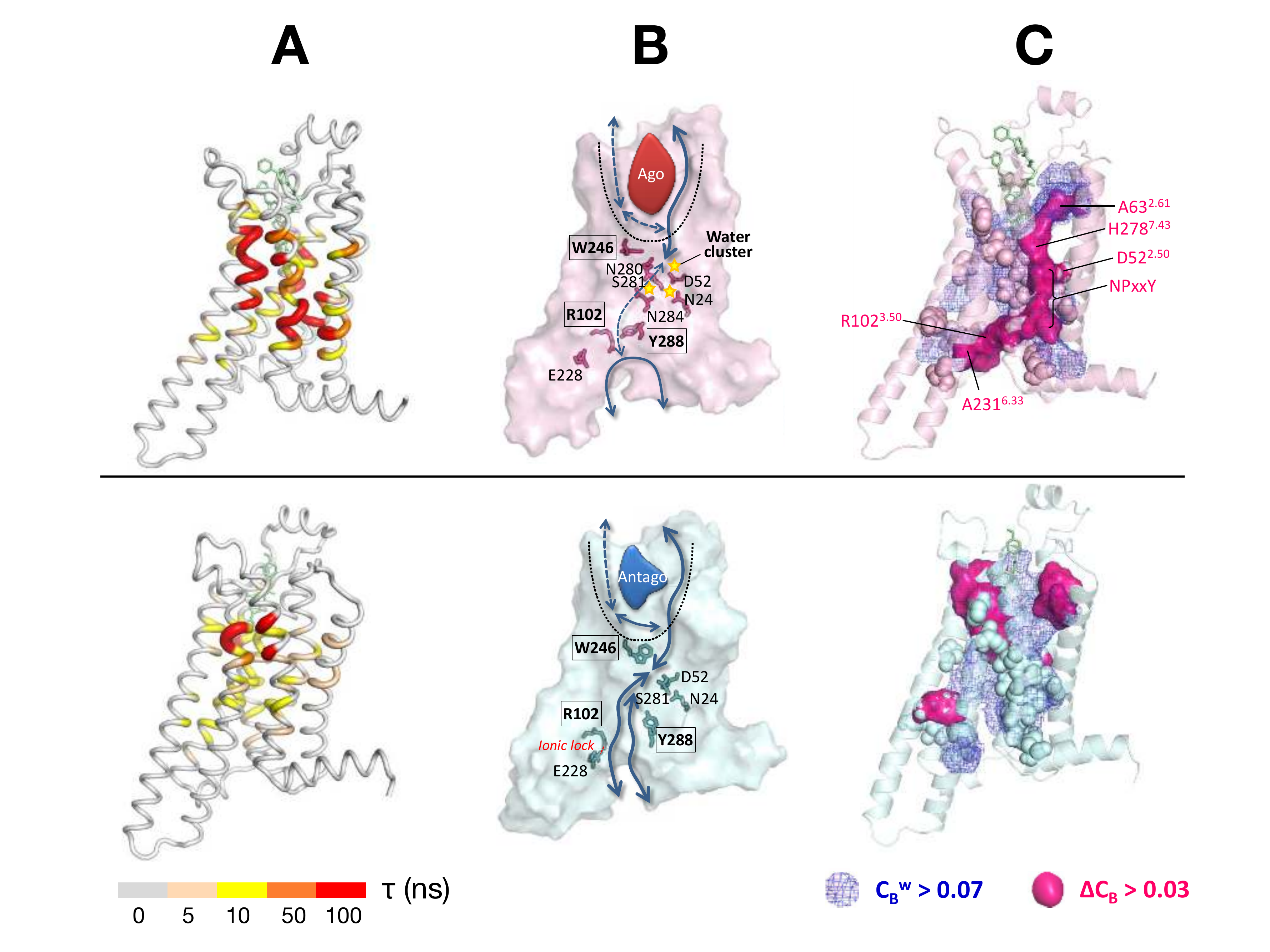}
   \caption{{\bf Role of water in allosteric signaling of A$_{\text{2A}}$AR.} Panels on the top and the bottom are for the agonist-bound active state and for the antagonist-bound inactive states.
(A) Relaxation time of water mapped onto the receptor structure. 
(B) Water flux map. Together with the key microswitches along the water channel, shown in solid and dashed lines, are the major and minor paths of water flux, respectively.  
When the ionic-lock in the inactive state breaks, R102$^{3.50}$ comes closer to Y288$^{7.53}$ in TM7, blocking the entry of water from the intracellular domain and reducing the water flux. 
(C)  
The values of betweenness centrality calculated by taking into account the inter-residue contact with and without taking into account water-mediated contact ($C_B^{\text{w}}$ and $C_B^o$. See \cite{Lee2016BiophysJ} for details), and their difference $\Delta C_B=C_B^{\text{w}}-C_B^o$, especially, $C_B^{\text{w}}$ and $\Delta C_B$   
are overlaid on the structures. 
High $\Delta C_B$ value area demarcated on the structure offers clear indication that the allosteric interface is extended by water-mediated interactions. The figure was adapted from Reference \cite{Lee2016BiophysJ}. 
\label{water}} 
\end{figure}

There are studies on GPCRs (e.g., rhodopsin \cite{Leioatts2014Biochemistry}), which proposed that a continuous stream of internal water is important for GPCR activation.
However, the continuous stream of water could be formed in both the active and the inactive receptor states.  
Water-mediated contacts bridging the key allosteric residues is more relevant for GPCR activation than the existence of a continuous water stream. 
In the agonist-bound active state, water molecules inside the TM channel are almost stagnant, displaying minimal flux and mobility; 
they stably hydrate the microswitches aligned along the TM7 helix. 
The water-mediated residue network extends from the extracellular domain to the intracellular part of the TM6 helix via the TM3 helix.
The water molecules around TM microswitches, some of which constitute the water cluster, stabilize the relative orientation and distance between TM helices by bridging them together. 
The interactions of internal water with microswitches, which contribute to extending and reinforcing the allosteric interface of GPCRs. 
These interactions are especially critical for the functional fidelity of the GPCR activity.

ÒThe number of water inside TM domainÓ and Òthe water flux across TM domain,Ó are two distinct physical quantities; one describes the statistics and the other dynamics. 
Both the active and inactive states of A$_{\text{2A}}$AR  accommodate comparable amount of water molecules at steady state; yet, the water flux in the inactive form at the steady state is greater than that in the active form by three times. 
The water molecules with slow relaxation time, stagnant around the TM channel in the vicinity of the microswitches (Fig.~\ref{water}A and B, top panels), allow the microswitches along the TM7 helix to form water-mediated contacts, and help extend the allosteric interface and contributes to maintaining the active configurations of the receptor (Fig.~\ref{water}C, top panel). 
The water-mediated AWD spanning the TM domain is unique in the active state (compare the top and bottom panels of Fig.~\ref{water}C), which underscores the importance of slow water molecules in the activation of GPCRs.       
The slow water dynamics at the core of TM domain and around microswitches are essential for the robust activation mechanism of A$_{\text{2A}}$AR.

\begin{figure}[t]
\includegraphics[width=0.7\textwidth]{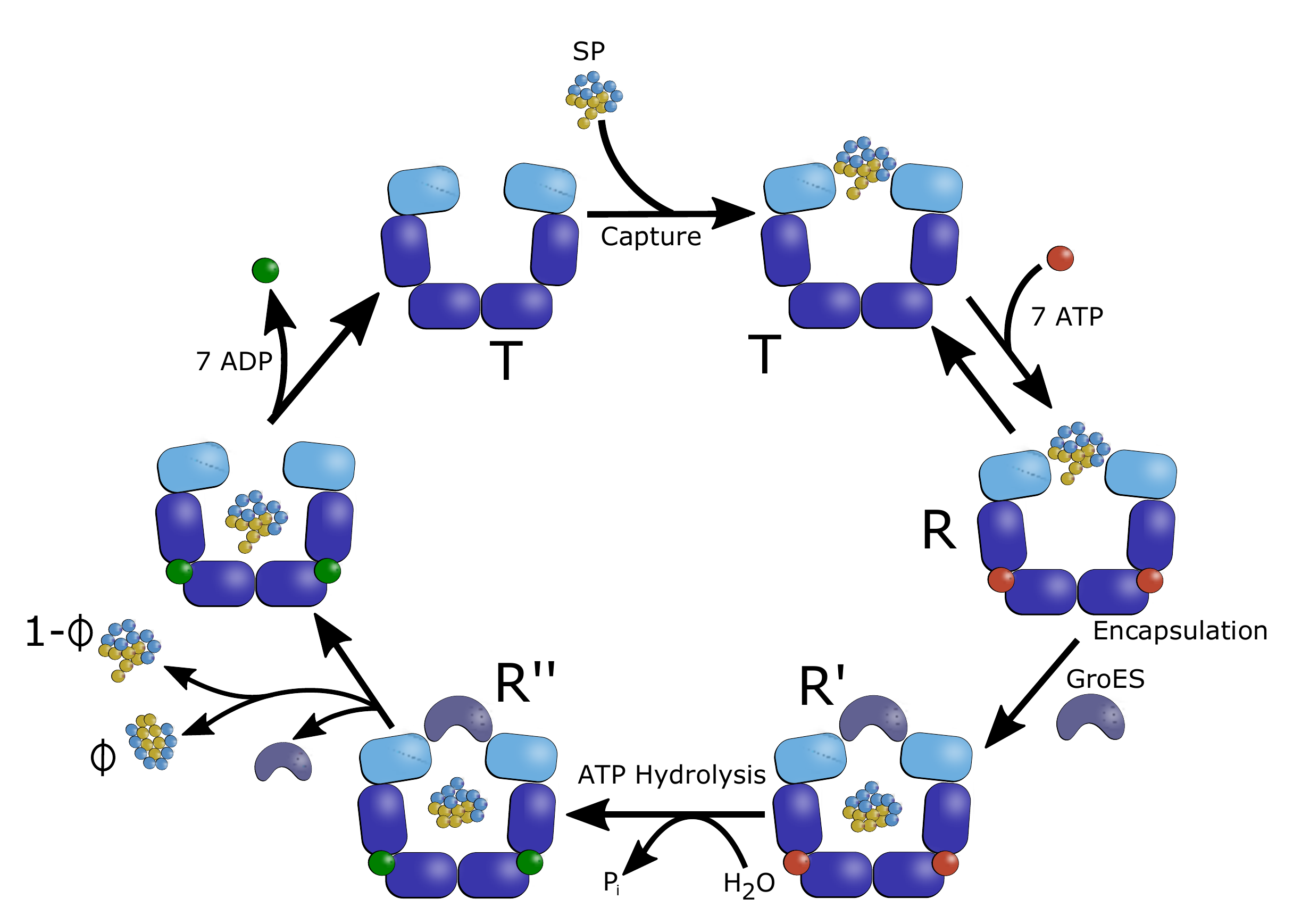}
\caption{Schematic illustration of the IAM illustrated using the hemicycle of GroEL and the kinetic partitioning mechanism \cite{Guo95BP}. The $T \rightleftharpoons R$ transition starts upon binding of ATP and the substrate protein (SP). GroES binding and ATP hydrolysis drives the $R^\prime \rightarrow R^{\prime\prime}$ transition with a fraction of the SP partitioning to the folded structure. Subsequently ADP, the inorganic phosphate, and the SP (folded or not)  are released and the $R^{\prime\prime} \rightarrow T$ transition completes the cycle. In the presence of the SP the machine turns over in about a second and ADP release is accelerated about a hundred fold. The rapid turn over is in accord with the predictions of the Iterative Annealing Mechanism (IAM) \cite{Todd96PNAS}. 
\label{fig:hemicycle}}
\end{figure}

\subsection{Allosteric transitions in the Bacterial Chaperonin GroEL}
GroEL, along with the co-chaperonin GroES constitute a stochastic parallel processing machine that rescues substrate proteins (SPs) that are otherwise destined for aggregation\cite{ThirumalaiARBBS01}. Although the GroEL-GroES machinery facilitates the folding of only $\sim$ (5-10)\% of the {\it E. Coli.} \cite{lorimer1996FASEBJ} proteome, deletion of the GroEL gene is lethal for the bacterium. The active mechanism by which GroEL helps the SPs fold, which is succinctly and quantitatively described by the Iterative Annealing Mechanism (IAM) \cite{Todd96PNAS}, involves a dynamic interplay between a series of allosteric transitions that the GroEL particle undergoes triggered by GroES, SP, and ATP binding. The $T$, $R$, and the  $R^{\prime\prime}$ are the three important allosteric states. High resolution structures of the $T$ state \cite{Braig94Nature}, the $R$ state \cite{Fei13PNAS}, and the $R^{\prime\prime}$ state \cite{Xu97Nature} have been determined. The $T$ state is predominantly populated in the absence of ATP, binding of ATP to the equatorial domain of the GroEL subunits drives the reversible $T \rightleftharpoons  R$ transition, and the $R^{\prime\prime}$ state is formed by GroES binding followed by ATP hydrolysis. The transition between the  states $R$ and $R^{\prime\prime}$ is irreversible, thus driving the GroEL-GroES machine far from equilibrium \cite{chakrabarti2017PNAS}. Indeed, the hydrolysis of ATP and the release of ADP and the inorganic phosphate competes the catalytic cycle, poising the machine to begin anew another cycle (Fig. \ref{fig:hemicycle}). 

GroEL is a homo oligomer with two heptamers that are stacked back-to-back. 
The subunits, which are identical, confer GroEL an unusual seven fold symmetry in the resting ($T$ or taut) state. Large scale conformational changes between the allosteric states of GroEL, $T\rightarrow R$ and $R \rightarrow R^{\prime\prime}$ transitions (see Fig. \ref{fig:hemicycle} for a schematic of the reaction cycle in a single ring), could be triggered solely by ATP binding and hydrolysis without the SP. This allows for the study of allosteric signaling in GroEL without the additional complication due to interactions with the SP. The ATP binding sites are localized in the  equatorial (E) domain in which about two-thirds of the inertial mass of GroEL resides. This makes the equatorial plate as the anchor, which likely serves as a base in the transmission of both intra and inter ring signaling. 

The nature of the reversible $T \rightleftharpoons R$ transition was first elucidated experimentally in pioneering studies by Yifrach and Horovitz \cite{Horovitz01JSB,Yifrach95Biochem} who also established an inverse relation, predicted earlier by computations \cite{BetancourtJMB99},  between the extent of cooperativity in this transition and the folding rates of certain SPs \cite{Yifrach00PNAS}. The irreversible $R\rightarrow R^{\prime\prime}$ transition is driven by ATP hydrolysis. In both these transitions strain due to ATP binding and hydrolysis at the catalytic site propagates through a network of inter-residue contacts \cite{Tehver09JMB}, thus inducing  large scale conformational changes. That such changes must occur during the reaction cycle of GroEL is evident by comparing  the static crystal structures in different allosteric states, such as the $T$ and $R^{\prime\prime}$ states \cite{Xu97Nature}. However, the static structures do not provide any information about the network of residues that carry allosteric signals, the dynamics of transition between the key states in the GroEL reaction cycle, the domains that undergo the largest changes, and most importantly a link to the function of GroEL.

The allosteric transitions involve remarkably large conformational changes in GroEL resulting in nearly two fold increase (upon completion of the $T \rightarrow R^{\prime\prime}$ transition) in the volume of the cavity\cite{Xu97Nature} that encapsulates the SPs for a brief period when the machine is active\cite{ThirumalaiARBBS01,Fei13PNAS}. Understanding the dynamics of the allosteric transitions and the network of residues that transmit signals across GroEL is the key to describing  its function from a structural perspective. The bacterial chaperonin is an outstanding example of a molecular machine in which binding and hydrolysis of ATP and GroES binding leads to large allosteric transitions\cite{Gruber16ChemRev,Sharon15COSB}, which are vital in its ability to facilitate the folding of SPs. One cannot account for its function by being seduced  into thinking that GroEL-GroES machine merely serves as an Anfinsen cage in which the SP is encapsulated till it folds, as some researchers  erroneously insist because   the static structure of GroEL has a cavity. Here, we focus first on some aspects of the spectacular allosteric transitions, focussing largely on the $T \rightarrow R$ transition, before describing the link to function.   

We should note that the $T \rightarrow R$ transition is reversible and is triggered by ATP binding. Based on the MWC-like models it might be tempting to conclude  that after a few ATP molecules bind to the $T$ state, GroEL ought to be predominantly in the $R$ state. This appears not to be the case, at least in a single ring construct of GroEL. Using  fluorescence correlation spectroscopy it has been shown \cite{Frank10PNAS} that even under saturating ATP conditions the $R$ state is only populated at 50\% level, which implies that this construct visits the $T$ state frequently, which the authors interpret an out of equilibrium reaction. This interesting result, which has not attracted much attention, given its general importance in allosteric signaling, should be investigated further.  Of course, from the point of view of  function, which for stringent substrates requires GroES, it is unlikely that the $R \rightarrow T$ transition would occur to ant measurable extent. 
\\

\begin{figure}[t]
\includegraphics[width=0.45\textwidth]{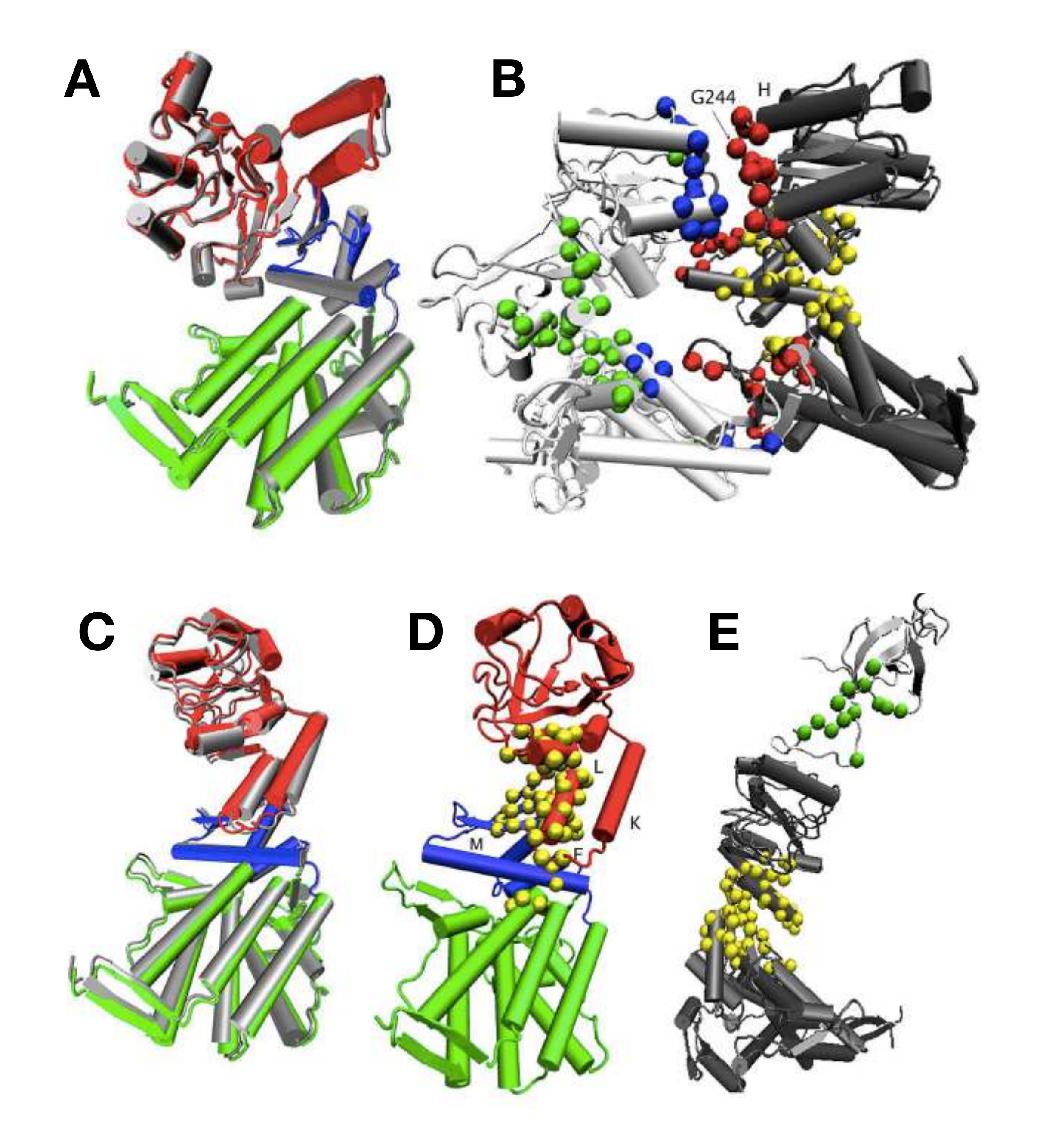}
\caption{
Illustration of the $T \rightarrow R$ transition and the associated allostery wiring diagram determined using the SPM. (a) Single-subunit structure in the $T$ state. The equatorial, the intermediate, and the apical domains are shown in green, blue, and red, respectively. The motions of the structural elements due to the dominant mode are in gray. (b) Structure of two adjacent subunits of GroEL (the chains are shown in dark and light gray) in the $T$ state. The  residues in the AWD are highlighted in color. The critical interface residues are in red and blue, and the other hot-spot residues are shown in yellow and green. (c) Same as (a) except this describes the $R \rightarrow R^{\prime\prime}$ transition. (d) The AWD  for the transition from the $R^{\prime\prime}  \rightarrow T$ state are in yellow. Helices K, L, F, and M  are labeled. The domains are colored as in (c). (e) GroEL (dark gray)--GroES (light gray) model. The AWD  is shown in yellow (GroEL) and green (GroES). This figure provides insights into the residues that signal the disassociation of GroES, a key event in the catalytic cycle of  chaperonin (see Fig. \ref{fig:hemicycle}). 
The figure adapted from Reference \cite{Tehver09JMB}. 
\label{GroELStruct}}
\end{figure}

{\bf AWD for the $T \rightarrow R$ transition using SPM:} The first step in the SPM is to  perform a normal mode analysis using the energy
function in Eq. \ref{ENM} in order to generate the spectrum of frequencies for
the normal modes along with the corresponding eigenvectors. 
Applications of  ENM to a large number of systems including GroEL \cite{Bahar2007COSB,Zheng06PNAS,Zheng07BJ,Zheng09CurrProtScience} have shown that typically  only a few of the lowest-frequency normal modes are required to characterize the allosteric transitions.  
In order to identify the modes that best describe the transition between two allosteric states an overlap function \cite{zheng03PNAS} between $T$ and $R$ is calculated using the eigenvectors of a specific mode. The modes with large overlap values (typically only one or two are needed) describe the $T \rightarrow R$ transition. Response to perturbations to the modes with high overlap is used to calculate the AWD using the SPM.  

\begin{figure}[t]
\includegraphics[width=0.5\textwidth]{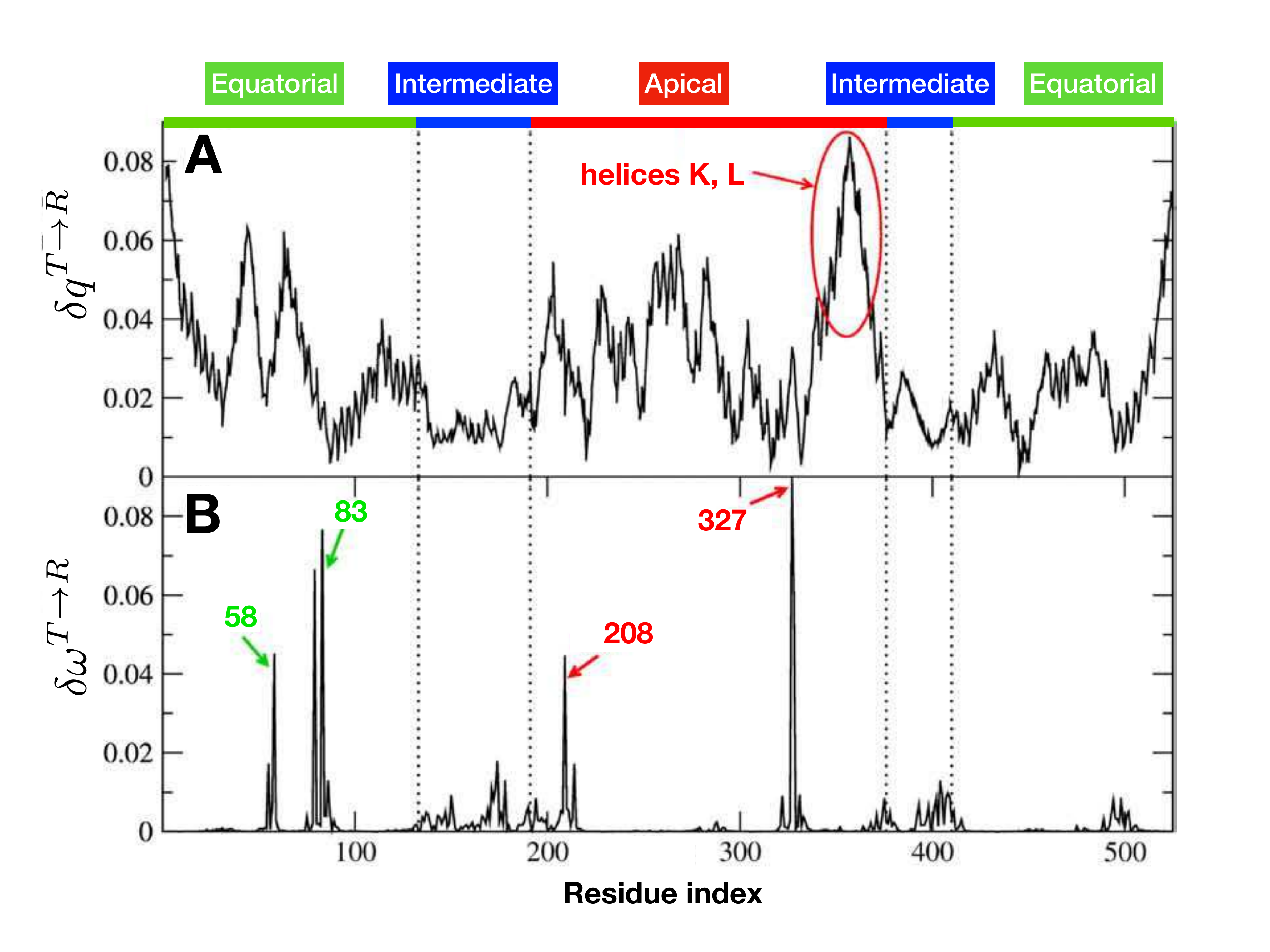}
\caption{SPM results for the GroEL with two adjacent subunits. (A) The amplitudes of motion in the dominant normal modes. The region with the highest amplitude corresponds to helices K and L. (B) Residue-dependent $\delta \omega$ for the dominant modes. The residues with the largest allosteric signal transmitting  values $\delta \omega$ are identified. The figure was adapted from Reference \cite{Tehver09JMB}.
\label{SPMTR}}
\end{figure}

 It is known that the intra ring $T \rightleftharpoons R$ transition is cooperative and likely concerted, which implies that changes in one subunit could spread to the neighboring rings. In order to describe the cooperative nature of the interring ATP-driven allosteric transitions, it is important to determine the interface residues in the AWD that are also involved in the transmission of allosteric signals. The most significant interface residues were identified by constructing two subunits of GroEL in the $T$ state (Fig. \ref{GroELStruct}A). 
 The $T\rightarrow R$ transition in an interacting two adjacent subunits is best described by two modes 7 (overlap 0.49) and 13 (overlap 0.35) \cite{Tehver09JMB}. Helices K and L (residues 339--371) exhibit the largest amplitudes of motion for the  two modes (Fig. \ref{SPMTR}A). 
The SPM result for the modes in Fig. \ref{SPMTR}B shows that residues D83 and K327 as well as G244 have the largest $\delta \omega$ values. The identification of G244 as playing a key role is noteworthy because it has been noted \cite{Brocchieri00ProtSci} that the number of contacts involving G244, located at the end of helix H and at the interface between the two subunits in the apical domain in $T$ the state, changes significantly during the $T \rightarrow R^{\prime\prime}$ transition. This residue is also located next to a set of highly conserved residues (246--253).

\begin{figure}[h!]
\includegraphics[width=0.6\textwidth]{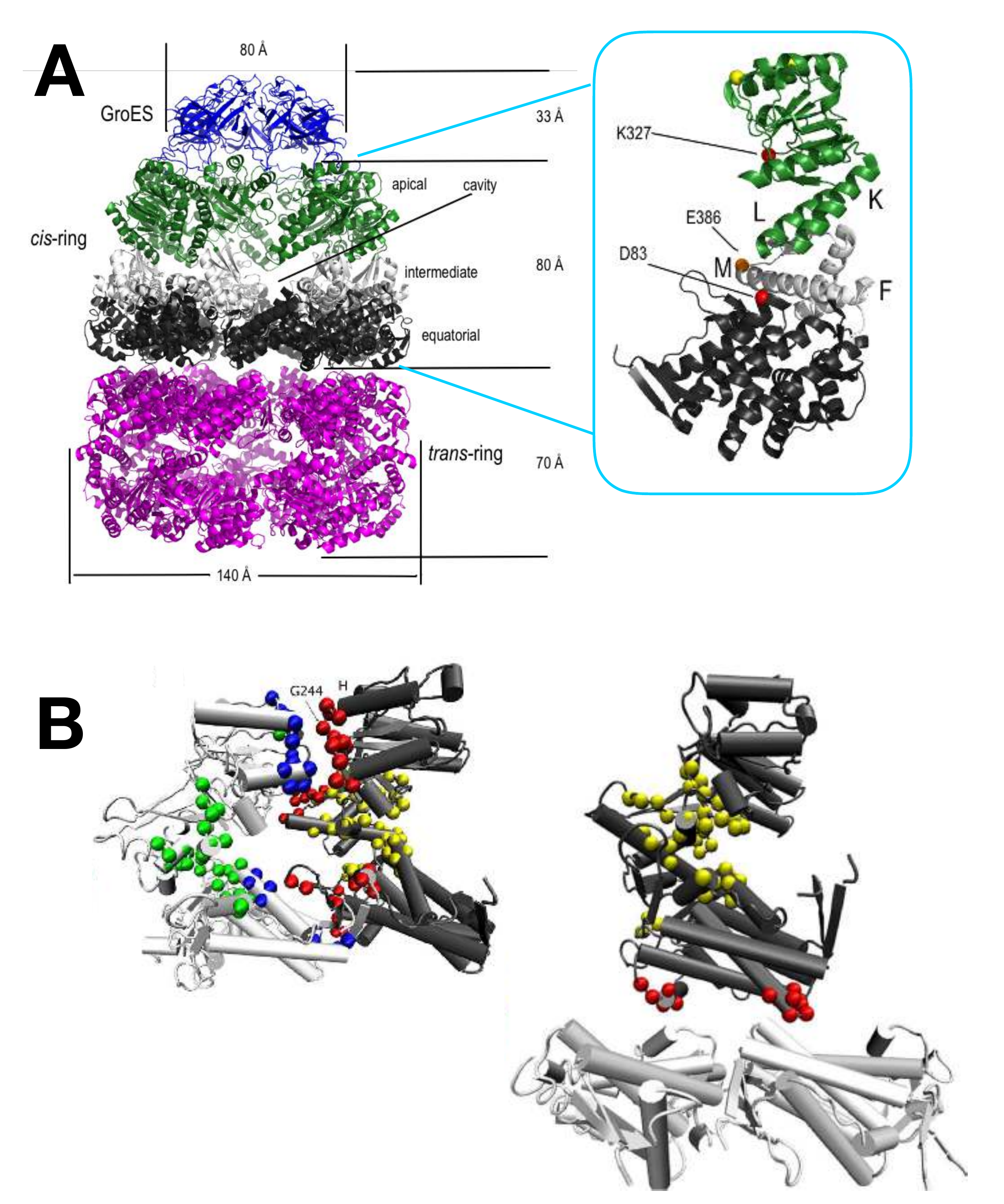}
\caption{
(A) Cartoon representation of the $R^{\prime\prime}$ structure is on the left. On the right  a single 548 residue subunit of GroEL from the \textit{cis}-ring is shown. The domains are color coded: the apical domain is green, the intermediate domain is light gray, and the equatorial domain is dark grey. The important salt bridge disrupted in the $T \rightarrow R$ transition is formed by residues D83 and K327 (highlighted in red) as determined by Brownian dynamics using the Self-Organized Polymer model simulations \cite{Hyeon06PNAS}. SPM indicates these two residues as the `hottest' in for the $T \rightarrow R$ transition. The tip of helix M (highlighted in orange)  also participates in several salt bridges. The resides highlighted in yellow form an inter subunit salt bridge, which is also disrupted in $T \rightarrow R$ transition.
(B) Allosteric hotspot residues of GroEL detected by the SPM. On the left is the structure of two adjacent subunits of GroEL (the chains are shown in dark and light gray) in the $T$ state. The AWD residues are highlighted in color. The critical interface residues are in red and blue, and the other hot-spot residues are in yellow and green. The interface residue G244 (see the text) is explicitly labeled. On the right is the $T$ state structure (dark gray) with adjacent {\it trans} equatorial domain (light gray). The interface hot-spot residues, which are likely involved in the inter ring communication, are shown in red.  The rest are shown in yellow.
\label{GroEL_SPM}}
\end{figure}

By mapping the hot-spot residues (listed in Table 1 for mode 7 in \cite{Tehver09JMB}) onto their structures, we find that 33 of the 85 hot-spot residues of chain H (per the chain labeling in the PDB structure 1AON) and 24 of the 62 hot-spot residues of chain I belong to the inter-subunit interface.  The interface hot-spot residues, highlighted in green, red and blue in Fig. \ref{GroEL_SPM} (left), 
show that the large number of interface residues in the AWD, and is the possible foundation for the strong intra-ring positive cooperativity.

A calculation of the normal modes of GroEL with the $T$ state structure with additional two {\it trans} ring equatorial domains (Fig. \ref{GroEL_SPM}B (right)) leads a single dominant normal mode (mode 10) with a significant overlap \cite{Tehver09JMB}. 
The residues in the apical domain have the  largest fluctuations, as also shown  in recent crystal structures\cite{Fei13PNAS}. Interestingly, there appears to be an asymmetry in the amplitudes in the two equatorial domains. Perturbations of the residues in this mode reveal that the residues with the highest $\delta \omega$ values are D83, E209, and K327 and the 12 hot-spot residues that are in the inter-ring interface as shown in Fig. \ref{SPMTR}B. Based on the AWD and the locations of the hot-spot residues, the interring interface interactions seem to play a less critical role in the inter ring $T\rightarrow R$ transition implying that cooperativity arises due to  intra ring transitions. Similar analyses has been carried out for the more spectacular $R \rightarrow R^{\prime\prime}$ transition (see the study by Tehver \cite{Tehver09JMB}).
\\

{\bf Dynamics of allosteric transitions in GroEL:}  Does the AWD also play an important role in the dynamics of allosteric signaling in GroEL? To answer this question requires generating a number of trajectories connecting the distinct allosteric states for long times, which given the size of GroEL is currently difficult to execute using atomically detailed MD simulations even using special purpose computers. As a result coarse-grained models\cite{hyeon2011NatureComm,Whitford12RepProgPhys} have been used to monitor the molecular events involved in large systems. Using Brownian dynamics simulations of the Self-Organized Polymer (SOP) model \cite{Hyeon06Structure} both the $T\rightarrow R$ and $R \rightarrow R^{\prime\prime}$ transitions were investigated by generating multiple trajectories \cite{Hyeon06PNAS}.
\\

\begin{figure}[h!]
\includegraphics[width=0.8\textwidth]{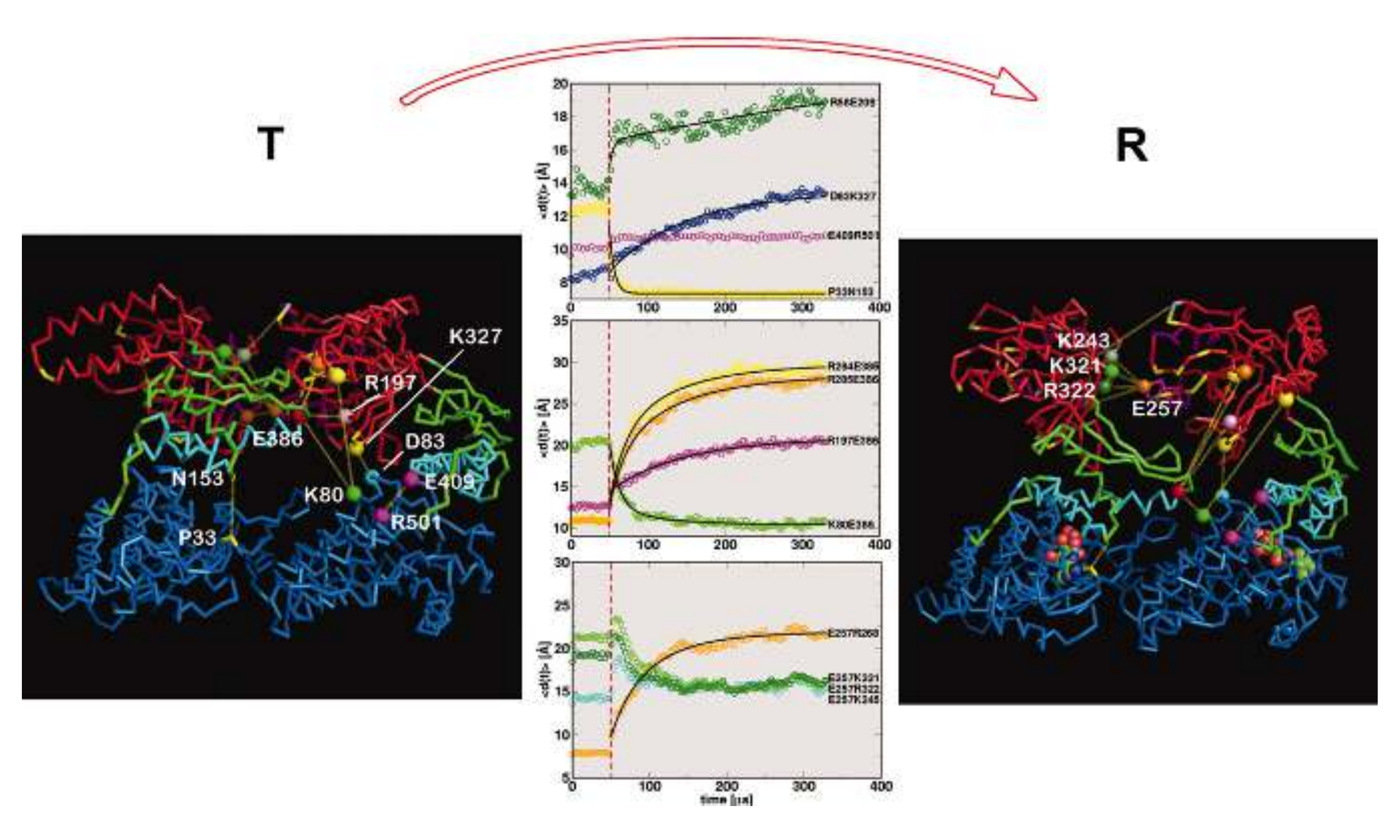}
\caption{$T\rightarrow R$ GroEL dynamics monitored using Brownian dynamics simulations of two interacting subunits. 
Side views from outside to the center of the GroEL ring and top views are presented for the $T$ (left panel) and $R$ (right panel) states. 
Few residue pairs  are annotated and connected with dotted lines.  
The ensemble average kinetics of a number of salt-bridges and contacts between few other residues are shown in the middle panel.  
Relaxation dynamics of distance between some of two residues of interest are fit to the multi-exponential kinetics \cite{Hyeon06PNAS}. Note that the distance criterion to define a salt-bridge in the SOP model with a single bead for each residue, located at the $C_{\alpha}$ position, is longer than it would be in an all-atom representation by a few \AA. 
The figure taken from Reference \cite{Hyeon06PNAS}. 
\label{saltbridgeanalysisTR}}
\end{figure}

{\it  Dynamics of the $T \rightarrow R$:} Although the initial event in the $T \rightarrow R$ transition involves ATP-binding-induced downward tilt of the F and M helices, which closes the ATP binding pocket in the equatorial domain\cite{Hyeon06PNAS}, here we focus on the dance of the salt bridges, which act as switches in the GroEL allosteric transitions. The rupture of the intra-subunit salt-bridge  D83-K327 occurs nearly simultaneously with the
breakage of the E386-R197 inter-subunit interaction (the top panel in Fig.~\ref{saltbridgeanalysisTR}). 
K80-E386 salt-bridge is formed around the same time as the rupture of the R197-E386 interaction.
 In the $T \rightarrow R$ transition a network of salt-bridges breaks and new ones form (see below). 
At the residue level, the reversible formation and disruption of the D83-K327 salt-bridge, in concert with
the inter-subunit salt-bridge switch associated with E386 \cite{Ranson01Cell} and E257 \cite{Stan05JMB,Danziger06ProtSci}, are among the most significant  events that dominate the $T \rightarrow R$ transition. It must be emphasized that the residues that have been identified to be important in the dynamics  are also part of the AWD, which was predicted using only  the static structure.

\begin{figure}[h!]
\includegraphics[width=0.70\textwidth]{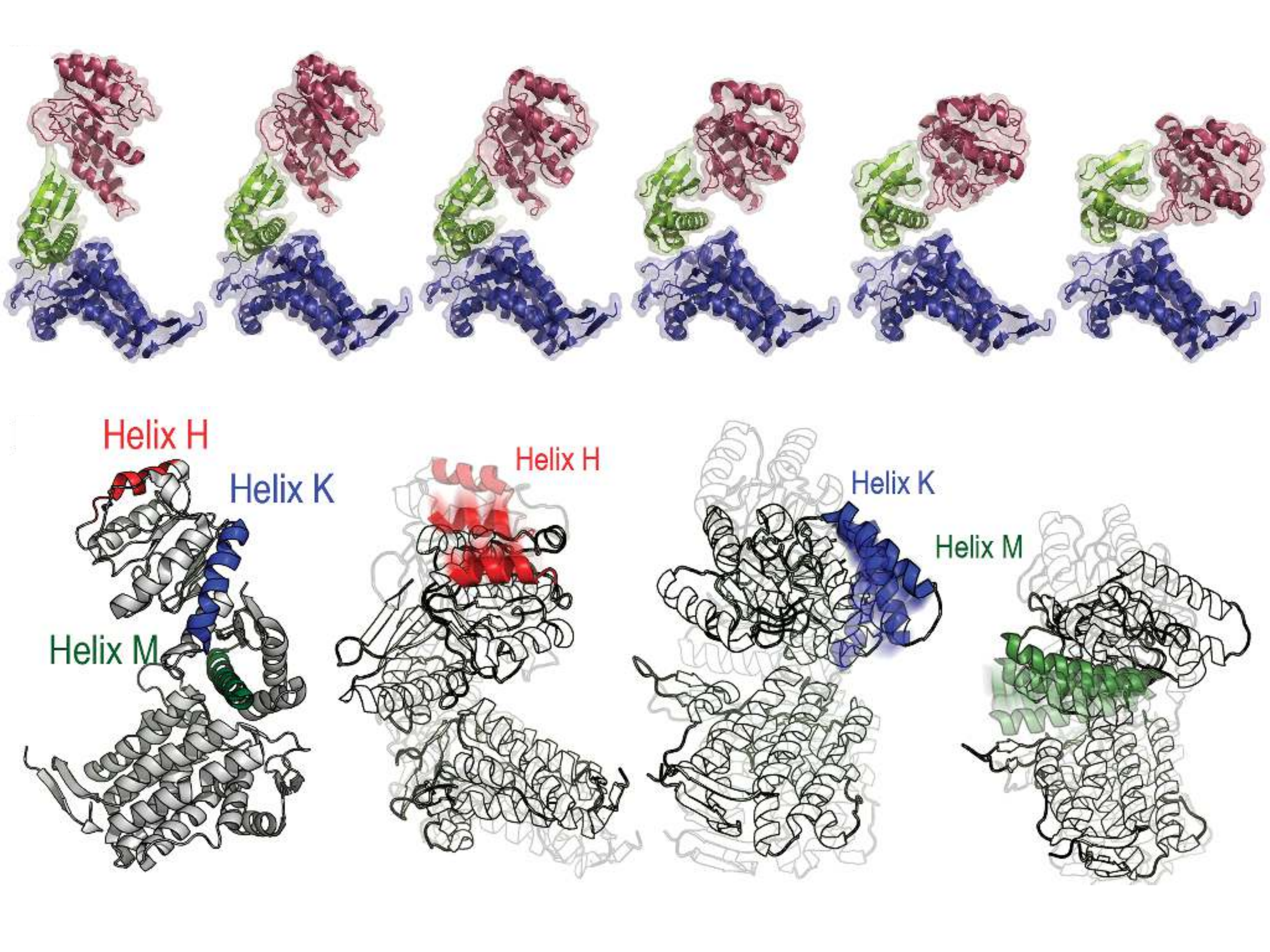}
\caption{(Top row) GroEL conformations along the reaction coordinate. 
Each domain (equatorial, intermediate, and apical domains) is colored in blue, green, and red, respectively. 
(Bottom row) Conformational transitions of H, K, M helices along the transition pathway. From the lightest to the darkest color, shown is the transition from $R^{\prime\prime}$ to T states. 
The figure taken from Reference \cite{Yang09PlosCompBiol}. 
\label{GroEL_Bahar}}
\end{figure}

The robustness of the findings reported in \cite{Hyeon06PNAS} was subsequently validated by Yang, Majeck, and Bahar \cite{Yang09PlosCompBiol} using an adaptive Anisotropic Network (aANM) model. Although the dynamics in this model is somewhat artificial,  the sequence of events at the molecular level could be predicted by using the observed time-dependent changes in the BD simulations\cite{Hyeon06PNAS} as a reference\cite{Yang09PlosCompBiol}. Remarkably, both these totally independent methods produce a consistent picture of the sequence of internal rupture and formation of contacts in the GroEL allosteric transition (see Table 6 in \cite{Yang09PlosCompBiol}. A series of conformations that are sampled, starting from the $R \rightarrow R^{\prime\prime}$ state, generated using aANM, vividly illustrates the large scale changes in a single subunit of GroEL (Fig. \ref{GroEL_Bahar}).
  \\

{\bf Salt bridges as molecular switches:} The coordinated global motion is orchestrated by a multiple salt-bridge switching mechanism.  
The movement of the  A domain results in the dispersion  of the SP binding sites  also  leads to the rupture of the  E257-R268 inter-subunit salt-bridge.  To maintain a stable configuration in the $R$ state, E257 engages in salt-bridge formation with positively charged residues that are initially buried at the interface
of the inter-apical domain in the $T$ state.  Three positively  charged residues at the interface of the apical domain in the $R$ state 
(K245, K321, and R322) are the potential candidates for the salt-bridge with E257.  During the $T \rightarrow R$ transitions E257 interacts partially with K245, K321, and R322 as evidenced by the decrease in their distances (the last panel in the middle column of Fig.~\ref{saltbridgeanalysisTR}). 
Similarly, the $R \rightarrow R^{\prime\prime}$ transition, which is accompanied by the doubling of the volume of the cavity, is also driven by rupture and formation of many salt bridges\cite{Hyeon06PNAS}. 

Interestingly, it was noted sometime ago that the rupture of salt bridges provides a structural explanation for allosteric transition between the $T$ and $R$ states in hemoglobin. The consequences of the Perutz mechanism \cite{Perutz70Nature,Bettati98JMB}  involving salt bridges as a switches, which are present in the $T$ state with low affinity for oxygen but are disrupted in the deoxygenated $R$ state, was put on a firm theoretical footing by Szabo and Karplus \cite{Szabo72JMB}. Together these studies showed how the coupling between tertiary structures and changes in the quaternary structures might occur in the process of the $T \rightarrow R$ transition. Although it is still unclear if these changes are related to oxygen transport and subsequent deoxygenation, it appears that nature utilizes salt bridges as switches in allosteric signaling and function even though folding of globular proteins may be dominated by hydrophobic interactions. To ascertain the generality of the role salt bridges might play as a universal switches in biology would require in depth structural and dynamical studies of many allosteric proteins. 
\\

\begin{figure}[h!]
\includegraphics[width=0.8\textwidth]{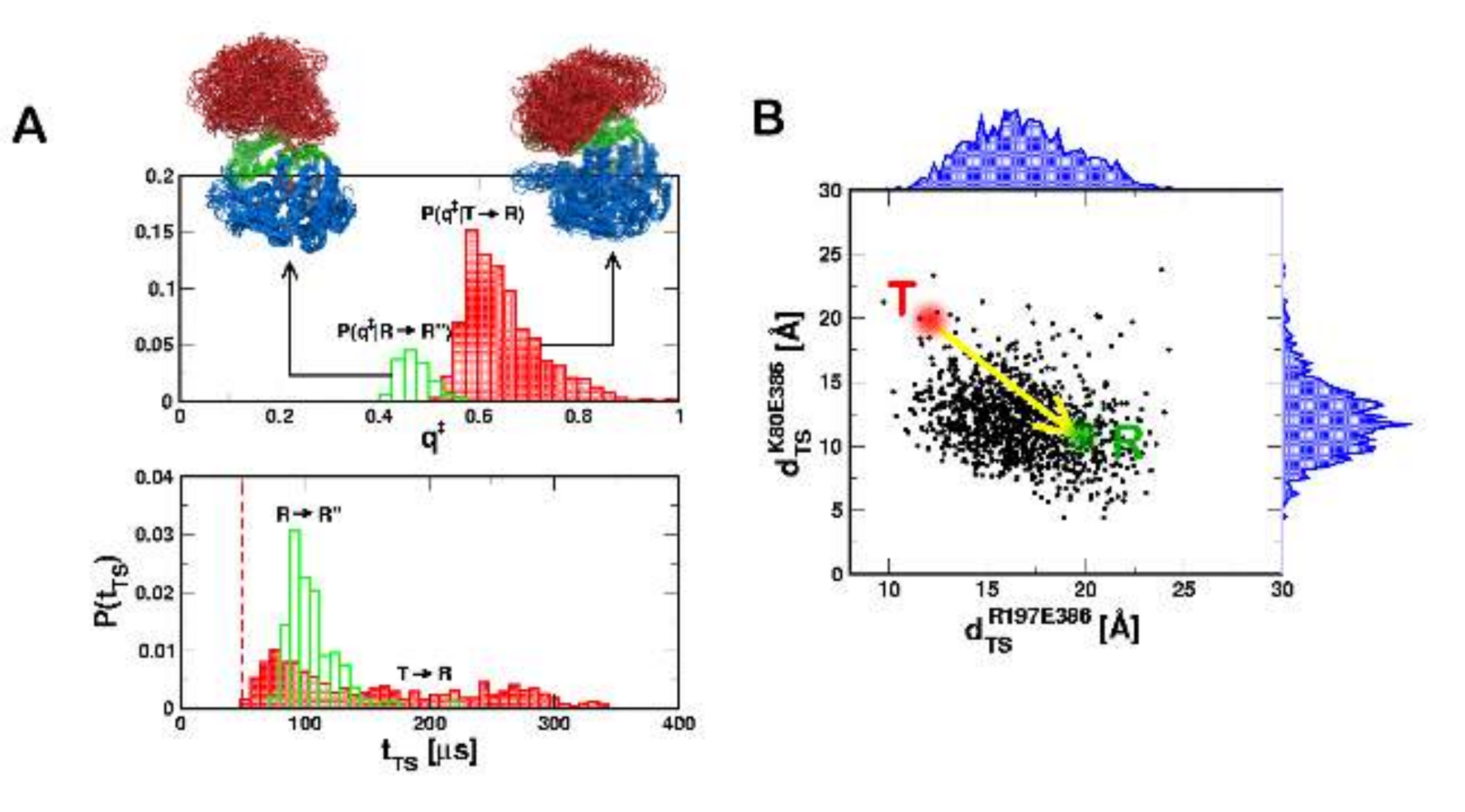}
\caption{Transition state ensembles (TSE) in the allosteric transitions in GroEL. 
{\bf A}. TSEs are represented in terms of the distributions $P(q^{\ddagger})$ where  
$q^{\ddagger}\equiv(\Delta^{\ddagger}-\min(\text{RMSD}/X))/(\max(\text{RMSD}/X)-\min(\text{RMSD}/X))$.  
Histograms the normalized $P(q^{\ddagger})$ are for $T\rightarrow R$ (red) and $R\rightarrow R^{\prime\prime}$ (green).   
Twenty overlapped TSE structures  for the two transitions are displayed.   
In the bottom panel, the distributions of $t_{TS}$ that satisfy  $\delta^{\ddagger} < 0.2$ \AA, 
are plotted for the $T\rightarrow R$ and the $R\rightarrow R''$ transitions. 
{\bf B}. TSE for the $T\rightarrow R$ transition represented by the pair of two salt-bridge distances $(d^{R197-E386}_{TS},d^{K80-E386}_{TS})$ (black dots). 
The equilibrium distances 
$(\langle d^{R197-E386}_{TS}\rangle,\langle d^{K80-E386}_{TS}\rangle)$ in the $T$ and the $R$ states are shown using the red and the green dots, respectively. 
The distance distributions for the TSE are shown in blue. The figure taken from Reference \cite{Hyeon06PNAS}. 
\label{TSEGroEL}}
\end{figure}

{\bf Partial unfolding and Transition State Ensembles (TSE)s:} 
The Brownian dynamics trajectories have been used to obtain the structures in the  TSEs connecting the $T$, $R$, and $R^{\prime\prime}$ states by using the RMSD as a surrogate reaction coordinate. 
It is assumed that the TS location is reached if 
$\delta^{\ddagger} = |(\text{RMSD}/T)(t_{TS}) - (\text{RMSD}/R)(t_{TS})|< r_c$, where $r_c=0.2$ \AA\ and $t_{TS}$ is the time at which $\delta^{\ddagger} <  r_c$. 
If the RMSD at the TS is $\Delta^{\ddagger} = 0.5|(\text{RMSD}/T)(t_{TS}) + (\text{RMSD}/ R)(t_{TS})|$, a Tanford-like $\beta$ parameter $q^{\ddagger}$  may be defined (see the caption  Fig. \ref{TSEGroEL} for definition).  The mean values of $q^{\ddagger}$ for the two transitions show that the most probable TS is located close to the $R$ state in both the $T \rightarrow R$ and the $R \rightarrow R^{\prime\prime}$ transitions. This is, of course, in accord with the Hammond postulate.

It is worth noting that disorder in the TSE structures (Fig. \ref{TSEGroEL}) is largely localized in the apical domain domain, which shows that the substructures in this domain partially unfold as the barrier crossings occur. By comparison the equatorial domain remains more or less structurally intact even at the transition state,  which suggests that the relative immobility of this domain is crucial to the function of this biological nanomachine. The dispersions in the TSE are also reflected in the heterogeneity of the distances between various salt-bridges in the transition states. The values of the contact distances, in the $T \rightarrow R$ transition among the residues involved in the salt-bridge switching between K80, R197, and E386 at the TS has a very broad distribution (Fig. \ref{TSEGroEL}B), which also shows that the R197-E386 is at least partially disrupted in the TS and the K80-E386 is partially formed \cite{Danziger03PNAS}.

\subsection{Link between allosteric transitions and function: Iterative Annealing Mechanism}
The GroEL allosteric dynamics  reveals that by breaking a number of salt bridges  the microenvironment that the SP experiences  changes continuously during the catalytic cycle. Upon ensnaring the SP,  the SP-GroEL complex is (marginally) stabilized predominantly by hydrophobic interactions although electrostatic interactions also are relevant. However, during the subsequent ATP-consuming and the irreversible step  R$\rightarrow$$R^{\prime\prime}$ transition not only does the volume double but also the microenvironment of the SP is largely polar. Thus, during a single catalytic cycle the microenvironment that the SP experiences changes from being hydrophobic to polar. The change in the SP microenvironment is the molecular  mechanism by which  GroEL anneals (resolves the incorrect interactions in the misfiled states) the SP stochastically by placing it from one region, in which the misfolded SP is trapped, to another region in the energy landscape.  In this process the SP could, with some probability ($\propto \Phi$ in Fig. \ref{fig:hemicycle}), reach the folded state \cite{Todd96PNAS}. The cycle of hydrophobic to polar change takes place with each catalytic cycle, and hence the GroEL-GroES machine iteratively anneals the misfolded SP enabling it to fold.

The physical picture of the IAM in which the coupling between allosteric transitions and the ability to facilitate folding, described above qualitatively,  has been translated into a set of kinetic equation with the express purpose of quantitatively describing the kinetics of chaperonin-assisted folding of the stringent {\it in vitro} substrates, such as Rubisco \cite{Tehver08JMB}. According to the IAM theory (see Fig. \ref{fig:hemicycle}) in each cycle, corresponding to the completion of $T\rightarrow R$ and $R\rightarrow R^{\prime\prime}$ transitions, the SP folds by the Kinetic Partitioning Mechanism (KPM) \cite{Guo95BP}. The KPM shows that a fraction, $\Phi$ (Fig. \ref{fig:hemicycle}), referred to as the partition factor, reaches the native state with each turnover. In the context of assisted folding it implies that with each round of folding, the fraction of folded molecules is $\Phi$ and the remaining fraction gets trapped in one of the many misfolded structures. 
After $n$ such cycles or iterations the yield of the native state is \cite{chakrabarti2017PNAS},
\begin{align}
\Psi (n)&= \frac{\Phi-(1-\kappa)^n(1-\Phi)^n\Phi}{\kappa+(1-\kappa)\Phi}\nonumber\\
&=\Lambda_{ss}(\kappa)[1-(1-\kappa)^n(1-\Phi)^n],
\end{align}
where $\kappa$ is approximately the ratio of two rate constants associated with chaperonin-induced unfolding of native and misfolded states, $\kappa\approx k_{\text{N}\rightarrow \text{I}}/k_{\text{M}\rightarrow \text{I}}$, and $\Lambda_{ss}(\kappa)=\Phi/(\kappa+(1-\kappa)\Phi)$ is the steady state yield. For the simple case of $\kappa \approx 0$, when the chaperone recognizes only the misfolded states, it follows that  
\begin{align}
\Psi (n) \approx 1-(1-\Phi)^n. 
\end{align}
The mathematical model \cite{Tehver08JMB}, based on the IAM, accounts for all the available experimental data, and shows that  for Rubisco the partition factor $\Phi \approx 0.02$, which means that only about 2\% of the SP reaches the folded state in each cycle.  It appears that  the GroEL chaperonin is the only molecular machine in which the details of the allosteric transitions and how that affects function has been fairly completely worked out. We note {\it en passant} that other models, which ignore allosteric transitions, cannot account for experimental data at all.

\begin{figure}[h!]
\centering
 \includegraphics[width=0.85\textwidth]{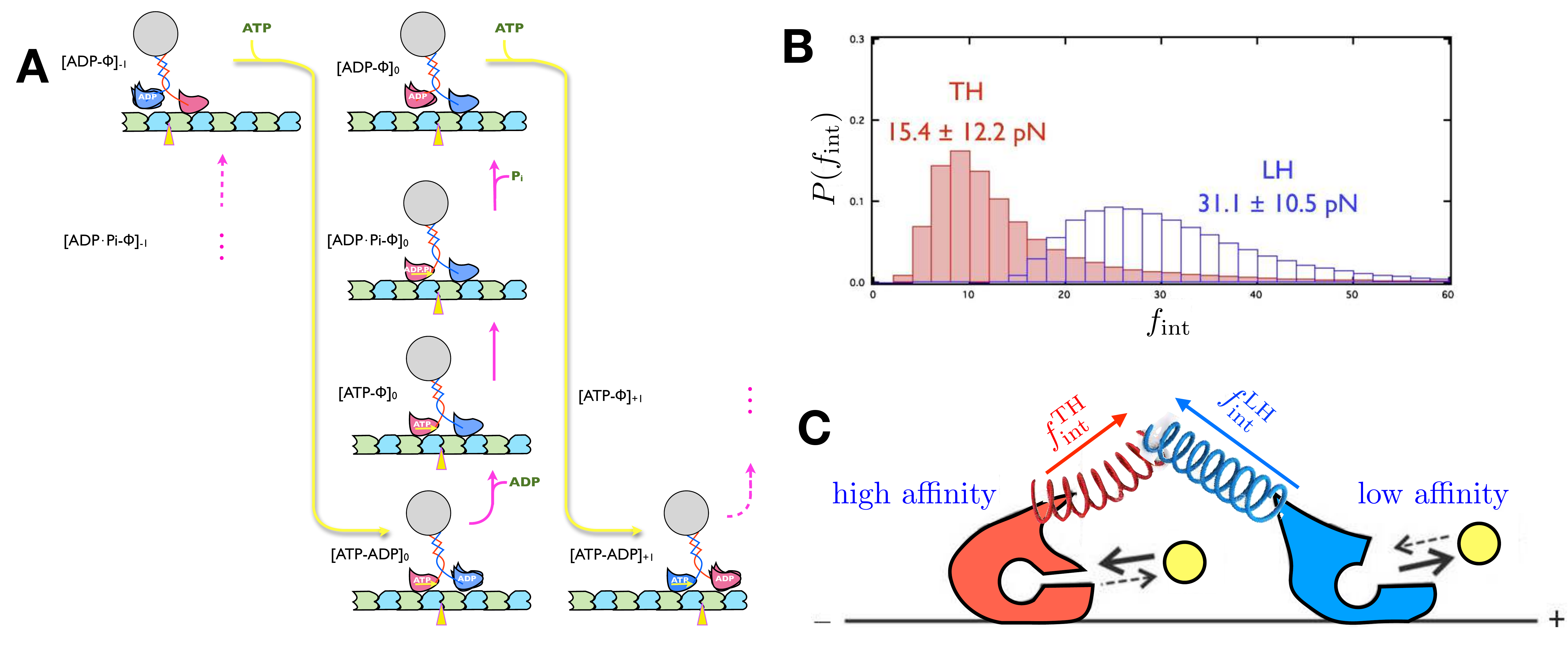}
   \caption{{\bf Inter-motor domain allosteric coordination of kinesin-1 motor by internal mechanical tension.} 
   (A) Chemomechanical cycle of kinesin-1. The perpendicular and horizontal directions are aligned with chemical and spatial coordinates, respectively.  The panel was adapted from Reference \cite{Hyeon11BJ}.
   (B) The distribution of tension in the  TH and the LH, calculated from Brownian dynamics simulations. This panel was adapted from Reference \cite{Zhang2012Structure}
   (C) The perturbation to the catalytic sites, induced by the mechanical tension resulting particularly in the neck-linker of the leading head (LH) inhibits a premature binding of ATP to the leading head (LH). The trailing head (TH), shown in red color, preserves a more native like conformation, whereas the catalytic site of the leading head (blue color) is disrupted. 
   \label{kinesin}} 
\end{figure}

\subsection*{ATP-consuming Molecular Motors}
Seminal examples of allostery, which are only starting to be investigated in detail, are found in a number of transport motors. 
Among them kinesin-1, known as the conventional kinesin, is a prototype of a bipedal transport motor in which the allosteric coordination between the  two motor domains is essential for maintaining a highly processive motion as it walks hand-over-hand predominantly on a single protofilament of the microtubules (MTs).  In order to take several steps without disengaging from the motor (maintain processivity), one of the two heads (motor-head domains) must remain bound to the MT surface at all times, while the other takes the roughly 8.2 nm step towards the + end of the MT consuming in the process one ATP molecule (Fig.~\ref{kinesin}A).  The binding affinity of kinesin motor domain to MTs changes as ATP is hydrolyzed in the catalytic site. 
Dissociation constants measured for a two-headed recombinant kinesin enzymes indicate that 
the nucleotide-free state binds most tightly, and the MT binding of the intermediate analogues, such as  AMPPNP, ADP$\cdot$AlF$_4$, ADP$\cdot$BeF$_4$, ADP$\cdot$Pi, and ADP, is weakened during the ATP processing with the ADP state being the weakest binding state \cite{Cross00PTRSL}.
To avoid a situation where both motor heads are in the ADP state, there must be a physical (gating) mechanism that keeps the enzymatic cycle of two heads always out of phase from each other. This too is an example of allostery. A quest to understand the molecular underpinnings and design principles of this head-head coordination is a recurring research topic in the  biological motor field.

While the tension on neck-linker was originally suggested for the out-of-phase coordination between the two motor head domains, two competing hypotheses could be considered: one is the facilitated detachment of the rear head induced by forward tension \cite{Hancock99PNAS}; the other is the rearward tension-induced inhibition of ATP binding to the leading head \cite{Uemura03NSB,BlockPNAS06}.  
A computational study of kinesin-1 suggests that, when both heads are bound to the MT binding site, the neck-linker of the leading head (LH) is stretched in the rearward direction and this in turn disrupts the catalytic site of the leading motor domain from its
native-like conformation \cite{Hyeon07PNAS}. Interestingly, the catalytic site of the trailing head (TH) maintains its native-like conformation.   
Given the extension of NL ($\delta x\approx 3.1\pm 0.8$ nm) inferred from molecular modeling and simulations and the contour length of NL ($\approx 5.7$ nm) consisting of 15 amino acid, 
the internal tension ($f_{\text{int}}$) built in the NL, 
based on the force-extension relationship of the worm-like chain model ($f_{\text{int}}=k_BT/l_p\times[1/4(1-\delta x/L)^2+\delta x/L-1/4]$), gives an estimate of $f_{\text{int}}\approx 8 - 15$ pN depending on the values of the
the persistence value ($l_p$) \cite{Hyeon07PNAS,Hyeon11BJ}. 
Zhang and Thirumalai estimated the NL tension for the leading and trailing heads, which are $\sim 31.1\pm 10.5$ pN and $\sim15.1\pm12.2$ pN, respectively \cite{Zhang2012Structure} (see Fig.~\ref{kinesin}B).  
Although the magnitude of the internal tension ($f_{\text{int}}$) clearly exceeds the stall force ($f_{\text{stall}}\approx 6-7$ pN), one should be aware that $f_{\text{stall}}$, externally applied to the motor through the tail domain, is discerned from the internal tension experienced by the neck-linker \cite{Hyeon11BJ}.  
The precise value of the shear force at the interface of the motor head domain and the MT surface in response to tension in the NL may be considerably smaller than the estimated value and is determined ultimately by the details of stress transmission from NL to the interface  \cite{Zhang2012Structure}. In executing the translational motion on MT, which is the function of kinesin, it is the tension that transmits allosteric signals in order to maintain processivity. Of course, the internal tension (a perturbation) would propagate along the structure (especially the neck linker), thus providing a structural basis for inter head communication.

The deformation of the catalytic site results in
the loss of ATP binding affinity to the catalytic site, thus inhibiting the premature binding of ATP to the LH until the chemical cycle of the TH reaches the ADP state, and allows the head to dissociate from the MT binding site via the change in chemical affinity for the MT from strong to weak. 
In addition to the inhibition of ATP binding, the disruption of the catalytic site in the leading head promotes the release of ADP \cite{Uemura03NSB}. 

In the case of kinesin-1, 
the out-of-phase head-head coordination is the outcome of the competition between the constraint arising from the interaction with MT surfaces and the driving force to shape its native state. 
As illustrated in Fig.~\ref{kinesin}C, the catalytic site of the leading head is in different conformation from that of the trailing head.  
The computational study by Hyeon and Onuchic \cite{Hyeon07PNAS} supported the experimental proposal \cite{Uemura03NSB,BlockPNAS06} that the rearward tension on the NL of the leading head regulates the out-of-phase coordination between the two motor heads.
This strategy adopted in kinesin-1 may be ubiquitous in other transport motors that consist of multiple domains \cite{purcell2005PNAS,baker2004PNAS,spudich2006Cell,Zhang2012Structure,mugnai2017PNAS,krementsova2017JBC}. 
These examples of inter head coordination in bipedal transport motors that are designed to function by exhibiting spectacular degree of conformational change provide a clear example that the allosteric communication in motors is mechanical in nature although force transmission is exquisitely sensitive to the molecular architecture. Although not discussed here, the molecular aspects of allostery in the cytoplasmic dynein, which occurs over a remarkably long length scale of $\approx$ 25 nm is just starting to be revealed using experiments \cite{Bhabha2014Cell} and computations. \cite{Cwik18Structure}

\section{Outlook}

The applications of computational tools along with comparisons to experiments described here show that allosteric signaling is complicated, especially if one seeks to understand the nuances in molecular terms. For this reason alone the venerable MWC and KNF models stand out as beacons in this important enterprise. Although going beyond these models, for which analytical solutions in terms of measurable binding constants could be developed, establishing their accuracy would require heroic experiments \cite{Viappiani14PNAS,Gruber16ChemRev}. It is probably for this reason one might insist that more could be done to quantitatively understand allostery even in Hemoglobin, which has been investigated for over fifty years. We conclude with the following additional comments:

\begin{itemize} 

\item
There may not be any general molecular principles for allosteric transition in biology, which might explain the many differences in the interpretation of mechanisms in specific systems. It does appear that the process of signaling very much depends on the architecture of the allosteric enzymes. Therefore, molecular understanding might arise only by studying one allosteric system at a time in detail. To decipher the molecular basis it is not only important to characterize the on and off rates of ligand binding  to various allosteric states but also obtain the dynamics of the structural changes between them.  For the latter, NMR experiments \cite{Lisi16ChemRev} seem most applicable although certain non-equilibrium aspects could be gleaned from spectroscopic methods as well.\cite{Buchli13PNAS} Although it is challenging to obtain complete molecular details of allostery, especially the dynamics of allosteric signaling, it is worth the effort to invent new methods because of potential uses of allostery concepts in drug design \cite{Wagner16ChemRev,Nussinov13Cell,Kumar15BJ}.

\item
It is inevitable that the experiments have to be paired with  computations in order to obtain molecular details. However, much needs to be done before the computations and simulations become reliable. Although uses of many methods, some of which are described here, have yielded qualitative insights in our understanding in allosteric transitions in specific enzymes, it is unclear if collectively they have made precisely testable predictions. The difficulty attests, perhaps, to the multifaceted aspects of allostery, which even in a single enzyme is complicated. 

\item
There are two general features that emerge from studying the dynamics of allosteric transitions using simulations.  (1) Computational studies probing the propagation of signals due to local strain show that the pathways connecting the allosteric states are highly heterogeneous involving substantial structural rearrangements even in monomeric enzymes with very little differences in the structures of the $T$ and $R$ states. 
 However, characterizing them in experiments might be hard, although high resolution cryo-EM experiments already hint at this possibility (see for example the recent report on GroEL \cite{Roh17PNAS}). Future studies are likely to clarify how prevalent this finding is and what the biological significance might be. 
(2) It is likely that, at least in multisubunit enzymes, salt-bridge formation and rupture is a universal structural mechanism governing allosteric signaling. In the two case studies (hemoglobin and GroEL) this appears to be the case. In GroEL the network of salt bridges serve as molecular switches whose rearrangements during the catalytic cycles are intimately related to function, and is the molecular basis of the IAM \cite{Todd96PNAS}. 
 Whether this is so in hemoglobin might still be a matter of debate \cite{Brunori15BJ}.

\item
We have focused exclusively on allosteric signaling in single enzymes. However, at the cellular level a cascade of reactions is triggered by changes in environmental cues, which transmit signals on length scales on the order of the cell size. The mitogen-activated protein kinases signaling pathway \cite{Chang01Nature,Qi05JCellSci}  is a well-known example in which cascade of reactions involving phophorylation by kinases and dephophorylation by phosphatases carry the signal down stream in order to regulate gene expression. This complex allosteric process occurs in a noisy environment. Although coarse-grained mathematical descriptions of the fidelity of information processing using network dynamics have been provided \cite{Lestas10Nature,Hinczewski14PRX}, the molecular underpinnings are completely unknown. It would be most interesting to investigate stochastic allosteric spreading in the simplest push-pull loop \cite{Koshland81PNAS}, which is a caricature of the more complicated cascade involving multiple enzyme reactions.     

\end{itemize}

{\bf Acknowledgements:} We acknowledge with gratitude discussions with Attila Szabo and William Eaton. We are grateful to Dr. Jie Chen, Prof. Ruxandra Dima, Prof. Zhenxing Liu, Prof. Riina Tehver, and Prof. Wenjun Zheng for collaborations and discussions.  This work was supported by a grant from the National Science Foundation (CHE 16-36424) and the Welch Foundation through the Collie-Welch Chair (F-0019).
	
\providecommand{\latin}[1]{#1}
\providecommand*\mcitethebibliography{\thebibliography}
\csname @ifundefined\endcsname{endmcitethebibliography}
  {\let\endmcitethebibliography\endthebibliography}{}

\end{document}